\documentclass[aps,prd,amsmath,floats,floatfix,
  twocolumn,
superscriptaddress,nofootinbib,showpacs]{revtex4-1}
\usepackage{soul}
\usepackage{newtxtext,newtxmath}
\usepackage{float}  
\usepackage{algorithm}
\usepackage{algpseudocode}
\usepackage{algorithmicx}
\usepackage{pifont}
\usepackage[pdfborder={0 0 0}, plainpages=false]{hyperref}
\usepackage{graphicx,adjustbox}
\usepackage{stmaryrd}
\usepackage{xspace}
\usepackage[usenames,dvipsnames]{xcolor}
\usepackage{amssymb}
\usepackage{tikz}
\usepackage[normalem]{ulem} 
\usepackage{mathtools}

\newcommand{\tv}[1]{\boldsymbol{#1}}

\usepackage{xparse}

\NewDocumentCommand{\dgal}{sO{}m}{%
  \IfBooleanTF{#1}
    {\dgalext{#3}}
    {\dgalx[#2]{#3}}%
}

\NewDocumentCommand{\dgalext}{m}{%
  \sbox0{%
    \mathsurround=0pt 
    $\left\{\vphantom{#1}\right.\kern-\nulldelimiterspace$%
  }%
  \sbox2{\{}%
  \ifdim\ht0=\ht2
    \{\kern-.45\wd2 \{#1\}\kern-.45\wd2 \}%
  \else
    \left\{\kern-.5\wd0\left\{#1\right\}\kern-.5\wd0\right\}%
  \fi
}

\NewDocumentCommand{\dgalx}{om}{%
  \sbox0{\mathsurround=0pt$#1\{$}%
  \sbox2{\{}%
  \ifdim\ht0=\ht2
    \{\kern-.45\wd2 \{#2\}\kern-.45\wd2 \}%
  \else
    \mathopen{#1\{\kern-.5\wd0 #1\{}
    #2
    \mathclose{#1\}\kern-.5\wd0 #1\}}
  \fi
}

\usepackage{booktabs}

\begin{document}

\title{
  A hp-adaptive discontinuous Galerkin solver for elliptic equations in numerical relativity
}

\newcommand{\Caltech}{\affiliation{TAPIR, Walter Burke Institute for Theoretical Physics, MC 350-17,
    California Institute of Technology, Pasadena, California 91125, USA}}
\newcommand{\Einstein}{\affiliation{NASA Einstein Fellow}}
\newcommand{\Cornell}{\affiliation{Cornell Center for Astrophysics and
    Planetary Science, Cornell University, Ithaca, New York, 14853, USA}}
\newcommand{\WSU}{\affiliation{Department of Physics \& Astronomy,
	Washington State University, Pullman, Washington 99164, USA}}
\newcommand{\CITA}{\affiliation{Canadian Institute for Theoretical 
    Astrophysics, University of Toronto, Toronto, Ontario M5S 3H8, Canada}}
\newcommand{\UofT}{\affiliation{Department of Physics,
    University of Toronto, Toronto, Ontario, M5S 3H5, Canada}}
\newcommand{\CIFAR}{\affiliation{Canadian Institute for Advanced Research, 180 Dundas St.~West, Toronto, ON M5G 1Z8, Canada}} %
\newcommand{\LBL}{\affiliation{Nuclear Science Division, Lawrence Berkeley National Laboratory,
1 Cyclotron Rd, Berkeley, CA 94720, USA}}
\newcommand{\NCSU}{\affiliation{Department of Physics, North Carolina State University, Raleigh, North Carolina 27695, USA; Hubble Fellow}}
\newcommand{\AEI}{\affiliation{Max Planck Institute for Gravitational Physics (Albert Einstein Institute), Am M\"uhlenberg 1, Potsdam 14476, Germany}}
\newcommand{\Maryland}{\affiliation{Department of Physics, University of Maryland, College Park, MD 20742, USA}}
\newcommand{\UCBAstro}{\affiliation{Astronomy Department and Theoretical Astrophysics Center, 
University of California, Berkeley, 601 Campbell Hall, Berkeley CA, 94720}}
\newcommand{\UCBPhysics}{\affiliation{Department of Physics, University of California, Berkeley, Le Conte Hall, Berkeley, CA 94720}}
\newcommand{\UNH}{\affiliation {Department of Physics, University of New Hampshire, 9 Library Way, Durham NH 03824, USA}}
\newcommand{\NCSA}{\affiliation{NCSA, University of Illinois at Urbana-Champaign, Urbana, Illinois, 61801, USA}} %
\newcommand{\RU}{\affiliation{Department of Astrophysics/IMAPP, Radboud University Nijmegen, P.O. Box 9010, 6500 GL Nijmegen, The Netherlands}}
\newcommand{\GRAPPA}{\affiliation{GRAPPA, Anton Pannekoek Institute for Astronomy and Institute of High-Energy Physics, University of Amsterdam, Science Park 904, 1098 XH Amsterdam, The Netherlands}}
\newcommand{\DeltaITP}{\affiliation{Delta Institute for Theoretical Physics, Science Park 904, 1090 GL Amsterdam, The Netherlands}}
\newcommand{\Nikhef}{\affiliation{Nikhef, Science Park 105, 1098 XG Amsterdam, The Netherlands}}

\author{Trevor Vincent} \CITA\UofT
\author{Harald P. Pfeiffer} \AEI\CITA
\author{Nils L. Fischer} \AEI

\date{\today}

\begin{abstract}
A considerable amount of attention has been given to discontinuous
Galerkin methods for hyperbolic problems in numerical relativity,
showing potential advantages of the methods in dealing with
hydrodynamical shocks and other discontinuities.  This paper
investigates discontinuous Galerkin methods for the solution of
elliptic problems in numerical relativity.  We present a novel
hp-adaptive numerical scheme for curvilinear and non-conforming
meshes.  It uses a multigrid preconditioner with a Chebyshev or
Schwarz smoother to create a very scalable discontinuous Galerkin code
on generic domains.  The code employs compactification to move the
outer boundary near spatial infinity.  We explore the properties of
the code on some test problems, including one mimicking Neutron stars
with phase transitions.  We also apply it to construct initial data
for two or three black holes.
\end{abstract}

\maketitle

\section{Introduction}

Discontinuous Galerkin (DG) methods~\cite{Reed.W;Hill.T1973,hesthaven2008nodal, Cock01,cockburn1998runge,Cockburn.B1998,Cockburn.B;Karniadakis.G;Shu.C2000}
have matured into standard numerical methods  to simulate a wide
variety of partial differential equations. In the context of numerical
relativity~\cite{baumgarte2010numerical}, discontinuous Galerkin
methods have shown advantages for relativistic hyperbolic problems
over traditional discretization methods such as finite difference, finite volume and
spectral finite elements
\cite{Field:2010mn,brown2012numerical,field2009discontinuous,zumbusch2009,Radice:2011qr,mocz:14,zanotti:14,endeve:15,teukolsky2015,Bugner:2015gqa,schaal2015astrophysicalfixed,zanotti2015,Miller:2016vik,kidder:16,fambri2018} by combining the best aspects of all three methods. As computers reach exa-scale power, new methods like DG are needed to tackle the biggest problems in numerical relativity such as realistic supernovae and binary neutron-star merger simulations on these very large machines~\cite{kidder:16}.

DG efforts in numerical relativity have so far
targeted evolutionary problems \cite{Field:2010mn,brown2012numerical,field2009discontinuous,zumbusch2009,Radice:2011qr,teukolsky2015,Bugner:2015gqa,schaal2015astrophysicalfixed,zanotti2015,Miller:2016vik,fambri2018}. Radice and Rezzolla \cite{Radice:2011qr} showed for spherically symmetric problems that the discontinuous Galerkin method could handle strong relativistic shock waves while maintaining exponential convergence rates in smooth regions of the flow. Teukolsky \cite{teukolsky2015} showed how to develop discontinuous Galerkin methods for applications in relativistic astrophysics. Bugner et al. \cite{Bugner:2015gqa} presented the first three-dimensional simulations of general relativistic hydrodynamics with a fixed spacetime background using a discontinuous Galerkin method coupled with a WENO algorithm. Kidder et al. \cite{kidder:16} showcased the first discontinuous Galerkin code to use a task-based parallelism framework for applications in relativisitic astrophysics. Kidder et al. tested the scalability and convergence of the code on relativistic magnetohydrodynamics problems. Miller et al. \cite{Miller:2016vik} developed a discontinuous Galerkin operator method for use in finite difference codes and used it to solve gauge wave problems involving the BSSN formulation of the Einstein field equations. Hebert et al. \cite{hebert:2018xbk} presented the first discontinuous Galerkin method for evolving neutron stars in full General Relativity. Finally, Fambri et al. \cite{fambri2018} used an ADER discontinuous Galerkin scheme to solve general relativistic ideal magnetohydrodynamics problems in fixed spacetimes. They compared their DG method to a finite volume scheme and showed that DG is much more efficient.

This paper explores DG for elliptic problems in numerical relativity.
We develop  an elliptic solver with the following primary features:
(i) It operates on curved meshes, with non-conforming elements.
(ii) It supports adaptive h and p refinement, driven by a
  posteriori error estimators.
(iii) It employs multi-grid for efficient solution of the resulting
  linear systems.
(iv) It uses compactified domains to treat boundary conditions at infinity.
  While each of these features has appeared in the literature before
\cite{arnold.d;brezzi.f;cockburn.b;marini.l2002,stiller2017robust,hesthaven2008nodal,kronbichler2018performance,fick2014interior,kozdon2018energy,kozdon2019robust}, to our knowledge, our solver for the first
time combines all these elements simultaneously and demonstrate their
effectiveness on difficult numerical problems.

Specifically, the present article is a step
toward a solver for the Einstein constraint
equations, which must be solved to
construct initial data for the evolution of compact binary
systems~\cite{pfeiffer:2005,cook2000,baumgarte2010numerical}.  The
constraint equations are generally rewritten as elliptic equations,
and depending on detailed assumptions, this results in one or more
coupled non-linear elliptic partial differential equations.
Construction of initial data is arguably the most important
elliptic problem in numerical relativity, but not the only one:
Elliptic equations also occur in certain gauge
conditions~\cite{baumgarte2010numerical} or for implicit
time-stepping to alleviate the computational cost of high-mass-ratio
binaries~\cite{laupfeiffer2008,lau:2011we}
.

The motivation for developing
a new solver is multi-fold. First, current spectral methods have
difficulty obtaining certain initial data sets, such as binaries at
short separation containing a neutron star, where the neutron star has
high compactness and a realistic equation of state
\cite{henriksson:2014tba}.  Furthermore, there is a need for a
solver which can routinely obtain high-accuracy. Errors from
inaccurate initial data sets creep into the evolutions with sizeable
effect: Ref.~\cite{tsokaros2016initialfixed} shows that despite only
global (local) differences of $0.02\%$ ($1\%$) in the initial data of
the two codes COCAL and LORENE for irrotational neutron-star binaries,
the gravitational wave phase at the merger time differed by 0.5~radians after 3 orbits. The design of a more accurate code requires adaptive mesh refinement, load-balancing and
scalability which a DG code potentially can
provide. The present work also complements the DG evolution code presented in Ref.~\cite{kidder:16},
leading to a complete framework for solving both
elliptic and hyperbolic PDEs in numerical relativity.

The organization of the paper is as follows: 
Section~\ref{sec:numericalalgorithm} presents the components of our
  discontinuous Galerkin code. Section~\ref{sec:testexamples} 
  showcases our hp-adaptive multigrid solver on  increasingly challenging test-problems, each
  illustrating the power of the discontinuous Galerkin method. This section includes a solution for the Einstein constraint equations in the case of a constant density star. This problem has a surface discontinuity which mimicks Neutron stars with phase transitions. Lastly, this section also presents solutions for
  initial-data of two orbiting non-spinning black-holes to
  showcase and compare it with the elliptic solver Spells~\cite{pfeiffer2003}
 as well as solutions for the initial data
  of three black-holes with random locations, spins and momenta.  We
  close with a discussion in Sec.~\ref{sec:Conclusions}.

\section{Numerical Algorithm}
\label{sec:numericalalgorithm}

\subsection{DGFEM discretization}
\label{sec:DGFEM}

We base our nomenclature on \cite{arnold.d;brezzi.f;cockburn.b;marini.l2002,sherwin20062d,di2011mathematical,fick2014interior,kozdon2018energy,kozdon2019robust}, which we summarize here for completeness.

\subsubsection{Mesh}
\label{sec:mesh}

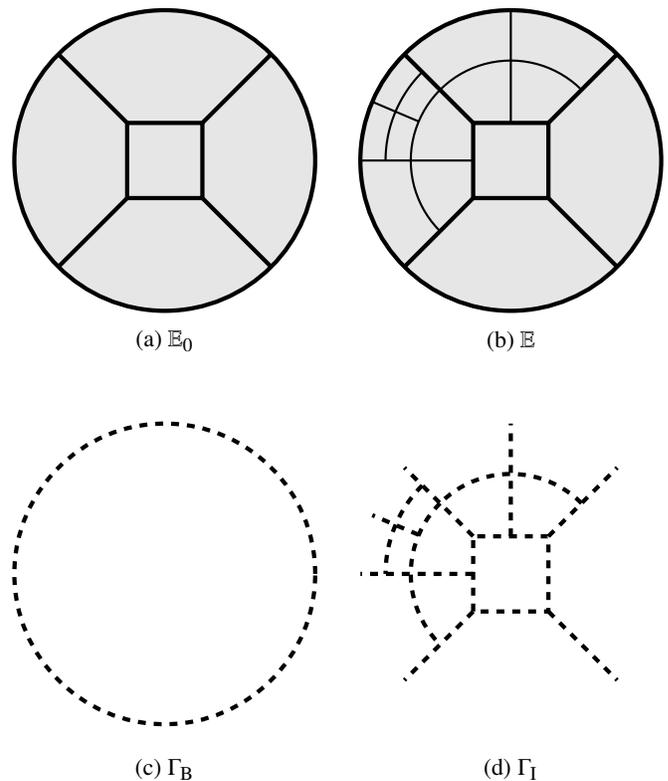
\begin{figure}
  \centering
  \begin{tikzpicture}[node distance=.25cm, auto]
    \draw [black, ultra thick, fill=gray!20] (-2,0) circle [radius=2];
    \draw [black, ultra thick, fill=gray!20] (2.6,0) circle [radius=2];
    \node[align=center,fill=white] at (+2.6,-2.4) {(b) $\mathbb{E}$};
    \draw [black, ultra thick] (-1.5,.5) -- (-.58578643763,1.41421356237);
    \draw [black, ultra thick] (-2.5,-.5) -- (-3.41421356237,-1.41421356237);
    \draw [black, ultra thick] (-2.5,+.5) -- (-3.41421356237,+1.41421356237);
    \draw [black, ultra thick] (-1.5,-.5) -- (-.58578643763,-1.41421356237);
    \draw [black, ultra thick] (-2.5,-.5) rectangle (-1.5,.5);
    \draw [black, ultra thick] (3.1,.5) -- (4.014213562369999,1.41421356237);
    \draw [black, ultra thick] (2.1,-.5) -- (1.1857864376299996,-1.41421356237);
    \draw [black, ultra thick] (2.1,+.5) -- (1.1857864376299996,+1.41421356237);
    \draw [black, ultra thick] (3.1,-.5) -- (4.014213562369999,-1.41421356237);
    \draw [black, ultra thick] (2.1,-.5) rectangle (3.1,.5);

    \draw [black , thick, domain=1.33:2] plot ({2.6 + \x*cos(157)}, {\x*sin(157)});
    \draw [black , thick, domain=135:180] plot ({2.6 + 1.665*cos(\x)}, {1.665*sin(\x)});
    
    \node[align=center,fill=white] at (-2,-2.4) {(a) $\mathbb{E}_0$};

    \draw [black, thick] (2.1,0) -- (.6,0);
    \draw [black , thick, domain=135:225] plot ({2.6 + 1.33*cos(\x)}, {1.33*sin(\x)});
    \draw [black , thick, domain=45:135] plot ({2.6 + 1.33*cos(\x)}, {1.33*sin(\x)});
    \draw [black , thick, domain=.5:2] plot ({2.6 + \x*cos(90)}, {\x*sin(90)});
    
    \draw [black, ultra thick, dashed] (2.1,-5.5) -- (.6,-5.5);
    \draw [black , ultra thick, domain=135:225, dashed] plot ({2.6 + 1.33*cos(\x)}, {-5.5 + 1.33*sin(\x)});
    \draw [black , ultra thick, domain=45:135, dashed] plot ({2.6 + 1.33*cos(\x)}, {-5.5 + 1.33*sin(\x)});
    \draw [black , ultra thick, domain=.5:2, dashed] plot ({2.6 + \x*cos(90)}, {-5.5 + \x*sin(90)});    
    
    \draw [black , ultra thick, domain=1.33:2, dashed] plot ({2.6 + \x*cos(157)}, {-5.5 + \x*sin(157)});
    \draw [black , ultra thick, domain=135:180, dashed] plot ({2.6 + 1.665*cos(\x)}, {-5.5 + 1.665*sin(\x)});
    
    \draw [black, ultra thick, dashed] (3.1,-5) -- (4.014213562369999,-4.08578643763);
    \draw [black, ultra thick, dashed] (2.1,-6) -- (1.1857864376299996,-6.91421356237);
    \draw [black, ultra thick, dashed] (2.1,-5) -- (1.1857864376299996,-4.08578643763);
    \draw [black, ultra thick, dashed] (3.1,-6) -- (4.014213562369999,-6.91421356237);
    \draw [black, ultra thick, dashed] (2.1,-6) rectangle (3.1,-5);
    \node[align=center,fill=white] at (2.6,-8.1) {(d) $\Gamma_{\rm I}$};
    \draw [black, ultra thick, dashed] (-2.0,-5.5) circle [radius=2];
        \node[align=center,fill=white] at (-2.0,-8.1) {(c) $\Gamma_{\rm B}$};
\end{tikzpicture} 
\caption{\label{fig:macromesh}  Ingredients into the setup of the domain-decomposition: (a) The
  macro-mesh $\mathbb{E}_0$ of a 2-dimensional disk consisting of five
  macro-elements. (b) A representative mesh derived from the
  macro-mesh by refining once in the left-most macro-element.  (c) The
  boundary mortar $\Gamma_{\rm B}$. (d) The interior mortar $\Gamma_{\rm I}$.
}
\end{figure}

Our purpose is to solve elliptic equations in a computational domain
$\Omega \subset \mathbb{R}^d$ in $d$ dimensions.  To allow for
non-trivial topology and shape of $\Omega$, we introduce a
\emph{macro-mesh} (also known as a multi-block mesh) $\mathbb{E}_0$ consisting of macro-elements (also known as blocks) $e_0 \in
\mathbb{E}_0$ such that (1) the macro-elements cover the entire
domain, i.e., $\cup_{e_0\in \mathbb{E}_0}e_0=\Omega$, (2) the
macro-elements touch each other at complete faces and do not overlap;
and (3) each $e_0 \in \mathbb{E}_0$ is the image of the reference cube
$[-1,1]^{d}$ under a mapping
  $\Phi_{e_0}:[-1,1]^d\rightarrow e_0$.  As an example,
Fig.~\ref{fig:macromesh} shows a macro-mesh of five elements covering a
two-dimensional disk.  The macro-mesh represents the coarsest level of
subsequent mesh-refinement.

The macro-mesh $\mathbb{E}_0$ is refined by subdividing macro-elements
into smaller elements along faces of constant reference cube
  coordinates. Refinement can be multiple levels deep (i.e.\ refined
elements can be further subdivided), and we do not assume uniform
refinement.  The refined mesh $\mathbb{E}$ is referred to as the
\emph{fine mesh}, which is exemplified in panel (b) of
Fig.~\ref{fig:macromesh}.  In contrast to the macro-mesh, the fine
mesh will generally be non-conforming at element boundaries, both
within one macro-element and at boundaries between macro-elements. We
assume that there is at most a 2:1 refinement difference at any element
boundary, i.e.\ the boundary of a coarse element faces at most two
smaller elements (per dimension). Each face of each element $e\in
\mathbb{E}$ is endowed with an outward-pointing unit-normal $\Hat n$.
The map $\Phi_{e_0}$ in macro-element $e_0\in \mathbb{E}_0$
  induces a map on each fine element $e$ within $e_0$, denoted
\begin{equation}\label{eq:Phi_e}
\Phi_e: [-1,1]^d\to e.
\end{equation}
The reference coordinates of $\Phi_e$ are
linearly related to the reference coordinates of $\Phi_{e_0}$.

Turning to boundaries, we define the set of all element boundaries (internal and external), $\Gamma = \cup_{e \in \mathbb{E}} \partial e $, where $\partial e$
is the boundary of element e of the fine mesh.  $\Gamma$ is called the \emph{mortar}, and it decomposes into a finite set of $(d-1)$-dimensional mortar
elements $m\in \mathbb{M}$, arising from the faces of each mesh element $e\in \mathbb{E}$, s.t. $\Gamma=\cup_{m\in\mathbb{M}}m$ and each mortar element intersects at most at the boundary of two elements.  
$\mathbb{M}$ splits into \textit{interior mortar elements}
$\mathbb{M}_{\rm I}$ and \textit{exterior mortar elements}
$\mathbb{M}_{\rm B}$, with $\mathbb{M}_{\rm I}\cap\mathbb{M}_{\rm B}=0$. 
Similarly, $\Gamma$ splits into
\textit{interior mortar} $\Gamma_{\rm I}=\cup_{m\in\mathbb{M}_I}m$ and \textit{exterior mortar} $\Gamma_{\rm B}=\cup_{m\in\mathbb{M}_{\rm B}}m$,
  cf.~Fig.~\ref{fig:macromesh}, which intersect
  at $(d-2)$-dimensional edges where the interior mortar touches the exterior boundary.
    We partition the exterior mortar elements further into elements where we
  apply Dirichlet boundary conditions, $\mathbb{M}_{\rm D}$, Neumann
  boundary conditions, $\mathbb{M}_{\rm N}$, and Robin boundary conditions, $\mathbb{M}_{\rm R}$, respectively.  This induces sets of
  boundary points via $\Gamma_{\rm D}=\cup _{m \in \mathbb{M}_{\rm D}} m$, $\Gamma_{\rm N}=\cup _{m \in \mathbb{M}_{\rm N}} m$ and $\Gamma_{\rm R}=\cup _{m \in \mathbb{M}_{\rm R}} m$.
  We assume $\Gamma_{\rm D}$, $\Gamma_{\rm N}$ and $\Gamma_{\rm R}$ are disjoint, i.e.\ one type of boundary condition is employed on each connected part of the boundary.
  Finally, we introduce two definitions to help us simplify equations later in the text. Firstly,
  because internal boundaries and external Dirichlet boundaries are
  often treated similarly, it is convenient to define $\mathbb{M}_{\rm
    ID}=\mathbb{M}_{\rm I}\cup \mathbb{M}_{\rm D}$ and $\Gamma_{\rm ID}=\Gamma_{\rm I}\cup\Gamma_{\rm D}$. 
Secondly, we refer to the set of mortar elements surrounding an element $e$ as $\mathbb{M}_e$.

\subsubsection{Weak Equations}
\label{sec:weakequations}

On the fine mesh $\mathbb{E}$, we wish to discretize the following model elliptic
problem with the symmetric interior penalty discontinuous Galerkin
method \cite{arnold.d;brezzi.f;cockburn.b;marini.l2002}:

\begin{subequations}\label{seq:modelproblem}
  \begin{align}
    \label{eq:Laplacian}
    \partial_i\partial_i u(\tv x) + f(\tv x)u &= g( \tv x )
    && \tv{x} \in \Omega, \\
           u &= g_D(\tv x)
    && \tv{x} \in  \Gamma_{\rm D},\\ 
           \label{eq:NeumannBC}
           \Hat n_i\partial_i u &= g_N(\tv x)
           && \tv{x}\in  \Gamma_{\rm N},  \\
           \Hat n_i\partial_i u + \gamma u &= g_R(\tv x) && \tv{x}\in  \Gamma_{\rm R}.
\label{eq:RobinBC}
  \end{align}
\end{subequations}
In Eqs.~(\ref{seq:modelproblem}) and in the following, we employ the
Einstein sum convention, so that the first term in
Eq.~(\ref{eq:Laplacian}) represents $\sum_{i=1}^d\partial_i\partial_iu$, and similar for the left hand side of Eqs.~(\ref{eq:NeumannBC})
and~(\ref{eq:RobinBC}).
In accordance with the underlying ideas of DG, we seek to approximate the
  solution of Eqs.~(\ref{seq:modelproblem}) with polynomials on each element $e\in \mathbb{E}$, without strictly enforcing continuity across elements. 
  Let $V_{h,e,p_e}$ denote the set of polynomials 
  on the reference cube $[-1,1]^d$ up to degree $p_e$ mapped to element $e\in \mathbb{E}$.
We assume the same maximum polynomial order along each dimension; it is straightforward to extend to different polynomial order along different dimensions.
  The functions in $V_{h,e,p_e}$ are understood to be extended by $0$ outside of $e$ (i.e.\ on other elements).   The global function space is the direct sum of the per-element polynomial spaces,
  \begin{equation}
    \label{eq:Vh}
    V_h = \bigoplus_{e \in \mathbb{E}}V_{h,e,p_e},
  \end{equation}
  where the polynomial order may vary between elements.
  Because
  neighboring element touch, on internal boundaries $\Gamma_{\rm I}$
  the discretized solution $u_h\in V_h$ will be represented
  twice on touching elements with generally different values on either element
  (this is origin of the term 'discontinuous' in DG).

The discretized solution $u_h\in V_h$ is determined, such that
  the residual of Eqs.~(\ref{eq:Laplacian}) is orthogonal to the function
  space $V_h$.  Within the symmetric interior penalty
  discontinuous Galerkin discretization \cite{arnold.d;brezzi.f;cockburn.b;marini.l2002,di2011mathematical}, this yields
\begin{equation}
  L_h(u_h,v_h) = F_h(v_h) \qquad  \forall v_h \in V_h,
\end{equation}
where
\begin{align} 
  \label{eq:lh}
 \nonumber
 L_h(u,v)
 =&
 \int\displaylimits_{\mathbb{E}}  \partial_i v\partial_i u \,\mathrm{d}\tv{x}
 - \int\displaylimits_{\Gamma_{\rm ID}} \llbracket u \rrbracket_i \dgal*{\partial_i v}\mathrm{d}\tv s \\ \nonumber
 &
 - \int\displaylimits_{\Gamma_{\rm ID}} \llbracket v \rrbracket_i \dgal*{\partial_i u} \mathrm{d}\tv s
+\int\displaylimits_{\Gamma_{\rm ID}}\sigma \llbracket u \rrbracket_i \llbracket v \rrbracket_i \mathrm{d}\tv s  \\
&
+ \int\displaylimits_{\mathbb{E}} f u v \,\mathrm{d}\tv{x} + \int\displaylimits_{\Gamma_R} \gamma u v \,\mathrm{d}\tv s ,
\end{align}
%
and

\begin{align}
  \label{equationFh}
\nonumber
F_h(v) =&
\int\displaylimits_\mathbb{E} gv \mathrm{d}x
- \int\displaylimits_{\Gamma_D}g_D \hat n_i \partial_i v \mathrm{d}\tv s + \int\displaylimits_{\Gamma_D}\sigma g_D v \mathrm{d}\tv s \\
&+ \int\displaylimits_{\Gamma_N}g_Nv\mathrm{d}\tv s -  \int\displaylimits_{\Gamma_{R}}g_Rv \mathrm{d}\tv s.
\end{align}
%

For internal boundaries, the operators $\llbracket\,.\, \rrbracket$ and $\dgal*{\,.\,}$ are
 defined by
\begin{align}
\label{eq:20}
  \llbracket q \rrbracket\; &= q^{+}\tv n^{+} + q^{-}\tv n^{-} \quad\mbox{on $\Gamma_I$}, \\
  \dgal*{q} &= \frac{1}{2}\left( q^{+} + q^{-}\right) \quad\;\;\mbox{on $\Gamma_I$}.
\end{align}
Here $q$ is a scalar function, '+' and '-' is an arbitrary labeling of the two elements $e^+$ and $e^-$ touching at the interface, and 
$q^\pm$ and $\tv n^\pm$ are the function values and the outward
pointing unit normal on the two elements that share the
interface. These operators are extended to external boundaries by
\begin{align}
  \llbracket q\rrbracket\; &= q\tv n\quad\;\mbox{on $\Gamma_B$},\\
  \dgal*{q} &= q\quad\;\;\;\,\mbox{on $\Gamma_B$}.
\end{align}

Breaking up the integrals in Eqs.~(\ref{eq:lh}) and~(\ref{equationFh}) into integrals over individual mesh-- and
mortar--elements, one finds

\begin{align} 
  \label{eq:lh2}
 \nonumber
 L_h(u,v)
 =&
 \sum_{e\in\mathbb{E}}\int\displaylimits_{e}  \partial_i v\partial_i u \,\mathrm{d}\tv{x}
 - \sum_{m \in \mathbb{M}_{\rm ID}}\int\displaylimits_m \llbracket u \rrbracket_i \dgal*{\partial_i v}\mathrm{d}\tv s \\ \nonumber
 &
 - \sum_{m \in \mathbb{M}_{\rm ID}}\int\displaylimits_m \llbracket v \rrbracket_i \dgal*{\partial_i u} \mathrm{d}\tv s  
 +\sum_{m \in \mathbb{M}_{\rm ID}}\int\displaylimits_m\sigma \llbracket u \rrbracket_i \llbracket v \rrbracket_i \mathrm{d}\tv s \\ & -\sum_{m \in \mathbb{M}_{R}}\int\displaylimits_m\gamma u v \mathrm{d}\tv s 
 + \sum_{e\in\mathbb{E}}\int\displaylimits_{e}f u v \,\mathrm{d}\tv{x},
\end{align}
and
\begin{align}
  \label{equationFh2}
\nonumber
F_h(v)
=& \sum_{e\in\mathbb{E}}\int\displaylimits_{e}\! gv \mathrm{d}x
- \!\sum_{m \in \mathbb{M}_{\rm D}}\int\displaylimits_m \!g_D \hat n_i \partial_i v \mathrm{d}\tv s
+\! \sum_{m \in \mathbb{M}_{\rm D}}\int\displaylimits_m\! \sigma g_D v \mathrm{d}\tv s \\
  &+
\sum_{m \in \mathbb{M}_{\rm N}}\int\displaylimits_m g_Nv\mathrm{d}\tv s - \sum_{m \in \mathbb{M}_{\rm R}}\int\displaylimits_m g_Rv\mathrm{d}\tv s.
\end{align}

For later reference, we will refer to the first four integrals in
Eq.~(\ref{eq:lh2}) as the stiffness, consistency, symmetry and
penalty integrals respectively and denote them
$L^{\text{stiff}}_h(u,v)$, $L^{\text{con}}_h(u,v)$,
$L^{\text{sym}}_h(u,v)$ and $L^{\text{pen}}_h(u,v)$ so that we may
write
\begin{align}
  \nonumber
  \label{eq:lh3}
  L_h(u,v) = &L^{\text{stiff}}_h(u,v) + L^{\text{con}}_h(u,v) + L^{\text{sym}}_h(u,v) 
  + L^{\text{pen}}_h(u,v)\\
  &+ \sum_{e\in\mathbb{E}}\int\displaylimits_{e}f u v \,\mathrm{d}\tv{x} - \int\displaylimits_{\Gamma_R} \gamma u v \,\mathrm{d}\tv s.
\end{align}
%

\subsubsection{Basis Functions}

So far, we have not yet specified a concrete basis for the polynomial spaces $V_{h,e,p_e}$ introduced in Sec.~\ref{sec:weakequations}. We do so now.
Recall that curvilinear elements $e \in \mathbb{E}$ are mapped to reference
cubic elements $[-1,1]^d$.
That is, each point $\tv x\in e$ corresponds to a reference cube coordinate $\tv\xi =\Phi_e^{-1}(\tv x)$, cf. Eq.~(\ref{eq:Phi_e}).

Along each dimension $\xi_i\in[-1,1]$, where the subscript
  $i\in\{1,\ldots,d\}$ denotes dimension, we first choose
$N_i + 1$ Legendre-Gauss-Lobatto collocation points
  $\xi^\mathrm{LGL}_{\alpha_i}$. The index
  $\alpha_i\in\{1,\ldots,N_i+1\}$ identifies the point along dimension
  $i$. We take $N_i\geq 1$ so that the points with $\xi_i=-1$ and
  $\xi_i=1$, that lie on the faces of the cube, are always collocation
  points. The collocation points in all $d$ dimensions form a tensor
  product grid of $\prod_{i=1}^d (N_i + 1)$ $d$-dimensional
  collocation points
  $\tv\xi^\mathrm{LGL}_\alpha=(\xi^\mathrm{LGL}_{\alpha_1},\ldots,\xi^\mathrm{LGL}_{\alpha_d})$
  that we index by $\alpha\in\{1,\ldots,\prod_{i=1}^d (N_i + 1)\}$
  that identifies a point regardless of dimension. With respect to
  these collocation points we can now construct the one-dimensional
  interpolating Lagrange polynomials
\begin{equation}
  l_{\alpha_i}(\xi) := \prod_{\substack{\beta_i=1\\\beta_i\neq\alpha_i}}^{N_i+1} \frac{\xi-\xi^\mathrm{LGL}_{\beta_i}}{\xi^\mathrm{LGL}_{\alpha_i} - \xi^\mathrm{LGL}_{\beta_i}}, \quad \xi\in[-1,1]
\end{equation}
and employ their tensor product to define the $d$-dimensional basis functions
\begin{equation}
\label{eq:Lagrange_Basis}
\psi_\alpha(\tv \xi) = \prod_{i=1}^d l_{\alpha_i}(\xi_i) \text{.}
\end{equation}

    When evaluating $\psi_\alpha$ in the physical coordinates $\tv x\in e$, one uses $\Phi_e^{-1}:\tv x\to \tv \xi$ to map the function argument. For instance, the set of test-functions on element $e\in \mathbb{E}$ becomes
  \begin{equation}
  V_{h,e,p_e}=\mbox{span}\left\{\,\Phi_e^{-1}\circ\psi_\alpha\,\right\}.
  \end{equation}
  Furthermore, because the $\psi_\alpha$ form a nodal basis, the expansion coefficients are the function values at the nodal points $\tv\xi^\alpha$, and each test-function can be written as
  \begin{equation}
    u^e_h = \sum_{\alpha}u^e_h\big(\tv \xi_\alpha^{\textrm{LGL}}\big)\psi_\alpha = \sum_{\alpha}u^e_\alpha\psi_\alpha,
    \end{equation}
where $u^e_\alpha:= u^e_h\big(\tv \xi_\alpha^{\textrm{LGL}}\big)$.
  \subsubsection{Semi-Discrete Global Matrix Equations}

The global solution over the entire mesh is the direct sum of the solutions on each element, that is
\begin{equation}\label{eq:GlobalExpansion}
  u_h = \bigoplus_{e \in \mathbb{E}} u^e_h = \bigoplus_{e \in \mathbb{E}} \sum_{\alpha}u^e_\alpha\psi^e_\alpha.
\end{equation}
Thus, with a chosen global ordering of elements $e \in \mathbb{E}$ and local ordering of basis functions $\psi^e_\alpha$, we may prescribe a global index $\alpha'$ to each of the local expansion coefficients $u^e_\alpha$.
With this, we may now turn Eqs.~\eqref{eq:lh2} and \eqref{equationFh2} into a global linear system of equations. Let $\mathbb{L} = L_h(\psi_{\alpha'},\psi_{\beta'}) =: L_{\alpha'\beta'}$, $\tv u = u_{\beta'}$ and $\tv F = F(\psi_{\alpha'}) =: F_{\alpha'}$, then the global linear system is

\begin{equation}\label{eq:semidiscrete}
\mathbb{L} \tv u = \tv F.
\end{equation}

Instead of forming the full global matrix $\mathbb{L}$ and performing the matrix-vector operator $\mathbb{L} \tv u$ over the global space, we can restrict the integrals in  \eqref{eq:lh2} and \eqref{equationFh2} to those pertaining to element $e$ and the mortar elements $m$ of $\partial e$, and perform elemental matrix-vector operations.

Equation \ref{eq:semidiscrete} contains integrals and therefore represents only a semi-discrete set of equations. In the next section we show how to make Eq.~(\ref{eq:semidiscrete}) fully discrete by using quadrature to approximate the integrals.

\subsubsection{Quadrature}
\label{sec:quadrature}

The integrals in Eqs.~(\ref{eq:lh2}) and~(\ref{equationFh2}) depend on various geometric objects:
We define the Jacobian matrix with respect to the mapping $\tv \phi_e: \tv \xi \rightarrow \tv x$ on an element as $e \in \mathbb{E}$
 \begin{equation}
    (\tv J)^i_j = \frac{\partial \tv x_i}{\partial \tv \xi_j},
 \end{equation}
with an inverse given by
 \begin{equation}
    (\tv J^{-1})^i_j = \frac{\partial \tv \xi_i}{\partial \tv x_j}.
 \end{equation}
The Jacobian determinant is
 \begin{equation}\label{eq:Jacobian}
   J = \text{det}\tv J
 \end{equation}
 Similarly, the surface Jacobian for a mortar element $m \in \mathbb{M}$ is

 \begin{equation}\label{eq:Surface-Jacobian}
   S^m = J\sqrt{\frac{\partial \xi_j}{\partial \tv x} \cdot \frac{\partial \xi_j}{\partial \tv x}},
 \end{equation}
 where $j$ is the index of the reference coordinate which is constant on the surface under consideration (no sum-convention).
 For a mortar on a macro-element boundary, $S^m$ may change depending on which of the two macro-element mappings are used to compute Eqn.~(\ref{eq:Surface-Jacobian}). This ambiguity is not problematic as long as all quantities in a mortar integrand are transformed to the macro-element coordinate system in which $S^m$ is being computed.
The normal is computed by
\begin{equation}
 \Hat{\tv n} = \text{sgn}(\xi_j)\frac{J}{S^m}\frac{\partial \xi_j}{\partial \tv x},
\end{equation}
where $\text{sgn}$ is the signum function, where $j$ labels the dimension that is constant on the face under consideration (no sum-convention).
 
Let us now consider the stiffness integral in Eq.~(\ref{eq:lh2}). Because basis-functions are local to each element, we can consider each element $e$ individually, and we will omit the superscript $e$ to lighten the notational load.   Substituting the expansion of the solution $u_h=u_\alpha\psi_\alpha$, as well as $v=\psi_\beta$, the stiffness integral becomes
\begin{align}
  \label{eq:Lstiff-in-x}
  L^{\text{stiff}}_{h,e}(u_h, \psi_\beta) & = \int_e u_\alpha \frac{\partial\psi_\alpha}{\partial x_k}\frac{\partial \psi_\beta}{\partial x_k}\,\mathrm{d}\tv x.
\end{align}
We recall that we employ the Einstein sum convention, i.e.\ Eq.~(\ref{eq:Lstiff-in-x}) has implicit sums over $\alpha$ and $k$. 
  The derivatives $\partial\psi_\alpha/\partial\xi_l$ are just
polynomials in $\tv\xi$, therefore, they can be re-expanded in our
nodal basis as $\partial\psi_\alpha/\partial\xi_l =
D^l_{\alpha\beta}\, \psi_\beta$ with
$D^l_{\alpha\beta}:=\partial\psi_\alpha/\partial\xi_l(\tv
\xi^\beta)$.  The physical derivative $\frac{\partial \psi_{\alpha}}{\partial x_k} $ is then $\frac{\partial \psi_\alpha}{\partial x_k} = (J^{-1})^l_kD^l_{\alpha\beta}\psi_{\beta} $. The tensor-grid in Eq.~(\ref{eq:Lagrange_Basis}) implies that 
the matrices $D^l_{\alpha\beta}$ are sparse for $d>1$.  Using these expressions, and converting to
an integral in the reference coordinates, we obtain
\begin{align}\label{eq:stiffness-3}
  L^{\text{stiff}}_{h,e}(u_h, v_h)
&  = u_\alpha \!
\int\displaylimits_{[-1,1]^{d}}
\!\!\!\!
    (J^{-1})^l_kD^l_{\alpha\gamma}\psi_{\gamma}
    (J^{-1})^m_kD^m_{\beta\delta}\psi_{\delta}
    \,J\,\mathrm{d}\tv\xi\\
\label{eq:temp-C}
    &  =u_\alpha L^{{\rm stiff},e}_{\alpha\beta}.
\end{align}
We evaluate the integral in Eq.~(\ref{eq:stiffness-3}) with Gauss-Legendre (GL) quadrature.
This choice follows~\cite{mengaldo2015dealiasing} in the use of a stronger set of quadrature points
 (higher order GL grids) to decrease the error in geometric aliasing.  We denote the GL abscissas and weights by $\xi_{\rm GL}^{(c)}$ and $w_{\rm GL}^{(c)}$, respectively, where $c=1,\ldots, N_{\rm GL}$.  In multiple dimensions these will be a tensor product of the 1-d GL abscissas and weights denotes by $\tv\xi^\sigma_{\rm GL}$ and $w^\sigma_{\rm GL}$, respectively.  All non-polynomial functions,
including geometric quantities such as $J, (J^{-1})^i_j, S^m, \Hat{\tv n}$, are evaluated directly at $\tv\xi^\sigma_{\rm GL}$, whereas polynomial functions like the trial
function $u_h$ and the test function $\psi_h$ ---which naturally are represented 
on the Legendre-Gauss-Lobatto grid--- are interpolated to
the GL--quadrature points. Denoting the interpolation matrix by $I_{\sigma\alpha}:=\psi_\alpha(\tv\xi^\sigma_{\rm GL})$, we find
\begin{align}
L^{\text{stiff},e}_{\alpha\beta}
      &\approxeq \sum_\sigma w_\sigma
    \left(J\,
    (\tv{J}^{-1})^l_kD^l_{\alpha\gamma}\psi_{\gamma}
    (\tv{J}^{-1})^m_kD^m_{\beta\delta}\psi_{\delta}
    \right)\Bigg|_{\xi^\sigma_{GL}} \\
    &=   D^l_{\alpha\gamma}I_{\sigma\gamma}(\tv{J}_\sigma^{-1})^l_k\, w_\sigma J_\sigma (\tv{J}_\sigma^{-1})^m_k I_{\sigma\delta} D^m_{\beta\delta}  \label{eq:harald-stiff}.
\end{align}

Here $I_{cm} = \psi_m(\xi^\zeta_{GL})$ is an interpolation matrix from
the GLL points to the GL points.
The other volume integrals in Eq.~(\ref{eq:lh3})
are computed in a similar manner.

  The mortar integrals are a bit more
  involved owing to the extra book-keeping arising from the two elements (named '+' and '-') that touch at the boundary $m\in\mathbb{M}$, each with their own local basis-functions,
  denoted by $\psi^-_\alpha$ and $\psi^+_\alpha$.  Taking the penalty-integral as an example, for test-functions $v_h$ that are a basis-function of the '-' element, the definition Eq.~(\ref{eq:20}) yields
\begin{align}
  L^{\text{pen}}_{m,h}(u_h, \psi^-_\alpha) 
  &= \int
  _{m}\sigma \llbracket u_h \rrbracket_k \llbracket \psi^-_\alpha \rrbracket_k \mathrm{d}\tv s \\
\label{eq:temp-E}
  &= \int
  _{[-1,1]^{d-1}}\sigma \left(u_h^{-} - u_h^{+}\right)\psi^-_\alpha\,S^m\,\mathrm{d}\tv\xi.
\end{align}
Equation~(\ref{eq:temp-E}) contains both $u_h^-$ and $u_h^+$;
therefore, when substituting in their respective local expansions,
$u_h^{\pm}=u_\alpha^\pm \psi_\alpha^\pm$, and pulling the coefficients
outside the integral, we see that the penalty term will result in
entries of the global matrix equation (\ref{eq:semidiscrete}) that
couple the two elements $e^\pm$.  Once the coefficients $u_\alpha^\pm$
are moved outside the integral, the remaining integral only depends on
basis-functions $\psi_\alpha^\pm$ and geometric quantities.  These integrals are
evaluated with GL-quadrature, which is expressed in terms of interpolation matrices $I_{\alpha\rho}^\pm :=\psi^\pm_\alpha(\tv\xi^\rho_{\rm GL})$, where  $\rho$ labels the GL collocation points on the boundary $[-1,1]^{d-1}$ and $\alpha$ labels the local basis-functions in $e^\pm$, which are to be evaluated in the suitably mapped $\tv\xi_{\rm GL}$ locations.

\begin{figure}
  \label{fig:mortar}
\centering
\begin{tikzpicture}[thick,scale=0.6, every node/.style={scale=0.6}]
\draw [fill=gray!20, draw=black] (13.000000,3.000000) rectangle (10.000000,0.000000);
  \draw [fill=black,draw=black] (10.041667,0.041667) circle [radius=0.050000];
\draw [fill=black,draw=black] (10.847814,0.041667) circle [radius=0.050000];
\draw [fill=black,draw=black] (12.152186,0.041667) circle [radius=0.050000];
\draw [fill=black,draw=black] (12.958333,0.041667) circle [radius=0.050000];
\draw [fill=black,draw=black] (10.041667,0.847814) circle [radius=0.050000];
\draw [fill=black,draw=black] (10.847814,0.847814) circle [radius=0.050000];
\draw [fill=black,draw=black] (12.152186,0.847814) circle [radius=0.050000];
\draw [fill=black,draw=black] (12.958333,0.847814) circle [radius=0.050000];
\draw [fill=black,draw=black] (10.041667,2.152186) circle [radius=0.050000];
\draw [fill=black,draw=black] (10.847814,2.152186) circle [radius=0.050000];
\draw [fill=black,draw=black] (12.152186,2.152186) circle [radius=0.050000];
\draw [fill=black,draw=black] (12.958333,2.152186) circle [radius=0.050000];
\draw [fill=black,draw=black] (10.041667,2.958333) circle [radius=0.050000];
\draw [fill=black,draw=black] (10.847814,2.958333) circle [radius=0.050000];
\draw [fill=black,draw=black] (12.152186,2.958333) circle [radius=0.050000];
\draw [fill=black,draw=black] (12.958333,2.958333) circle [radius=0.050000];
\draw [fill=red, draw=red] (8.000000,3.000000) rectangle (8.000000,0.000000);
\draw [fill=red,draw=red] (8.000000,0.041667) circle [radius=0.050000];
\draw [fill=red,draw=red] (8.000000,0.041667) circle [radius=0.050000];
\draw [fill=red,draw=red] (8.000000,0.041667) circle [radius=0.050000];
\draw [fill=red,draw=red] (8.000000,0.041667) circle [radius=0.050000];
\draw [fill=red,draw=red] (8.000000,0.041667) circle [radius=0.050000];
\draw [fill=red,draw=red] (8.000000,0.545297) circle [radius=0.050000];
\draw [fill=red,draw=red] (8.000000,0.545297) circle [radius=0.050000];
\draw [fill=red,draw=red] (8.000000,0.545297) circle [radius=0.050000];
\draw [fill=red,draw=red] (8.000000,0.545297) circle [radius=0.050000];
\draw [fill=red,draw=red] (8.000000,0.545297) circle [radius=0.050000];
\draw [fill=red,draw=red] (8.000000,1.500000) circle [radius=0.050000];
\draw [fill=red,draw=red] (8.000000,1.500000) circle [radius=0.050000];
\draw [fill=red,draw=red] (8.000000,1.500000) circle [radius=0.050000];
\draw [fill=red,draw=red] (8.000000,1.500000) circle [radius=0.050000];
\draw [fill=red,draw=red] (8.000000,1.500000) circle [radius=0.050000];
\draw [fill=red,draw=red] (8.000000,2.454703) circle [radius=0.050000];
\draw [fill=red,draw=red] (8.000000,2.454703) circle [radius=0.050000];
\draw [fill=red,draw=red] (8.000000,2.454703) circle [radius=0.050000];
\draw [fill=red,draw=red] (8.000000,2.454703) circle [radius=0.050000];
\draw [fill=red,draw=red] (8.000000,2.454703) circle [radius=0.050000];
\draw [fill=red,draw=red] (8.000000,2.958333) circle [radius=0.050000];
\draw [fill=red,draw=red] (8.000000,2.958333) circle [radius=0.050000];
\draw [fill=red,draw=red] (8.000000,2.958333) circle [radius=0.050000];
\draw [fill=red,draw=red] (8.000000,2.958333) circle [radius=0.050000];
\draw [fill=red,draw=red] (8.000000,2.958333) circle [radius=0.050000];
\draw [fill=blue, draw=blue] (8.000000,6.000000) rectangle (8.000000,3.000000);
\draw [fill=blue,draw=blue] (8.000000,3.041667) circle [radius=0.050000];
\draw [fill=blue,draw=blue] (8.000000,3.041667) circle [radius=0.050000];
\draw [fill=blue,draw=blue] (8.000000,3.041667) circle [radius=0.050000];
\draw [fill=blue,draw=blue] (8.000000,3.041667) circle [radius=0.050000];
\draw [fill=blue,draw=blue] (8.000000,3.041667) circle [radius=0.050000];
\draw [fill=blue,draw=blue] (8.000000,3.041667) circle [radius=0.050000];
\draw [fill=blue,draw=blue] (8.000000,3.041667) circle [radius=0.050000];
\draw [fill=blue,draw=blue] (8.000000,3.289257) circle [radius=0.050000];
\draw [fill=blue,draw=blue] (8.000000,3.289257) circle [radius=0.050000];
\draw [fill=blue,draw=blue] (8.000000,3.289257) circle [radius=0.050000];
\draw [fill=blue,draw=blue] (8.000000,3.289257) circle [radius=0.050000];
\draw [fill=blue,draw=blue] (8.000000,3.289257) circle [radius=0.050000];
\draw [fill=blue,draw=blue] (8.000000,3.289257) circle [radius=0.050000];
\draw [fill=blue,draw=blue] (8.000000,3.289257) circle [radius=0.050000];
\draw [fill=blue,draw=blue] (8.000000,3.816262) circle [radius=0.050000];
\draw [fill=blue,draw=blue] (8.000000,3.816262) circle [radius=0.050000];
\draw [fill=blue,draw=blue] (8.000000,3.816262) circle [radius=0.050000];
\draw [fill=blue,draw=blue] (8.000000,3.816262) circle [radius=0.050000];
\draw [fill=blue,draw=blue] (8.000000,3.816262) circle [radius=0.050000];
\draw [fill=blue,draw=blue] (8.000000,3.816262) circle [radius=0.050000];
\draw [fill=blue,draw=blue] (8.000000,3.816262) circle [radius=0.050000];
\draw [fill=blue,draw=blue] (8.000000,4.500000) circle [radius=0.050000];
\draw [fill=blue,draw=blue] (8.000000,4.500000) circle [radius=0.050000];
\draw [fill=blue,draw=blue] (8.000000,4.500000) circle [radius=0.050000];
\draw [fill=blue,draw=blue] (8.000000,4.500000) circle [radius=0.050000];
\draw [fill=blue,draw=blue] (8.000000,4.500000) circle [radius=0.050000];
\draw [fill=blue,draw=blue] (8.000000,4.500000) circle [radius=0.050000];
\draw [fill=blue,draw=blue] (8.000000,4.500000) circle [radius=0.050000];
\draw [fill=blue,draw=blue] (8.000000,5.183738) circle [radius=0.050000];
\draw [fill=blue,draw=blue] (8.000000,5.183738) circle [radius=0.050000];
\draw [fill=blue,draw=blue] (8.000000,5.183738) circle [radius=0.050000];
\draw [fill=blue,draw=blue] (8.000000,5.183738) circle [radius=0.050000];
\draw [fill=blue,draw=blue] (8.000000,5.183738) circle [radius=0.050000];
\draw [fill=blue,draw=blue] (8.000000,5.183738) circle [radius=0.050000];
\draw [fill=blue,draw=blue] (8.000000,5.183738) circle [radius=0.050000];
\draw [fill=blue,draw=blue] (8.000000,5.710743) circle [radius=0.050000];
\draw [fill=blue,draw=blue] (8.000000,5.710743) circle [radius=0.050000];
\draw [fill=blue,draw=blue] (8.000000,5.710743) circle [radius=0.050000];
\draw [fill=blue,draw=blue] (8.000000,5.710743) circle [radius=0.050000];
\draw [fill=blue,draw=blue] (8.000000,5.710743) circle [radius=0.050000];
\draw [fill=blue,draw=blue] (8.000000,5.710743) circle [radius=0.050000];
\draw [fill=blue,draw=blue] (8.000000,5.710743) circle [radius=0.050000];
\draw [fill=blue,draw=blue] (8.000000,5.958333) circle [radius=0.050000];
\draw [fill=blue,draw=blue] (8.000000,5.958333) circle [radius=0.050000];
\draw [fill=blue,draw=blue] (8.000000,5.958333) circle [radius=0.050000];
\draw [fill=blue,draw=blue] (8.000000,5.958333) circle [radius=0.050000];
\draw [fill=blue,draw=blue] (8.000000,5.958333) circle [radius=0.050000];
\draw [fill=blue,draw=blue] (8.000000,5.958333) circle [radius=0.050000];
\draw [fill=blue,draw=blue] (8.000000,5.958333) circle [radius=0.050000];
\draw [fill=gray!20, draw=black] (13.000000,6.000000) rectangle (10.000000,3.000000);
\draw [fill=black,draw=black] (10.041667,3.041667) circle [radius=0.050000];
\draw [fill=black,draw=black] (10.289257,3.041667) circle [radius=0.050000];
\draw [fill=black,draw=black] (10.816262,3.041667) circle [radius=0.050000];
\draw [fill=black,draw=black] (11.500000,3.041667) circle [radius=0.050000];
\draw [fill=black,draw=black] (12.183738,3.041667) circle [radius=0.050000];
\draw [fill=black,draw=black] (12.710743,3.041667) circle [radius=0.050000];
\draw [fill=black,draw=black] (12.958333,3.041667) circle [radius=0.050000];
\draw [fill=black,draw=black] (10.041667,3.289257) circle [radius=0.050000];
\draw [fill=black,draw=black] (10.289257,3.289257) circle [radius=0.050000];
\draw [fill=black,draw=black] (10.816262,3.289257) circle [radius=0.050000];
\draw [fill=black,draw=black] (11.500000,3.289257) circle [radius=0.050000];
\draw [fill=black,draw=black] (12.183738,3.289257) circle [radius=0.050000];
\draw [fill=black,draw=black] (12.710743,3.289257) circle [radius=0.050000];
\draw [fill=black,draw=black] (12.958333,3.289257) circle [radius=0.050000];
\draw [fill=black,draw=black] (10.041667,3.816262) circle [radius=0.050000];
\draw [fill=black,draw=black] (10.289257,3.816262) circle [radius=0.050000];
\draw [fill=black,draw=black] (10.816262,3.816262) circle [radius=0.050000];
\draw [fill=black,draw=black] (11.500000,3.816262) circle [radius=0.050000];
\draw [fill=black,draw=black] (12.183738,3.816262) circle [radius=0.050000];
\draw [fill=black,draw=black] (12.710743,3.816262) circle [radius=0.050000];
\draw [fill=black,draw=black] (12.958333,3.816262) circle [radius=0.050000];
\draw [fill=black,draw=black] (10.041667,4.500000) circle [radius=0.050000];
\draw [fill=black,draw=black] (10.289257,4.500000) circle [radius=0.050000];
\draw [fill=black,draw=black] (10.816262,4.500000) circle [radius=0.050000];
\draw [fill=black,draw=black] (11.500000,4.500000) circle [radius=0.050000];
\draw [fill=black,draw=black] (12.183738,4.500000) circle [radius=0.050000];
\draw [fill=black,draw=black] (12.710743,4.500000) circle [radius=0.050000];
\draw [fill=black,draw=black] (12.958333,4.500000) circle [radius=0.050000];
\draw [fill=black,draw=black] (10.041667,5.183738) circle [radius=0.050000];
\draw [fill=black,draw=black] (10.289257,5.183738) circle [radius=0.050000];
\draw [fill=black,draw=black] (10.816262,5.183738) circle [radius=0.050000];
\draw [fill=black,draw=black] (11.500000,5.183738) circle [radius=0.050000];
\draw [fill=black,draw=black] (12.183738,5.183738) circle [radius=0.050000];
\draw [fill=black,draw=black] (12.710743,5.183738) circle [radius=0.050000];
\draw [fill=black,draw=black] (12.958333,5.183738) circle [radius=0.050000];
\draw [fill=black,draw=black] (10.041667,5.710743) circle [radius=0.050000];
\draw [fill=black,draw=black] (10.289257,5.710743) circle [radius=0.050000];
\draw [fill=black,draw=black] (10.816262,5.710743) circle [radius=0.050000];
\draw [fill=black,draw=black] (11.500000,5.710743) circle [radius=0.050000];
\draw [fill=black,draw=black] (12.183738,5.710743) circle [radius=0.050000];
\draw [fill=black,draw=black] (12.710743,5.710743) circle [radius=0.050000];
\draw [fill=black,draw=black] (12.958333,5.710743) circle [radius=0.050000];
\draw [fill=black,draw=black] (10.041667,5.958333) circle [radius=0.050000];
\draw [fill=black,draw=black] (10.289257,5.958333) circle [radius=0.050000];
\draw [fill=black,draw=black] (10.816262,5.958333) circle [radius=0.050000];
\draw [fill=black,draw=black] (11.500000,5.958333) circle [radius=0.050000];
\draw [fill=black,draw=black] (12.183738,5.958333) circle [radius=0.050000];
\draw [fill=black,draw=black] (12.710743,5.958333) circle [radius=0.050000];
\draw [fill=black,draw=black] (12.958333,5.958333) circle [radius=0.050000];
\draw [fill=gray!20, draw=black] (6.000000,6.000000) rectangle (0.000000,0.000000);
\draw [fill=black,draw=black] (0.041667,0.041667) circle [radius=0.050000];
\draw [fill=black,draw=black] (1.063316,0.041667) circle [radius=0.050000];
\draw [fill=black,draw=black] (3.000000,0.041667) circle [radius=0.050000];
\draw [fill=black,draw=black] (4.936684,0.041667) circle [radius=0.050000];
\draw [fill=black,draw=black] (5.958333,0.041667) circle [radius=0.050000];
\draw [fill=black,draw=black] (0.041667,1.063316) circle [radius=0.050000];
\draw [fill=black,draw=black] (1.063316,1.063316) circle [radius=0.050000];
\draw [fill=black,draw=black] (3.000000,1.063316) circle [radius=0.050000];
\draw [fill=black,draw=black] (4.936684,1.063316) circle [radius=0.050000];
\draw [fill=black,draw=black] (5.958333,1.063316) circle [radius=0.050000];
\draw [fill=black,draw=black] (0.041667,3.000000) circle [radius=0.050000];
\draw [fill=black,draw=black] (1.063316,3.000000) circle [radius=0.050000];
\draw [fill=black,draw=black] (3.000000,3.000000) circle [radius=0.050000];
\draw [fill=black,draw=black] (4.936684,3.000000) circle [radius=0.050000];
\draw [fill=black,draw=black] (5.958333,3.000000) circle [radius=0.050000];
\draw [fill=black,draw=black] (0.041667,4.936684) circle [radius=0.050000];
\draw [fill=black,draw=black] (1.063316,4.936684) circle [radius=0.050000];
\draw [fill=black,draw=black] (3.000000,4.936684) circle [radius=0.050000];
\draw [fill=black,draw=black] (4.936684,4.936684) circle [radius=0.050000];
\draw [fill=black,draw=black] (5.958333,4.936684) circle [radius=0.050000];
\draw [fill=black,draw=black] (0.041667,5.958333) circle [radius=0.050000];
\draw [fill=black,draw=black] (1.063316,5.958333) circle [radius=0.050000];
\draw [fill=black,draw=black] (3.000000,5.958333) circle [radius=0.050000];
\draw [fill=black,draw=black] (4.936684,5.958333) circle [radius=0.050000];
\draw [fill=black,draw=black] (5.958333,5.958333) circle [radius=0.050000];

  \node at (8.000000,-0.400000) {\LARGE $\text{mortar}$};
\node at (7.000000,3.60000) {\LARGE $I_{hp}$};
\draw [->, line width=0.500000mm] (6.600000,3.000000) -- (7.400000,3.000000);
\node at (9.000000,4.940000) {\LARGE $I_{p}$};
\draw [->, line width=0.500000mm] (9.400000,4.500000) -- (8.600000,4.500000);
\node at (9.000000,1.940000) {\LARGE $I_{p}$};
\draw [->, line width=0.500000mm] (9.400000,1.500000) -- (8.600000,1.500000);
\end{tikzpicture}
\medskip
\caption{A representation of a 2:1 interface in two dimensions. On the
  left we have a 4th-order element and on the right we have two
  elements of degree 8 and degree 2 respectively. The data on the
  faces of these three elements will be used in the mortar integrals
  of equation \ref{eq:lh}, but since the interface is
  non-conforming, we need to interpolate the face data to a shared
  broken polynomial space on the mortar. To ensure that the face data
  on each element is in a polynomial space which a subset of the
  polynomial space on the mortar, we demand that the polynomial degree
  p on the mortar is the maximum of the polynomial degrees on either
  side. For the lower mortar face (red) this would be $\max(4,3) = 4$
  and for the upper mortar face (green) this would be $\max(4,6) = 6$.
  Since the left element must interpolate it's face data to a
  space containing two faces, we use the hp-interpolation operator
  $\mathcal{I}_{hp}$. On the right side, each element has to map its
  data from one face to one face, so we use the p-interpolation
  operator $\mathcal{I}_p$.}
\end{figure}
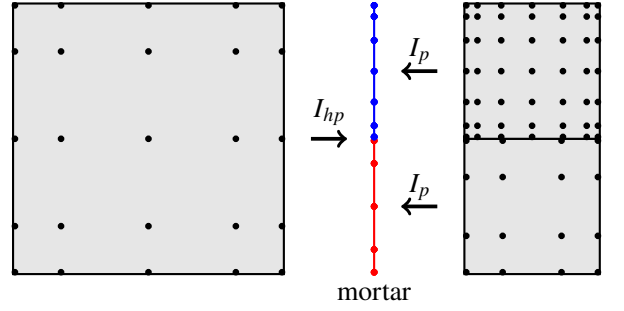

\subsection{Penalty Function}

The penalty parameter $\sigma$ in Eqs.~(\ref{eq:lh}) and~(\ref{eq:lh2}) is a spatially dependent function on boundaries $m\in\mathbb{M}$, defined by
\begin{equation}
\label{eq:penaltyparameter}
\sigma = C\frac{p_{m}^{2}}{h_{m}},
\end{equation}
where $p_m$ and $h_m$ represent a typical polynomial degree and a typical length-scale of the elements touching at the boundary, respectively,  and where the parameter $C>0$ is large enough such
that $L\left(\cdot, \cdot\right)$ is coercive. We choose $p_{m} = \max(p^+,p^-)$
and we set $h_{m} = \min(J^{+}/S^m,J^{-}/S^m)$ as this has yielded the best results empirically. Here $J^{\pm}$ is the
volume Jacobian on $e^\pm$ mapping $[0,1]^d$ to $e^\pm$, and $S^m$ is the
surface Jacobian mapping $[0,1]^{d-1}$ to the boundary face of $e^\pm$.

\subsection{Norms}

Throughout the following sections we will use the energy norm,
  \begin{equation}
  \label{eq:energynorm}
 ||\nu||_{*} = \left(\sum_{\substack{e \in \mathbb{E}}}\; \int_e |\nabla \nu |^{2} \mathrm{d}\tv {x} + \sum\limits_{m \in \Gamma}\; \int_m|\sqrt{\sigma} \llbracket \nu \rrbracket |^{2} \mathrm{d} \tv s \right)^{\frac{1}{2}},
 \end{equation}
and the $L_2$ norm,
 \begin{equation}
  \label{eq:l2norm}
  ||\nu||_{2}
  = \left(\int_{\Omega}\nu^2\mathrm{d}\tv x\right)^{\frac{1}{2}} 
 = \left(\sum_{\substack{e \in \mathbb{E}}}
  \int_{e}\nu^2\mathrm{d}\tv x\right)^{\frac{1}{2}}.
\end{equation}
Here, $\nu$ is a scalar function on $\Omega$.

\subsection{Multigrid-Preconditioned Newton-Krylov}

\subsubsection{Newton-Krylov}

In general, a set of elliptic partial differential equations can be written in the form
\begin{equation}\label{eq:Ru=0}
  R\left(\tv u \right) = 0,
\end{equation}
where $\tv u$ is the solution and R is a non-linear elliptic operator.
Defining the Jacobian of the system by
%
$\mathrm{J}_{R}(\tv u) \equiv \frac{\partial R}{\partial \tv u}(\tv u)$,
%
Newton-Raphson iteratively refines an initial guess for the solution
$\tv u^{(k)}$ by solving the following 
linear system for the correction $\delta \tv u^{(k)} = \tv u^{(k+1)} -
\tv u^{(k)}$:
\begin{equation}
\label{eq:newtoncorrectionsystem}
\mathrm{J}_R(\tv u^{(k)})\delta \tv u^{(k)} = -R(\tv u^{(k)}).
\end{equation}
Upon discretization as described in Sec.~\ref{sec:DGFEM}, Eq.~(\ref{eq:newtoncorrectionsystem}) results in
  a $\mathrm{N} \times \mathrm{N}$ linear system.
Once Eq.~(\ref{eq:newtoncorrectionsystem}) is solved, the improved solution is given by $\tv u^{(k+1)}=\tv u^{(k)}+\delta \tv u^{(k)}$.
  The Newton-Raphson iterations are continued until the residual $R(\tv u^{(k)})$ is sufficiently small. At each step, the linear system is solved by a linear solver we abbreviate as $LSOLVE$. The full algorithm is described in Algorithm~\ref{alg:newtonraphson}.

\begin{figure}
\begin{algorithm}[H]
  \caption{\label{alg:newtonraphson}
    Multigrid Preconditioned Newton-Krylov: Solves Eq.~(\ref{eq:Ru=0}).$N_{\text{iter,NK}}$ and $\text{tol}_{\text{NK}}$ are user-specified parameters.
 }
    \begin{algorithmic}[1]
      \Procedure{NK}{$u_0$}
      \State $k \leftarrow 0$
      \While {$k \leq N_{\text{iter,NK}}\;$  or $\;||R(u_k)|| \geq \text{tol}_{\text{NK}}$}
       \State  $\delta u_k\leftarrow $ LSOLVE($u_k$, $R(u_k)$, $\mathrm{J}_R(u_k)$) 
       \State $u_{k+1} \leftarrow u_k + \delta u_k$
       \State $k \leftarrow k + 1$
       \EndWhile
       \State \Return $u_{k}$
      \EndProcedure 
    \end{algorithmic}
\end{algorithm}
\end{figure}

For three-dimensional problems with a large number $\mathrm{N}$ of
degrees of freedom, the Jacobian $\mathrm{J}$ is too large to form fully and partition across processes. For
building a scalable solver, we are therefore limited to iterative
solvers, the most popular class of which are the Krylov
solvers~\cite{saad2003iterative} such as the Conjugate Gradient method or GMRES,
which only involve matrix-vector operations. In this paper we use the
flexible conjugate gradient Krylov~\cite{notay2000flexible} solver at each
Newton-Raphson step, as summarized in Algorithm~\ref{alg:FCG}, where
$u_0$ denotes the initial guess of the linear solver, and
$N_{\text{its}}$ is the number of iterations.

\begin{figure}
  \begin{algorithm}[H]
    \caption{\label{alg:FCG}
    Flexible Conjugate Gradients: Solves Eq.~(\ref{eq:newtoncorrectionsystem}), with the variable $u$ representing $\delta u^{(k)}$. $N_{\text{iter,FCG}}$ and $\text{tol}_{\text{FCG}}$ are user-specified parameters.}%
    \begin{algorithmic}[1]
      \Function{LSOLVE}{$u_0$,$R$,$\mathrm{J}$}
      \State $r_0 \leftarrow \mathrm{J}u_0 + R$
      \State $k \leftarrow 0$
      \While {$k$++ $\leq N_{\text{iter,FCG}}\;$ or $\;||r_k|| \geq \text{tol}_{\text{FCG}}$}
      \State $v_k \leftarrow \text{VCYCLE}(r_k,\mathrm{J}v_k)$
      \State $w_k \leftarrow \mathrm{J}v_k$
      \State $\alpha_k \leftarrow v_k \cdot r_k$ 
      \State $\beta_k \leftarrow v_k \cdot w_k$
      \If {k = 0}  
      \State $d_k \leftarrow v_k$
      \State $q_k \leftarrow w_k$
      \State $\rho_k \leftarrow \beta_k$
      \Else
      \State $\gamma_k \leftarrow v_k \cdot q_{k-1}$
      \State $d_k \leftarrow v_k - (\gamma_k/\rho_{k-1})d_{k-1}$
      \State $q_k \leftarrow w_k - (\gamma_k/\rho_{k-1})q_{k-1}$
      \State $\rho_k \leftarrow \beta_k - \gamma^2_k/\rho_{k-1}$
      \EndIf
      \State $u_{k+1} \leftarrow u_k + (\alpha_k/\rho_k)d_k$
      \State $r_{k+1} \leftarrow r_k - (\alpha_k/\rho_k)q_k$
      \EndWhile
      \State \Return $u_k$
      \EndFunction
    \end{algorithmic}
  \end{algorithm}
\end{figure}

A typical Krylov solver will take $\mathrm{O}(\sqrt{\kappa})$
iterations to reach a desired accuracy, where the condition number
$\kappa$ is defined as the ratio of the maximum eigenvalue and minimum
eigenvalue of $\mathrm{J}_R$. For the  discontinuous
galerkin method, the discretized Laplacian matrix has a condition
number that grows with p-refinement and h-refinement (see \cite{antonietti2011class,hesthaven2008nodal} for example) and therefore the number of iterations will grow with each AMR step unless Eq.~(\ref{eq:newtoncorrectionsystem}) is preconditioned.

\subsubsection{Multigrid preconditioner}

We use a multigrid V-cycle~\cite{briggs2000multigrid} as a preconditioner for each
FCG solve (called on line 4 in Algorithm~\ref{alg:FCG}). 
Multigrid is an multi-level iterative method aimed at achieving
mesh-independent error reduction rates through a clever method of
solving for the error corrections on coarser grids and prolonging the
corrections to the finer grids. For a detailed overview of multigrid
methods see~\cite{briggs2000multigrid}. The main drawback of multigrid
is its complexity. In this section we will briefly describe the
components of the multigrid algorithm employed in solving problems in
this paper with hp-grids in parallel with the interior penalty
discontinuous Galerkin method.

\subsubsection{Multigrid Meshes}
\label{sec:MultigridMeshes}

Multigrid uses a hierarchy of coarsened meshes, labeled with
$l=0, \ldots L$, where $l=L$
represents the fine mesh $\mathbf{E}$ on which the solution is
desired, and $l=0$ represents the coarsest mesh.  One V-cycle proceeds
from the fine grid to the coarsest grid, and back to the fine grid, as
indicated in Fig.~\ref{fig:vcycle}.  We construct the coarse meshes
$l=L-1, \ldots 0$ by successively coarsening. Coarsening by
  combining $2^d$ elements into one element can lead to interfaces
  with a 4:1 balance, even if the original mesh is 2:1 balanced.  We
  avoid such 4:1 balance with
surrogate meshes, which have been used before in large-scale FEM
multigrid solvers, see for example
\cite{sampath2008dendro,sundar2012parallel}.  A surrogate mesh is the
naively h-coarsened mesh, as indicated in blue in
Fig.~\ref{fig:vcycle}.  If the surrogate mesh has indeed interfaces
with 4:1 balance, the desired 2:1 balance is enforced by refining at
the unbalanced interfaces.  This results in the coarsened mesh on
Level $l=L-1$, and iteratively down to $l=0$.  It is easy to show that
at each coarsening the coarse level $l-1$ mesh is strictly
coarser than the level $l$ fine mesh.  Following this, we must make a decision as to what
polynomial degree we choose for the coarsened elements. We take the minimum polynomial degree of the $2^{d}$
children elements for each parent element of the coarsened mesh, ensuring that functions on the coarse grid can be represented exactly on the fine mesh, a property we will utilize below in Eq.~(\ref{eq:psi-interpolation}). Lastly, it is important to note that we never purely p-coarsen on any of the multigrid levels.
We do not store the multigrid meshes
because we have empirically shown that the coarsening, refining and
balancing is not the bottleneck for the multigrid algorithm.

%
%
\usetikzlibrary{arrows,positioning} 
\begin{figure}

  \centering
  \begin{adjustbox}{width=.45\textwidth,trim=15 0 0 0, clip=true}
\begin{tikzpicture}[node distance=.25cm, auto]  
\tikzset{
    myCnode/.style={rectangle,rounded corners,draw=black, top color=white, bottom color=yellow!50,very thick, inner sep=1em, minimum size=3em, text centered},
    mySnode/.style={rectangle,rounded corners,draw=black, top color=white, bottom color=cyan!50,very thick, inner sep=1em, minimum size=3em, text centered},
    myarrow/.style={->, line width=1.3mm},
    myDarrow/.style={-, line width=1.3mm, dashed},
    myDarrow2/.style={->, line width=1.3mm, dashed},
    myBarrow/.style={-, line width=1.3mm},
    mylabel/.style={text width=8em, text centered} 
  }

\node[myCnode] (l1) {\Large   \,Level L\,};
\node[below=1cm of l1] (dummy) {}; 

\node[mySnode, right=of dummy] (l2) {\Large  Surrogate};
\node[below=1cm of l2] (dummy1) {}; 
\node[myCnode, right=of dummy1] (l3) {\Large  Level L-1};
\node[mylabel, left=.9cm of l2] (label1) {\Large   Coarsen};
\node[mylabel, left=.9cm of l3] (label1) {\Large   Enforce \break \\\vspace{-.4cm} 2:1 balance};
\node[below=3cm of l3] (dummy2) {};

\node[myCnode, right=.625cm of dummy2] (l4) {\Large  Level 0};

\node[myCnode, right=.86cm of l3] (l5) {\Large  Level L-1};
\node[above=2cm of l4] (dummydummy1) {};
\node[above=.5cm of l4] (dummydummy2) {};
\node[above=1cm of l5] (dummy4) {}; 
\node[mySnode, right=4.15cm of l2] (l6) {\Large  Surrogate};
\node[above=1cm of l6] (dummy5) {}; 
\node[myCnode, right=of dummy5] (l7) {\Large  \,Level L\,};
\node[right=of l7] (dummy51) {};
\node[mylabel, right=.7cm of l5] (label1) {\Large  Undo \break \\\vspace{-.4cm} 2:1 balance};
\node[mylabel, right=.7cm of l6] (label1) {\Large  Refine};

\node[left=of l1] (dummy61) {}; 
\node[below=1cm of l3] (dummyf) {}; 
\node[left=of l3] (dummyf1) {};
\node[above=.1cm of l4] (dummy312) {};
\node[left=.7cm of dummy312] (dummy314) {};
\node[above=.1cm of l4] (dummy312R) {};
\node[right=.7cm of dummy312] (dummy314R) {};
\draw[myBarrow] (dummy61.south west) -- (dummyf1.south west);	
\draw[myDarrow2] (dummyf1.south west) -- (dummy314);

\node[right=of l7] (dummy612) {}; 
\node[below=1cm of l5] (dummyf2) {}; 
\node[right=of l5] (dummyf12) {};

\draw[myDarrow] (dummy314R) -- (dummyf12.south east);
\draw[myarrow] (dummyf12.south east) -- (dummy612.south east);
\draw[myDarrow] (dummydummy1.center) -- (dummydummy2.south);

\end{tikzpicture} 
\end{adjustbox}
\medskip
\caption{
  \label{fig:vcycle}
  A structural representation of a multigrid v-cycle. The nodes in yellow are actual grid-levels, while the nodes in blue represent surrogate grids which are not necessarily 2:1 balanced. In order for a multigrid v-cycle to represent a symmetric operation, the grid levels along the down arrow are exactly the same as the grid-levels along the up arrow. Level 0 represents the coarsest possible mesh.} 
\end{figure}
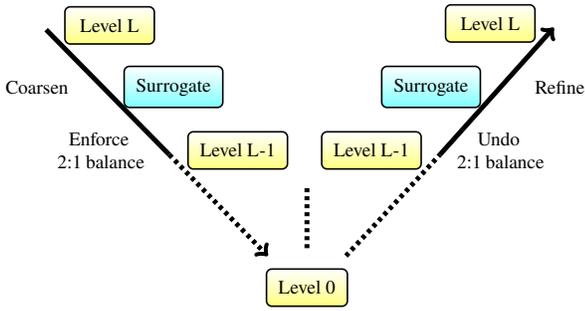
%
%

\subsubsection{Multigrid algorithm}
\begin{figure}
  \begin{algorithm}[H]
    \caption{\label{alg:Multigrid_Vcycle}
      Multigrid Preconditioner Vcycle.\\
      $L$ denotes the number of multigrid Levels.}
    \begin{algorithmic}[1]
      \Function{Vcycle}{$f_L,\mathrm{J}_L$}
      \For {$l = L, 1$}
      \State $v_l \leftarrow 0$
      \State $v_l \leftarrow \text{SMOOTHER}(v_l,f_l,\mathrm{J}_l)$ 
      \State Coarsen grid
      \State Balance grid
      \State $f_{l-1} \leftarrow I_l^T(f_l - J_lv_l)$ \Comment{Restriction}
      \EndFor
      \State $v_0 \leftarrow 0$
      \State $v_0 \leftarrow \text{SMOOTHER}(v_0,f_0,\mathrm{J}_0)$ 
      \For {$l = 1, L$}
      \State Refine grid
      \State $v_l \leftarrow I_l(v_{l-1}) + v_{l}$ \Comment{Prolongation}
      \State $v_l \leftarrow \text{SMOOTHER}(v_l,f_l,\mathrm{J}_l)$ 
      \EndFor
      \State \Return $v_L$
      \EndFunction
    \end{algorithmic}
  \end{algorithm}
\end{figure}
%
Having constructed the coarse meshes, we can now turn to the
  multigrid algorithm, summarized in
  Algorithm~\ref{alg:Multigrid_Vcycle}.  It consists of several
  important components most importantly smoothing and inter-mesh operators like coarsening,  balancing of the mesh, restriction and prolongation. We now look
  at each of these in turn.
%
\subsubsection{Multigrid Smoother}
%
We explore two different smoothers, Chebyshev smoothing and a Schwarz smoother.  Both of these avoid explicit storage of the matrix, as required by smoothers like the Gauss-Seidel method~\cite{shang2009distributed}.

Chebyshev smoothing~\cite{li2011chebyshev}, a type of polynomial
smoother~\cite{saad2003iterative}
requires only matrix-vector operations and has been shown to work
satisfactorily in scalable multigrid
solvers~\cite{ghysels2012improving}.  Our implementation of
Chebyshev smoothing \cite{li2011chebyshev} is presented in
Algorithm~\ref{alg:Chebyshev_Smoother}. For this algorithm there are two user-defined parameters: $N_\text{iter,Cheb}$ and $\Lambda$. Typical values we use for $N_\text{iter,Cheby}$ are in the range $8-15$ and $\Lambda$ is usually set in the range $10-30$.
\begin{figure}
  \begin{algorithm}[H]
    \caption{Chebyshev Smoothing\\ $N_\text{iter,Cheb}$ and $\Lambda$ are user-defined parameters.
}\label{alg:Chebyshev_Smoother}
    \begin{algorithmic}[1]
      \Function{SMOOTHER}{$x, b, \mathrm{J}$}
      \State $p \leftarrow 0$
      \State  $\lambda_{max}, x \leftarrow \text{CGEIGS}(x,b,\mathrm{J})$
      \State $\lambda_{min} \leftarrow \lambda_{max}/\Lambda$
      \State $c \leftarrow (\lambda_{max} - \lambda_{min})/2$
      \State $d \leftarrow (\lambda_{max} + \lambda_{min})/2$
      \For{$k = 1...N_{\text{iter,Cheb}}$}
      \State $r \leftarrow b - \mathrm{J}x$
\State  $\alpha \leftarrow 
  \begin{cases}
  d^{-1} & k = 1\\
  2d(2d^2-c^2)^{-1} & k = 2\\
  (d - \alpha c^2/4)^{-1} & k \neq 1,2
  \end{cases}
  $\;
  \State $\beta \leftarrow \alpha d - 1$
  \State $p \leftarrow \alpha r - \beta p$
  \State $x \leftarrow x + p$
  \EndFor
  \State \Return $x$
      \EndFunction
    \end{algorithmic}
  \end{algorithm}
\end{figure}
Chebyshev polynomial smoothers require a spectral bound on the
eigenvalues of the linear operator. We use conjugate gradients to estimate the eigenvalues of a general symmetric linear operator, as implemented in Algorithm~\ref{alg:CG_Spectral_Bound_Solver}.  It can be shown~\cite{scales1989use} that each iteration of conjugate gradients obtains one row of the underlying linear operator in tri-diagonal form:

\begin{equation}\label{eq:tridiag}
    \left(
    \begin{array}{ccccc}
      \frac{1}{\alpha_1} & - \frac{\sqrt{\beta_2}}{\alpha_1}  & 0 & \cdots & 0 \\
      - \frac{\sqrt{\beta_2}}{\alpha_1} & \frac{1}{\alpha_2} + \frac{\beta_2}{\alpha_1} & - \frac{\sqrt{\beta_3}}{\alpha_2} & & \vdots \\
0 &  & \quad\quad\ddots &  & 0 \\
      \vdots &  &-\frac{\sqrt{\beta_{k-1}}}{\alpha_{k-2}} & \frac{1}{\alpha_{k-1}} + \frac{\beta_{k-1}}{\alpha_{k-2}}  & -\frac{\sqrt{\beta_k}}{\alpha_{k-1}} \\
               0 & \cdots & 0 &\frac{-\sqrt{\beta_k}}{\alpha_{k-1}} &  \frac{1}{\alpha_k} + \frac{\beta_k}{\alpha_{k-1}}
    \end{array}
    \right)
\end{equation}
The values $\alpha_k$ and $\beta_k$ in this expression are obtained
  from the conjugate gradients algorithm~\ref{alg:CG_Spectral_Bound_Solver} on lines 6 and 9. Furthermore,
the eigenvalues of each of the sub tri-diagonal matrices is a subset
of the eigenvalues of the full matrix. So at each iteration we can get
an estimate of the bound by using the Gershgorin circle theorem ~\cite{bell1965gershgorin}. Our Alg.~\ref{alg:CG_Spectral_Bound_Solver} combines the CG steps with the estimation of the bound $\lambda_{\rm max}$ of the largest eigenvalue.

\begin{figure}
  \begin{algorithm}[H]
    \caption{\label{alg:CG_Spectral_Bound_Solver}
      CG Spectral Bound Solver\\ $N_{\rm{iter,eigs}}$ is a user-defined parameter.
    }
    \begin{algorithmic}[1]
      \Function{CGEIGS}{$x_1, b, \mathrm{J}$} 
      \State $r_1 \leftarrow b - \mathrm{J}x_1$  
      \State $p_1 \leftarrow r_1$
      \State $\lambda_{\rm{max}} \leftarrow 0$
      \For{$k = 1,N_{\rm{iter,eigs}}$}
      \State $\alpha_k \leftarrow (r_{k} \cdot r_{k})/(p_k \cdot Jp_k)$
      \State $x_{k+1} \leftarrow x_{k} + \alpha_k p_k$
      \State $r_{k+1} \leftarrow r_{k} - \alpha_kJ p_k$
      \State $\beta_k \leftarrow (r_{k+1} \cdot r_{k+1})/(r_{k} \cdot r_{k})$ 
      \State $p_{k+1} \leftarrow r_{k+1} + \beta_kp_{k}$
      \State  $\lambda_k\leftarrow 
\begin{cases}
      \frac{1}{a_1} + \frac{\sqrt{\beta_1}}{a1} & \text{if} \,\,\,\, k = 1 \\[10pt]
    \frac{1}{a_{k}} + \frac{\beta_{k-1}}{a_{k-1}} + \frac{\sqrt{\beta_{k-1}}}{\alpha_{k-1}}  & \text{if} \,\,\,\, k \neq 1\\[10pt]
  \end{cases}
$\; 
\State  $\lambda_{\rm{max}} = \max(\lambda_k, \lambda_{\rm{max}})$
\EndFor\\
\Return $\lambda_{\rm{max}},x_{k + 1}$ 
  \EndFunction
    \end{algorithmic}
  \end{algorithm}
\end{figure}

Besides estimating the spectral radius, we also use the CG iterations to further smooth the
solution. This adds robustness to multigrid algorithms \cite{elman2001multigrid}.

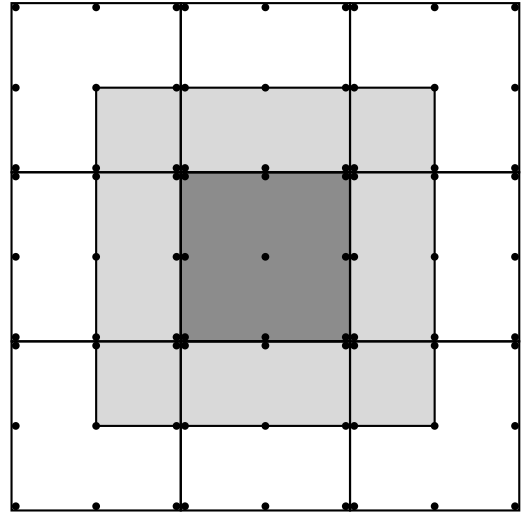
\begin{figure}
\centering
\begin{tikzpicture}[thick,scale=0.75, every node/.style={scale=0.8}]
\draw [fill=gray!30, draw=black] (7.500000,7.500000) rectangle (1.500000,1.500000);
\draw [fill=gray!90, draw=black] (6.000000,6.000000) rectangle (3.000000,3.000000);
\draw [fill=none, draw=black] (3.000000,3.000000) rectangle (0.000000,0.000000);
\draw [fill=black,draw=black] (0.075000,0.075000) circle [radius=0.050000];
\draw [fill=black,draw=black] (1.500000,0.075000) circle [radius=0.050000];
\draw [fill=black,draw=black] (2.925000,0.075000) circle [radius=0.050000];
\draw [fill=black,draw=black] (0.075000,1.500000) circle [radius=0.050000];
\draw [fill=black,draw=black] (1.500000,1.500000) circle [radius=0.050000];
\draw [fill=black,draw=black] (2.925000,1.500000) circle [radius=0.050000];
\draw [fill=black,draw=black] (0.075000,2.925000) circle [radius=0.050000];
\draw [fill=black,draw=black] (1.500000,2.925000) circle [radius=0.050000];
\draw [fill=black,draw=black] (2.925000,2.925000) circle [radius=0.050000];
\draw [fill=none, draw=black] (3.000000,6.000000) rectangle (0.000000,3.000000);
\draw [fill=black,draw=black] (0.075000,3.075000) circle [radius=0.050000];
\draw [fill=black,draw=black] (1.500000,3.075000) circle [radius=0.050000];
\draw [fill=black,draw=black] (2.925000,3.075000) circle [radius=0.050000];
\draw [fill=black,draw=black] (0.075000,4.500000) circle [radius=0.050000];
\draw [fill=black,draw=black] (1.500000,4.500000) circle [radius=0.050000];
\draw [fill=black,draw=black] (2.925000,4.500000) circle [radius=0.050000];
\draw [fill=black,draw=black] (0.075000,5.925000) circle [radius=0.050000];
\draw [fill=black,draw=black] (1.500000,5.925000) circle [radius=0.050000];
\draw [fill=black,draw=black] (2.925000,5.925000) circle [radius=0.050000];
\draw [fill=none, draw=black] (3.000000,9.000000) rectangle (0.000000,6.000000);
\draw [fill=black,draw=black] (0.075000,6.075000) circle [radius=0.050000];
\draw [fill=black,draw=black] (1.500000,6.075000) circle [radius=0.050000];
\draw [fill=black,draw=black] (2.925000,6.075000) circle [radius=0.050000];
\draw [fill=black,draw=black] (0.075000,7.500000) circle [radius=0.050000];
\draw [fill=black,draw=black] (1.500000,7.500000) circle [radius=0.050000];
\draw [fill=black,draw=black] (2.925000,7.500000) circle [radius=0.050000];
\draw [fill=black,draw=black] (0.075000,8.925000) circle [radius=0.050000];
\draw [fill=black,draw=black] (1.500000,8.925000) circle [radius=0.050000];
\draw [fill=black,draw=black] (2.925000,8.925000) circle [radius=0.050000];
\draw [fill=none, draw=black] (6.000000,3.000000) rectangle (3.000000,0.000000);
\draw [fill=black,draw=black] (3.075000,0.075000) circle [radius=0.050000];
\draw [fill=black,draw=black] (4.500000,0.075000) circle [radius=0.050000];
\draw [fill=black,draw=black] (5.925000,0.075000) circle [radius=0.050000];
\draw [fill=black,draw=black] (3.075000,1.500000) circle [radius=0.050000];
\draw [fill=black,draw=black] (4.500000,1.500000) circle [radius=0.050000];
\draw [fill=black,draw=black] (5.925000,1.500000) circle [radius=0.050000];
\draw [fill=black,draw=black] (3.075000,2.925000) circle [radius=0.050000];
\draw [fill=black,draw=black] (4.500000,2.925000) circle [radius=0.050000];
\draw [fill=black,draw=black] (5.925000,2.925000) circle [radius=0.050000];
\draw [fill=none, draw=black] (6.000000,6.000000) rectangle (3.000000,3.000000);
\draw [fill=black,draw=black] (3.075000,3.075000) circle [radius=0.050000];
\draw [fill=black,draw=black] (4.500000,3.075000) circle [radius=0.050000];
\draw [fill=black,draw=black] (5.925000,3.075000) circle [radius=0.050000];
\draw [fill=black,draw=black] (3.075000,4.500000) circle [radius=0.050000];
\draw [fill=black,draw=black] (4.500000,4.500000) circle [radius=0.050000];
\draw [fill=black,draw=black] (5.925000,4.500000) circle [radius=0.050000];
\draw [fill=black,draw=black] (3.075000,5.925000) circle [radius=0.050000];
\draw [fill=black,draw=black] (4.500000,5.925000) circle [radius=0.050000];
\draw [fill=black,draw=black] (5.925000,5.925000) circle [radius=0.050000];
\draw [fill=none, draw=black] (6.000000,9.000000) rectangle (3.000000,6.000000);
\draw [fill=black,draw=black] (3.075000,6.075000) circle [radius=0.050000];
\draw [fill=black,draw=black] (4.500000,6.075000) circle [radius=0.050000];
\draw [fill=black,draw=black] (5.925000,6.075000) circle [radius=0.050000];
\draw [fill=black,draw=black] (3.075000,7.500000) circle [radius=0.050000];
\draw [fill=black,draw=black] (4.500000,7.500000) circle [radius=0.050000];
\draw [fill=black,draw=black] (5.925000,7.500000) circle [radius=0.050000];
\draw [fill=black,draw=black] (3.075000,8.925000) circle [radius=0.050000];
\draw [fill=black,draw=black] (4.500000,8.925000) circle [radius=0.050000];
\draw [fill=black,draw=black] (5.925000,8.925000) circle [radius=0.050000];
\draw [fill=none, draw=black] (9.000000,3.000000) rectangle (6.000000,0.000000);
\draw [fill=black,draw=black] (6.075000,0.075000) circle [radius=0.050000];
\draw [fill=black,draw=black] (7.500000,0.075000) circle [radius=0.050000];
\draw [fill=black,draw=black] (8.925000,0.075000) circle [radius=0.050000];
\draw [fill=black,draw=black] (6.075000,1.500000) circle [radius=0.050000];
\draw [fill=black,draw=black] (7.500000,1.500000) circle [radius=0.050000];
\draw [fill=black,draw=black] (8.925000,1.500000) circle [radius=0.050000];
\draw [fill=black,draw=black] (6.075000,2.925000) circle [radius=0.050000];
\draw [fill=black,draw=black] (7.500000,2.925000) circle [radius=0.050000];
\draw [fill=black,draw=black] (8.925000,2.925000) circle [radius=0.050000];
\draw [fill=none, draw=black] (9.000000,6.000000) rectangle (6.000000,3.000000);
\draw [fill=black,draw=black] (6.075000,3.075000) circle [radius=0.050000];
\draw [fill=black,draw=black] (7.500000,3.075000) circle [radius=0.050000];
\draw [fill=black,draw=black] (8.925000,3.075000) circle [radius=0.050000];
\draw [fill=black,draw=black] (6.075000,4.500000) circle [radius=0.050000];
\draw [fill=black,draw=black] (7.500000,4.500000) circle [radius=0.050000];
\draw [fill=black,draw=black] (8.925000,4.500000) circle [radius=0.050000];
\draw [fill=black,draw=black] (6.075000,5.925000) circle [radius=0.050000];
\draw [fill=black,draw=black] (7.500000,5.925000) circle [radius=0.050000];
\draw [fill=black,draw=black] (8.925000,5.925000) circle [radius=0.050000];
\draw [fill=none, draw=black] (9.000000,9.000000) rectangle (6.000000,6.000000);
\draw [fill=black,draw=black] (6.075000,6.075000) circle [radius=0.050000];
\draw [fill=black,draw=black] (7.500000,6.075000) circle [radius=0.050000];
\draw [fill=black,draw=black] (8.925000,6.075000) circle [radius=0.050000];
\draw [fill=black,draw=black] (6.075000,7.500000) circle [radius=0.050000];
\draw [fill=black,draw=black] (7.500000,7.500000) circle [radius=0.050000];
\draw [fill=black,draw=black] (8.925000,7.500000) circle [radius=0.050000];
\draw [fill=black,draw=black] (6.075000,8.925000) circle [radius=0.050000];
\draw [fill=black,draw=black] (7.500000,8.925000) circle [radius=0.050000];
\draw [fill=black,draw=black] (8.925000,8.925000) circle [radius=0.050000];
\end{tikzpicture}
\caption{A simple 2-D Schwarz subdomain, with no h-nonconforming or
  p-nonconforming boundaries. In grey is the element $e$ which is the
  center of the subdomain. The light grey area is the overlap (of size
  $\delta_\xi$) into the other elements. The subdomain is composed of
  everything in light and dark grey. 
}
\label{fig:schwarz_subdomain}
\end{figure}

While the Chebyshev smoother is easy to implement, it is reliant on a robust estimate of the largest eigenvalue and this may not be always true in our case. Thus we also implement an additive Schwarz smoother in the manner of \cite{stiller2017robust}, which is much more robust. The Schwarz smoother works by performing local solves on element-centered subdomains and then adding a weighted sum of these local solves to obtain the smoothed solution. A simple example of one of these subdomains (where the grid is both p and h-uniform) is shown in Figure~\ref{fig:schwarz_subdomain}.
More generally, the Schwarz subdomain centered on element $e_c$
  of the mesh is constructed as follows: Starting with all collocation
  points on $e_c$, one adds all boundary points of neighboring elements
  which coincide with faces, vertices or corners of $e_c$.  Around these
  collocation points, one then
  adds $N_{\rm overlap}-1$ layers of additional collocation points (in Fig.~\ref{fig:schwarz_subdomain}, $N_{\rm overlap}=2$).  If the mesh has non-uniform $h$ or $p$ refinement, the resulting set of collocation points will have ragged boundaries.

The solutions on the individual Schwarz subdomains are
  combined as a weighted sum.  The weights differ with each
  collocation point of each subdomain. In 1-D, for a Schwarz
  subdomain centered on element $e_c$ with left and right neighbours
  $e_{c-1}$ and $e_{c+1}$ we define an extended LGL coordinate $\xi_{\text{ext}}$
  for a collocation point $x$ as follows:
  \begin{equation}
\xi_{\text{ext}}(\tv x) = \begin{cases} 
      \xi^* & \tv x \in e_c, \\
      \xi^* \pm 2 & \tv x \in e_{c\pm1},
\end{cases}
  \end{equation}
where $\xi^*$ is the LGL coordinate of the collocation point $x$ in the reference coordinate system of the macro-element containing $e_c$. This definition takes care of the case when a Schwarz subdomain contains a face that is on a tree boundary.
  
\begin{figure}
  \includegraphics[width=0.45\textwidth,trim=15 0 25 25,clip=true]{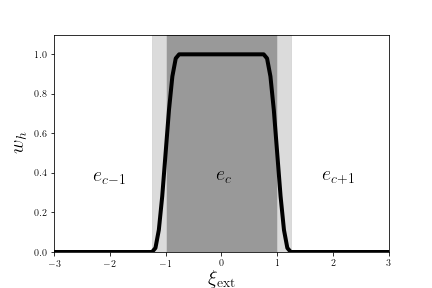}
\caption{
  \label{fig:schwarz_weight}
  A plot of the Schwarz weighting function defined by Eq.~(\ref{eqn:schwarz_weight}). The dark grey shaded region is the center element of the Schwarz subdomain  $e_c$  and the light grey shaded region is the overlap of width $\delta_\xi$. For this plot, we set $\delta_\xi = .25$.
}
\end{figure}

  We denote the overlap size as $\delta_\xi$ and compute it as the width
  of the Schwarz subdomain overlap in the coordinate $\xi_{\text{ext}}$. With
  these definitions we compute the weights for this 1-D subdomain using
  the function $w_h:\mathbb{R}\to\mathbb{R}$ defined as
  
\begin{equation}
  \label{eqn:schwarz_weight}
  w_h(\xi_{\text{ext}}) = \frac{1}{2}\left(\phi\left(\frac{\xi_{\text{ext}}+1}{\delta_\xi}\right) - \phi\left(\frac{\xi_{\text{ext}}-1}{\delta_\xi}\right)\right),
\end{equation}
where the $\phi$ function is given by
\begin{equation}
\phi(\xi) = \begin{cases} 
      \text{sgn}(\xi), & |\xi| > 1, \\
      \frac{1}{8}(15\xi - 10\xi^3 + 3\xi^5), & |\xi| \leq 1.
   \end{cases}
\end{equation}
We have plotted the weighting function $w_h$ in
Figure~\ref{fig:schwarz_weight}.

For Schwarz subdomains in d-dimensions, the weights are computed
as a product of 1-dimensional weights along each dimension,
\begin{equation}
  W = \prod_{i=1}^{d} w_h(\xi^i_{ext}).
\end{equation}

For the Schwarz subdomain solves we
need to define the following operators: $R_s$ is the restriction
operator for a Schwarz subdomain, it reduces the data on the mesh to
the nodes of the subdomain. $R_s^T$ is the transpose of this
operator. $W_s$ are the weights for a Schwarz subdomain, computed by
evaluating Eq.~(\ref{eqn:schwarz_weight}) on the nodes of the
subdomain. With these definitions, the Schwarz smoother algorithm is
listed in Alg.~\ref{alg:schwarz_smoother}. In
Alg.~\ref{alg:schwarz_smoother}, $N_s$ indicates the number of
smoothing cycles (typically, $N_s=3$), $N_{subs}$ indicates the number of subdomains which for our implementation is equal to the number of elements. The linear
system on line 6 of Alg.~\ref{alg:schwarz_smoother} is of the size of the Schwarz-subdomain; we solve it with conjugate gradients to a relative tolerance of $10^{-3}$.

\begin{figure}
  \begin{algorithm}[H]
    \caption{\label{alg:schwarz_smoother}
      Schwarz Smoother\\ $N_{\text{iter,Sch}}$ is a user-defined parameter.
    }
    \begin{algorithmic}[1]
      \Function{SMOOTHER}{$x, b, \mathrm{J}$}  
      \For{$k = 1...N_{\text{iter,Sch}}$}
      \State $r \leftarrow b - {\rm J}x$  
      \For{$s = 1...N_{\text{subs}}$} 
      \State $r_s \leftarrow R_s r$
      \State Solve $R_s \mathrm{J} R^T_s \delta x_s = r_s$
      \EndFor
      \State $x \leftarrow x + \sum_{s}R_s^TW_s\delta x_s$
      \EndFor
      \State \Return $x$
      \EndFunction
    \end{algorithmic}
  \end{algorithm}
\end{figure}
%
%
\subsubsection{Multigrid Inter-mesh Operators}

Let us now turn to the implementation of restriction and prolongation operators.
For clarity, we use the
following notation: a superscript lowercase h refers to the children
elements and an uppercase H refers to the parent element.

We recall our requirement that the polynomial order of a coarse element is smaller than or equal to the polynomial orders of its children elements (cf.~Sec.~\ref{sec:MultigridMeshes}).
This ensures that a parent element's Lagrange
polynomial space is always embedded in the broken Lagrange polynomial space of
its children. In particular, the coarse-mesh basis-functions can be written, exactly, as
\begin{equation}\label{eq:psi-interpolation}
  \psi_\alpha^{H}=(I_{h}^{H})_{\beta\alpha}\psi_\beta^{h},
\end{equation}
where $(I_{h}^{H})_{\beta\alpha}=\psi_\alpha^H(\tv \xi^\beta_h)$ is the interpolation-matrix
of the coarse-mesh basis-functions to the fine-mesh collocation
points, and where we utilized our choice of a \emph{nodal} basis.

For
$h$-refinement, each $\psi_\beta^{h}$ has support in only one of the
child-elements and, consequently, the implicit sum over $\beta$ in
Eq.~(\ref{eq:psi-interpolation}) goes over the basis-elements of all
children elements.

The preceding paragraph immediately suggests the natural definition of the prolongation operator from a coarse-grid function $u^{H}\!=\!u^{H}_\alpha\psi_\alpha^{H}$ to a fine-grid
  function $u^{h}\!=\!u^{h}_\beta\psi_\beta^{h}$ such that $u^{h}\equiv u^{H}$.   This yields
  \begin{equation}
    u^{h}_\beta= u^{H}(\tv \xi^\beta_h)    = (I_{h}^{H})_{\beta\alpha} u^{H}_\alpha,
    \end{equation}so that $(I_{h}^{H})$ represents the prolongation operator.

Because the left hand side operator for the weak equations ---e.g. $L(\cdot,\cdot)$ in Eqn.~(\ref{eq:lh3})--- is a bilinear form, the residual will also be a bilinear form, which we will denote $B(\cdot,\cdot)$; on the coarse grid.  Equation~(\ref{eq:psi-interpolation}) therefore implies
  \begin{equation}\label{eq:temp-G}
    B(\,\cdot\,, \,\psi_\alpha^{H})=(I_{h}^{H})_{\beta\alpha}B(\,\cdot\,,\psi_\beta^{h}),
  \end{equation}
  which defines our coarse grid residual.

Lastly, let us consider restriction of the operators itself.  After linearization  around the current solution $u_0$, the problems we consider here take the form
\begin{equation}\label{eq:temp-F}
\nabla^2\delta u + f(u_0)\delta u = 0,
\end{equation}
for some function $f$. The associated weak form reads
\begin{equation}
L_{h}( \delta u, \psi_\alpha) + \int f(u_0)\psi_\alpha \delta u\,\mathrm{d}\tv x = 0\qquad\forall\; \psi_\alpha.
\end{equation}
The weak Laplacian operator ($L_{h}(\cdot,\cdot)$) can be computed on any coarse grid via Eqn.~(\ref{eq:lh3}) without any need for a restriction operator. The second term will require either a restriction operator on $f(u_0)$, or a restriction operator on $O^{h}_{\alpha\beta}\equiv\int f(u_0)\psi^h_\beta\psi^h_\alpha \mathrm{d} \tv x$.  We choose the latter. By Eq.~(\ref{eq:psi-interpolation}), it holds that 
  \begin{equation}\label{eq:temp-H}
\int f(u_0) \psi_\alpha^{H}\psi_\beta^{H}\,\mathrm{d}\tv x
    =(I_{h}^{H})_{\gamma\alpha}(I_{h}^{H})_{\delta\beta}\int f(u_0)\psi_\gamma^{h}\psi_\delta^{h}\,\mathrm{d}\tv x,
  \end{equation}
  so that the restricted operator in matrix form becomes
  \begin{equation}
    \label{eqn:operator_oh}
      O^{H} = (I_{h}^{H})^T\;O^{h}\;(I_{h}^{H}).
    \end{equation}

\subsection{hp-Adaptivity}
\label{sec:hp-adaptivity}

One of the great advantages of the discontinuous Galerkin method is
that it naturally allows for two different methods of refining the
grid, h-adaptivity, where one subdivides the element into
smaller sub-elements, and p-adaptivity, where one
increases the order $p_e$ of the polynomial basis on an element. The
combination of both h-adaptivity and p-adaptivity is called
hp-adaptivity. If one strategically p-refines in smooth regions of the
mesh and h-refines in discontinuous regions of the mesh, then it is
possible to regain exponential convergence in particular error norms
for problems with non-smooth functions. In the next four sections we
will discuss the four components of our hp-adaptive scheme. These are:
(1) The expected convergence of the solution on a series of adaptively refined hp-meshes (2) an a posteriori error estimator to decide which elements will be refined; (3) a driving strategy that determines based
on the convergence of the error estimator whether to h-refine
or p-refine; (4) an efficient method to apply discrete operators in multi-dimensions on an hp-grid. We discuss these four items in order in the follow subsections.

\subsubsection{Expected Convergence}
\label{sec:expectedconvergence}

 For quasi-linear problems with piecewise-analytic solutions $u$ on polygonal domains, the convergence of the energy norm (Eq. \ref{eq:energynorm}) of the analytical error for the numerical solution $u_h$ is
 \begin{equation}
 \label{eq:expectedconvergence}
 || u - u_h ||_* \leq C\left( \sum\limits_{e \in \mathbb{E}} \frac{ h_e^{2s_e - 2}}{ p_e^{2k_e -3}} ||u||^2_{H^{k_e}} \right)^{1/2}.
 \end{equation}
 Here ${H^{k_{e}}}$ is the Sobolev space on element $e$ with Sobolev order $k_{e}$ and $1 \leq s_{e} \leq \text{min}(p_{e}+1, k_{e})$ \cite{houston2007energy, houston2008posteriori,houston2005discontinuous, bi2015posteriori}. For uniform refinement and uniform Sobolev order k across elements, we expect $|| u - u_h ||_* \leq Ch^{\text{min}(p+1, k) - 1}$. Thus, for smooth problems we expect $|| u - u_h ||_* \leq Ch^{p}$.

 One can show~\cite{schotzau2014exponential} that there exists a series of hp-adaptive refinement steps where the convergence of the energy norm is asymptotically bounded by
 
  \begin{equation}
 \label{eq:expconvergence}
 ||u-u_{h}||_{*} \leq C_{1}\exp(-C_{2}\text{DOF}^{1/(2d-1)}). 
\end{equation}

Here $C_{1}$ and $C_{2}$ are constants, $d$ is the dimension of the mesh and DOF is the number of degrees of freedom, in other words the number of grid points on the mesh.

\subsubsection{A Posteriori Error Estimator}

 Ref.~\cite{bi2015posteriori} derives a local a posteriori
error estimator for an interior penalty hp-DG discretization of the
following class of strongly nonlinear elliptic problems with Dirichlet
boundary conditions
\begin{equation}
\label{eq:classofproblems}
-\nabla \cdot \tv a(\tv x, u, \nabla u) + f(\tv x,\nabla u) = 0 .
\end{equation}
The error estimator can be computed locally on each element by the
following prescription.  Given a discretized solution $u_h$, first define  quantities on each element $e$ or mortar element $m$ by
\begin{subequations}
\label{eq:Eta2_Ingredients}
\begin{align}
R_{e} &\equiv f(u_{h}, \nabla u_{h}) - \nabla \cdot \tv a(u_{h}, \nabla u_{h}),
&& e\in \mathbb{E},\\
     J_{m,1} &\equiv \llbracket \tv a(u_{h}, \nabla u_{h}) \rrbracket_{m},
&&     m\in \mathbb{M_I},\\
     J_{m,2} &\equiv \llbracket u_h \rrbracket_{m},
     &&     m\in \mathbb{M_I},\\
     J_{m,3} &\equiv \llbracket  u_{h} - g_D \rrbracket_{m},
     &&     m\in \mathbb{M}_{\rm D}.
\end{align}
\end{subequations}

From these quantities we compute the followings integrals on an
element e and mortar m:
\begin{subequations}
\label{eq:Eta2_Integrals}
\begin{align}
\eta^{2}_{e} &= h^{2}_{e}p^{2}_{e}||R_{e}||^{2}_{0,e}, \\
\eta^{2}_{m,1} &= h_{m}p^{-1}_{m}||J_{m,1}||^{2}_{0,m }, \\
\eta^{2}_{m,2} &= \sigma h_{m}^{-1}p^{2}_{m}||J_{m,2}||^{2}_{0,m},\\
\eta^{2}_{m,3} &= \sigma h_{m}^{-1}p^{2}_{m}||J_{m,3}||^{2}_{0,m},
\end{align}
\end{subequations}
Here, $|| . ||^2_{0,e} = \int_e ( . )^2 \mathrm{d}\tv x = \int_{[-1,1]^d} J ( . )^2
\mathrm{d}\tv\xi$ and $|| . ||^2_{0,m} = \int_{m} ( . )^2 \mathrm{d} \tv s =
\int_{[-1,1]^{d-1}}S^m ( . )^2 \mathrm{d}\tv\xi$.
  Furthermore, $h_e$ is the diameter of the element $e$, $p_m, h_m$
  and $\sigma$ are defined in Eq.~(\ref{eq:penaltyparameter}) and $J,S^m$ are defined in Eq.~(\ref{eq:Jacobian}) and Eq.~(\ref{eq:Surface-Jacobian}).
  
Following \cite{bi2015posteriori}, we take the local estimator on the element to be
\begin{equation}
\label{eq:Eta2_Local}
\eta^2(e) =\eta^{2}_{e} + \sum_{m \subset \mathbb{M}_e}\eta^{2}_{m,1} + \sum_{m \subset \mathbb{M}_e\setminus \mathbb{M}_D}\eta^{2}_{m,2} + \sum_{m \subset \mathbb{M}_e \cap \mathbb{M}_D}\eta^{2}_{m,3},
\end{equation}

where $\mathbb{M}_e$ and $\mathbb{M}_d$ are the sets of mortar elements touching the element $e$ and  $\partial \Omega_d$ respectively (see Sec.\ref{sec:mesh}). 
The estimator provides an estimate of the error in the energy norm, $ || u - u_h ||_* $, see Eq.~(\ref{eq:energynorm}).  Similar error estimators for various classes of linear and non-linear elliptic PDEs can be found in \cite{houstonschotzau05,hansbo2011energy,zhu2011energy,houston2007energy,schotzau.d;zhu.l2009,houstonperugia05,lovadina.c;marini.l2009}.
  Equations~(\ref{eq:Eta2_Integrals}) are typically computed on the physical grid on which the elliptic PDE is solved.
  For highly stretched grids (for instance, when a mapping inverse in
  radius is used to push the outer boundary to very large radius
  $R\sim 10^3 \cdots 10^{10}$), the geometric factors in
  Eqs.~(\ref{eq:Eta2_Integrals}) can distort the error estimates.  For
  those cases, we will sometimes introduce a fiducial grid of
  identical structure and connectivity, that avoids or mitigates
  excessive grid-stretching.  Once a fiducial grid is chosen, its geometric properties can be used in Eqs.~(\ref{eq:Eta2_Integrals}) and the corresponding norms.  Below, we demonstrate that such a ``fiducial-grid'' error-estimator allows
  efficient hp-refinement even in highly stretched computational domains.  This problem is encountered below in
Sec.~\ref{sec:CompactifiedLorentizan} and discussed there in greater
detail.

\subsubsection{Driving Strategy}

The a posteriori estimator
Eq.~(\ref{eq:Eta2_Local}) indicates which elements to refine.  Then the choice must be made for each cell, whether to h-refine or p-refine elements with a large error $\eta(e)$. In the survey paper \cite{mitchell2011survey}, Mitchell and McClain looked at 15 different hp-adaptive strategies with a finite elements scheme. Their results indicate that the strategy known as smooth-pred, first introduced in~\cite{melenk2001residual} performs quite well, if not the best under
the example problems they tested. For this reason alone we use it, but
our code is flexible enough to use any of strategies studied in
Mitchell and McClain's paper.

The idea behind the smooth-pred strategy is based on the observation that for a locally smooth solution, the
energy norm Eq.~(\ref{eq:energynorm}) $\eta(e)$ will converge as
$h^{p}$, (see~Sec.~\ref{sec:expectedconvergence}).  To take advantage of this observation, we first predict the error in the next refinement step and test if this prediction satisfies the smooth error convergence law, if so we p-refine, otherwise we h-refine. The full algorithm is given in Algorithm~\ref{alg:hpamr}.

\begin{figure}
  \begin{algorithm}[H]
    \caption{\label{alg:hpamr}
      hp-AMR Driving Strategy\\
      $\gamma_{h}$ and $\gamma_{p}$ are user-defined parameters.
    }
    \begin{algorithmic}[1]
      \Procedure{SMOOTH-PRED}{}
        \If{$\eta^2(e)\text{ is large}$}
          \If{$\eta^2(e) > \eta^2_{\rm pred}(e)$}
           \State h-refine element
          \State $\eta^2_{\rm pred}(e_{\textrm{children}}) \gets \gamma_{h}\,\big(\frac{1}{2}\big)^{d}\big(\frac{1}{2}\big)^{2p_e}\eta^{2}(e)$
          \Else
                     \State p-refine element
          \State $\eta^2_{\rm pred}(e) \gets \gamma_{p_e}\eta^2(e)$
          \EndIf
        \EndIf
      \EndProcedure
    \end{algorithmic}
  \end{algorithm}
\end{figure}

In Algorithm~\ref{alg:hpamr}, $\eta^2(e)$ is the square of the error
estimator, Eq.~(\ref{eq:Eta2_Local}) of the element under
  consideration, and $\eta_{\text{pred}}^2(e)$ is the predicted error
  estimator from the last AMR step assuming the solution on the
  element is smooth. We always start with $\eta_{\rm pred}(e)=0$, so that
  each element will first be $p$-refined, before $h$-refinement is
  considered.

  The parameters $\gamma_p$ and $\gamma_h$
      influence the behavior of Alg.~\ref{alg:hpamr} as follows: As
      long as the actual $\eta^2(e)$ is \textit{small} compared to the
      predicted $\eta^2_{\rm pred}(e)$ , Line 3 implies continued
      $p$-refinement.  Since Line 8 reduces the predicted $\eta^2_{\rm
        pred}$ by a factor $\gamma_p$, this means, that $p$-refinement
      continues as long as each increment in $p_e$ reduces the
      error-estimator $\eta^2(e)$ by a factor $\gamma_p$, i.e.\ as
      long as exponential convergence is obtained with convergence
      rate better than $\sqrt{\gamma_p}$.  If this convergence-rate is
      not observed, the driver switches to $h$-refinement, and
      $\gamma_h$ begins to matter.  In this case, a large $\gamma_h$
      will preferably switch back to $p$-refinement, whereas a small
      $\gamma_h$ will prefer continued $h$-refinement.  All runs shown
      in the remainder of this paper use $\gamma_p=0.1$.  The runs
      differ in $\gamma_h$, which is used to tune h- vs p-refinement,
      and in how many elements are refined in each AMR-iteration,
      which is used to control how quickly AMR increases the number of
      degrees of freedom. However for the majority of the runs we find that $\gamma_h=0.25$ is a good choice.
  
In the numerical examples of this paper we use the
  following two alternatives for the conditional on line 2 of
  Alg.~\ref{alg:hpamr}, i.e.\ to decide which elements should be
  refined: In the first test-problem, we refine if $\eta^2$ is
greater than some constant factor times the mean $\bar \eta^2$ across all elements $e$. This criterion was used in the original description of the hp-AMR scheme (See \cite{melenk2001residual}).  In
the remaining problems we refine a percentage of elements with the largest $\eta(e)$. We change criterion because the estimator may vary over orders of magnitude and this variation may change at each refinement step. In such cases, the percentage criterion robustly captures more of the elements with a large estimator than thresholding on a constant times the average estimator). However, the constant factor times the
mean method has the advantage that it is computationally cheap and
does not require a global sort of the estimator over all cores like the
percentage method does. When we use the percentage indicator, we make use of the highly parallel global sort outlined in \cite{feng2015mp}.

\subsubsection{On-the-fly hp Tensor-Product Operations}
\label{sec:tensorproduct}
To solve the elliptic equations with multigrid on grids that are hp-adaptively refined, a large number of different operators must be generated at run-time. In no specific order, we need at least the following linear operators

\begin{itemize}
\item interpolation operators on faces, edges and volumes of size $(p'+1) \times (p+1)$ in 1-D, where p is the order of the original data and p' is the order of the new interpolated data
\item restriction operators on faces, edges and volumes of size $(p+1) \times (p'+1)$ in 1-D, where p' is the order of the original data and p is the order of the new restricted data
\item derivative operator of order p on the reference element, which has size $(p+1) \times (p+1)$ in 1-D.
\end{itemize}

We can take advantage of the tensor product nature of the reference element in order to efficiently generate these operators on demand. All of the operators
listed can be reduced to a tensor product of 1-D operators when applied in $d\ge 2$ dimensions. Upon requiring a certain operator in a calculation, the process is then as follows

\begin{enumerate}
\item Check if the 1-D version of the operator has already been computed in the database, if it has been computed, retrieve and goto step 3, if not goto step 2.
\item Compute the 1-D version of the operator and store in the MPI process-local database.
\item Apply the 2-D or 3-D version of the operator on a vector by using optimized Kronecker product matrix vector operation on the nodal vector v: $(A_{1D} \otimes B_{1D})\vec{v}$ for $d=2$, or $(A_{1D} \otimes B_{1D} \otimes C_{1D})\vec{v}$ for $d=3$. Here $A,B, C$ denote the 1-D matrices of the respective operator, as applied to the first, second, and third dimension.
\end{enumerate}

A Kronecker product matrix vector operation can be efficiently performed using a series of BLAS matrix-multiply calls, see \cite{buis1996efficient} for example.

\subsection{Implementation}

To implement the multi-block adaptive mesh refinement we use the p4est library, which has been shown to scale to $\mathcal{O}$(100,000) cores \cite{burstedde2011p4est}. We use PETSc \cite{petsc_home_page} for the Krylov subspace linear solves and the Newton Raphson iterations. The components of the multigrid algorithm are written by the authors and do not use PETSc.

\section{Test Examples}

\label{sec:testexamples}
We examine the components of our code through three test examples, the first a linear Poisson problem solved on a square grid where the solution is only C$^{2}$-smooth at the $(0,0)$ grid point. The second example is a non-linear elliptic problem, where we solve the Einstein constraint equations for the initial data of a constant density star. We then end the test examples section with a linear problem on a cubed-sphere with stretched outer boundary and a solution that falls off as $r^{-1}$ with radial coordinate r as $r \rightarrow \infty$. Each test is aimed at isolating different aspects of the puncture black-hole problem, whose solution contains points that are C$^{2}$-smooth and falls off as $r^{-1}$. The puncture black-hole problem further requires us to solve nonlinear Einstein constraint elliptic equations on a cubed-sphere mesh.

\subsection{Poisson with $H^{4-\epsilon}$ solution}

The first test problem we will investigate numerically is taken from \cite{stamm2010hp}, where the authors solve $\nabla^{2}u = f$ on $\Omega = (0,1)^{2}$ with the solution chosen as

\begin{equation}
    \label{eq:testexample1solution}
 u = x\left(1-x\right)y\left(1-y\right)\left[\left(x-\frac{1}{2}\right)^2 + \left(y-\frac{1}{2}\right)^2\right]^{3/2}.
\end{equation}

\begin{figure}
  \centering
  \includegraphics[width=.47\textwidth,trim=10 0 40 30, clip=true]{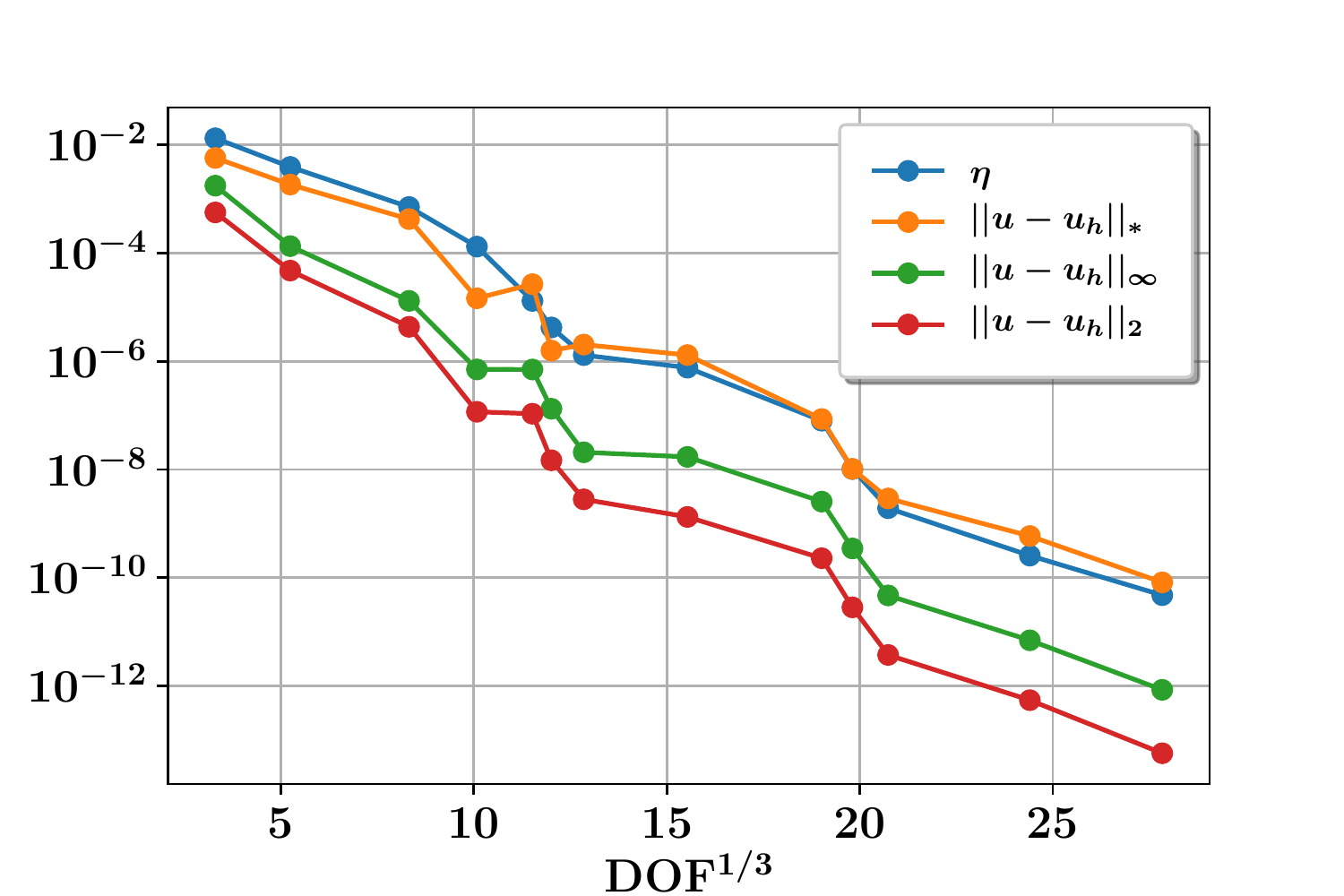}
  \caption{Problem A, Eq.~(\ref{eq:testexample1solution}): Convergence of the energy norm estimator $\eta^2$ and the error between the numerical solution $u_h$ and the analytic solution $u$ in the energy norm $||\cdot||_*$ (Eq. \ref{eq:energynorm}) and the $L_2$ norm $||\cdot||_2$ (Eq. \ref{eq:l2norm}).}
  \label{fig:test_example_1_u_convergence} 
\end{figure}

\begin{figure}
  \includegraphics[width=.47\textwidth,trim=500 450 400 680,clip=true]{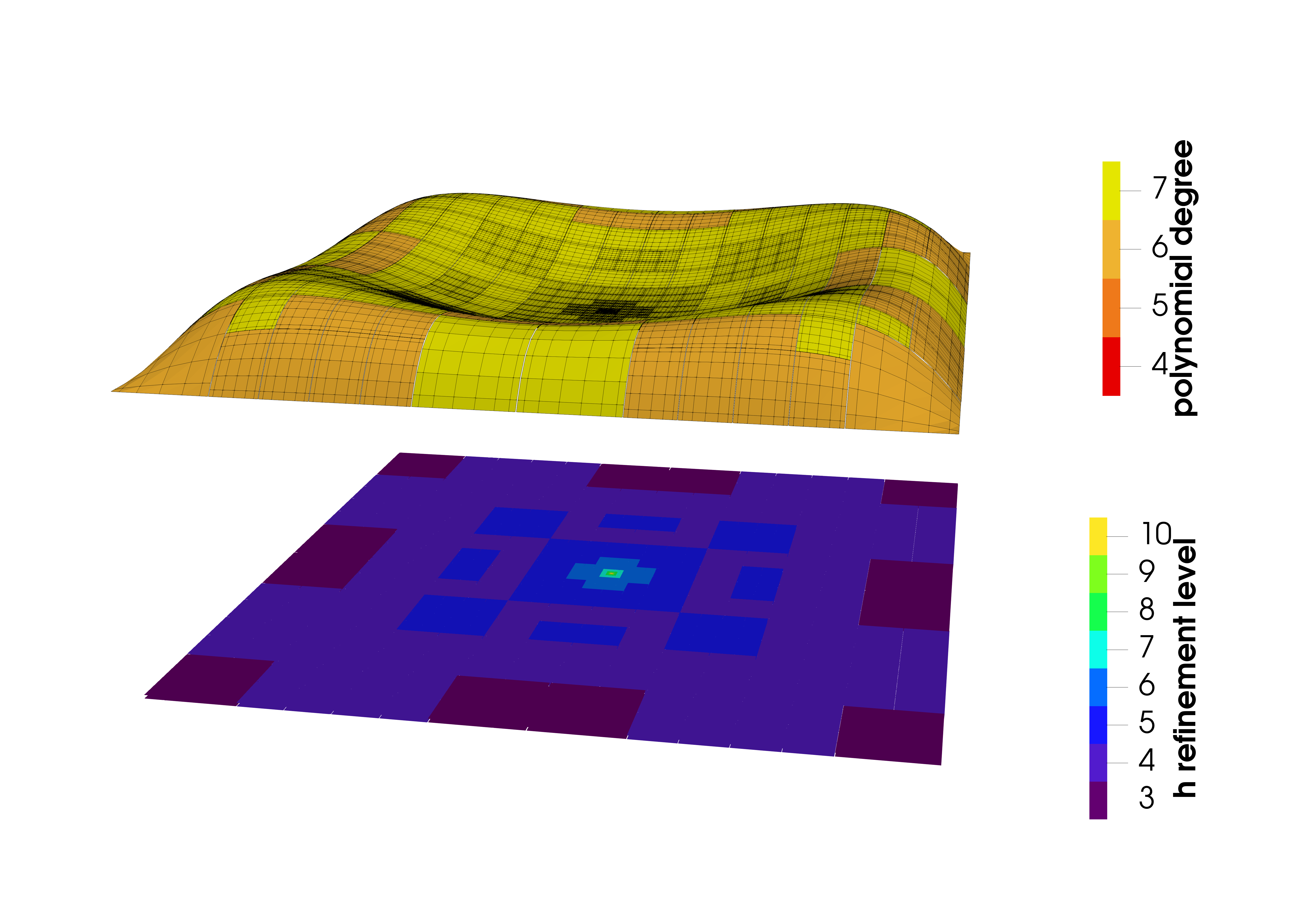}
  
  \caption{ \label{fig:test_example_1_mesh} Problem A,
    Eq.~(\ref{eq:testexample1solution}): Visualization of the
      solution and the hp-refined computational mesh.  {\bf Top
        portion:} The computational grid, color-coded by polynomial
      degree, with height representing the solution $u$.  {\bf Bottom
        portion:} Color-coded by $h$-refinement level.  A cell on
      level $l$ has size $2^{-l}$ of the overall computational
      domain.}
\end{figure}

Here $u \in H^{4-\epsilon}$, where $\epsilon > 0$ so from
Eq.~(\ref{eq:expectedconvergence}) we expect third order convergence
for uniform refinement. We have confirmed this numerically. However,
with hp-adaptivity, it is possible to achieve $||u-u_{h}||_{*} \sim
\exp(\text{DOF}^{\frac{1}{3}})$, as we show in
Fig.~\ref{fig:test_example_1_u_convergence}. Notice also the close
agreement between the estimator and energy norm $||u-u_{h}||_*$. The final mesh is shown in Fig.~\ref{fig:test_example_1_mesh}. Of
note is the fact that the elements with the lowest polynomial order
(p=4) are only in the vicinity of the $C^{2}$-smooth $(0,0)$
grid-point and the refinement level of the mesh is locally much higher
here as well. Here we
used AMR parameters $\gamma_h = 10$, $\gamma_p = 0.1$; at each AMR
iterations all cells were refined which have $\eta^2(e)$ larger than
$1/4$ of the overall mean of all $\eta^2(e)$. For this
problem, we performed a survey of different choices for $\gamma_h$
and $\gamma_p$.  We found rather modest dependence on $\gamma_h$
within $0.25\lesssim \gamma_h\lesssim 10$, with $\gamma_h\!=\!10$ leading to the good balance between
$h-$ and $p-$refinement shown in Fig.~\ref{fig:test_example_1_mesh}.
Our preferred value for the rest of the paper ($\gamma_h=0.25$) also
yields good convergence, but leads to a more uniform polynomial
degree through the mesh.

\begin{figure}
  \centering
  \includegraphics[width=.45\textwidth,trim=10 0 0 10,clip=true]{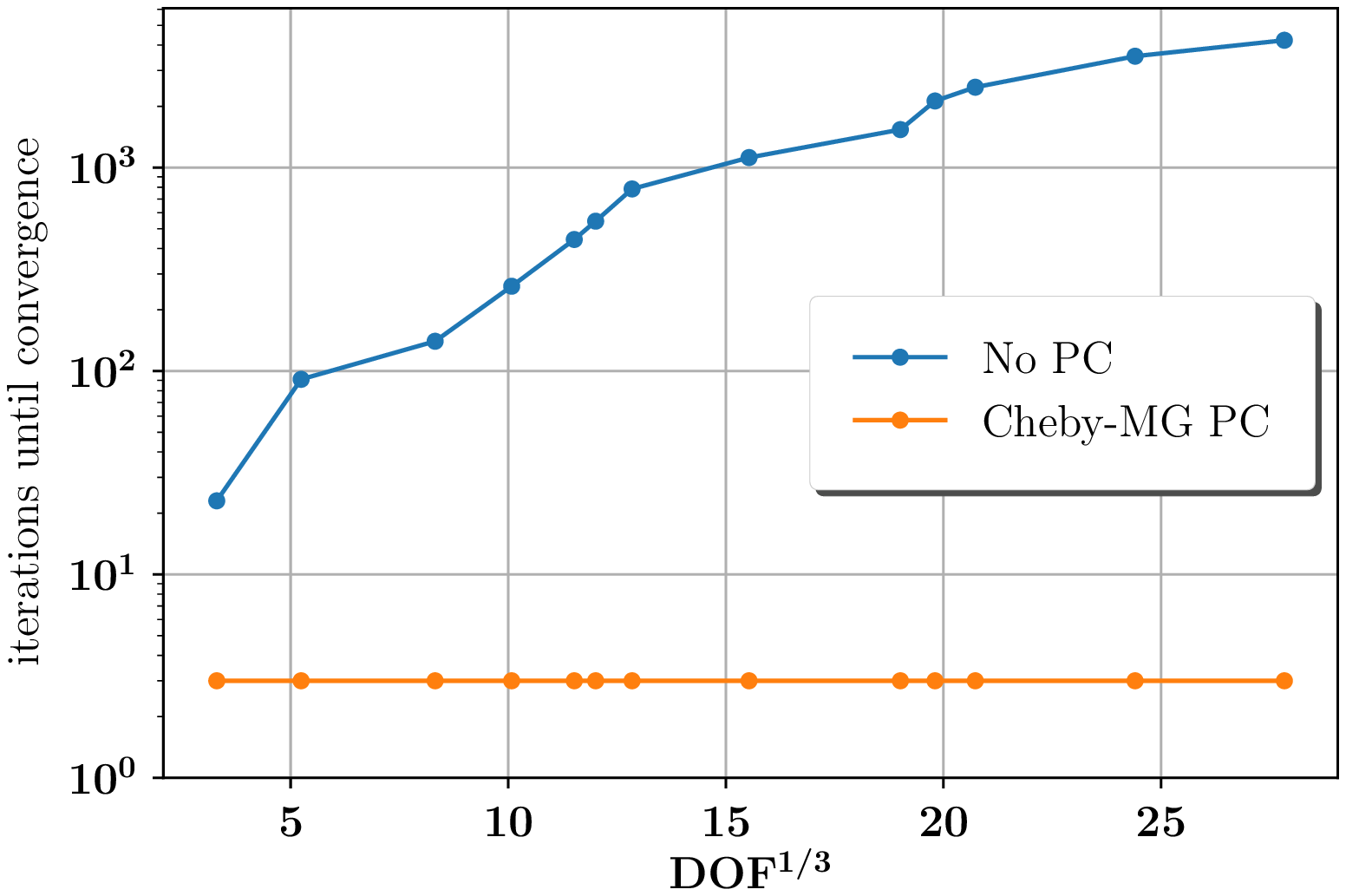}
  \caption{Problem A, Eq.~(\ref{eq:testexample1solution}): Iteration-count of the flexible conjugate gradient (FCG) Krylov subspace solver versus the Multigrid preconditioned FCG solver (MG-FCG). The MG-FCG completes all solves in $\leq 3$ iterations.}
  \label{fig:mgfcg_vs_fcg}
\end{figure}

\subsection{Constant density star}

For the next test problem, taken from ~\cite{baumgarte2007},  we solve the Einstein constraints in the simplest possible scenario, a constant density star. The goal of this test problem is to investigate how the elliptic solver behaves for problems that contain surface discontinuities that mimick surface and phase transition discontinuities in a Neutron star. The Einstein constraint equations for the case of a constant density star reduce to
\begin{equation}
\label{eq:Constant_Density_Star_PDE}
 \nabla^{2}\psi + 2\pi \rho \psi^{5} = 0,
\end{equation}
where $\rho$ is the density of the star and $\psi$ is the conformal factor which describes the deviation of the space from flat space. In \cite{baumgarte2007} the authors solve the above equation for the case of a star with radius $r_0$ and mass-density
\begin{equation}
\rho = \begin{cases} 
      \rho_{0} & r\leq r_0 \\
      0 & r > r_0,
   \end{cases}
\end{equation}
where r is the radial spherical polar coordinate. Since the star is in isolation, the boundary condition at infinity is $\psi = 1$, corresponding to a asymptotically-flat space. For such a problem, there is an analytic solution given by
\begin{equation}
\psi = \begin{cases} 
      Cu_{\alpha}(r) & r\leq r_0 \\
     \frac{\beta}{r} + 1 & r > r_0.
   \end{cases}
\end{equation}
with $C=(2\pi\rho_{0}/3)^{-1/4}$ and

\begin{equation}
u_{\alpha}(r) \equiv \frac{(\alpha r_0)^{1/2}}{(r^{2} + (\alpha r_0)^{2})^{1/2}}.
\end{equation}

The parameters $\alpha$ and $\beta$ are determined from the continuity of $\psi$ and it's first derivative at the surface of the star, and are given by

\begin{align}
 &\beta = r_0(Cu_{\alpha}(r_0) - 1) \\
 &\rho_{0}r_0^{2} = \frac{3}{2\pi}\frac{\alpha^{10}}{(1+\alpha^{2})^{3}}
\end{align}

\begin{figure}
  \centering
  \includegraphics[width=.45\textwidth,trim=10 0 30 34, clip=true]{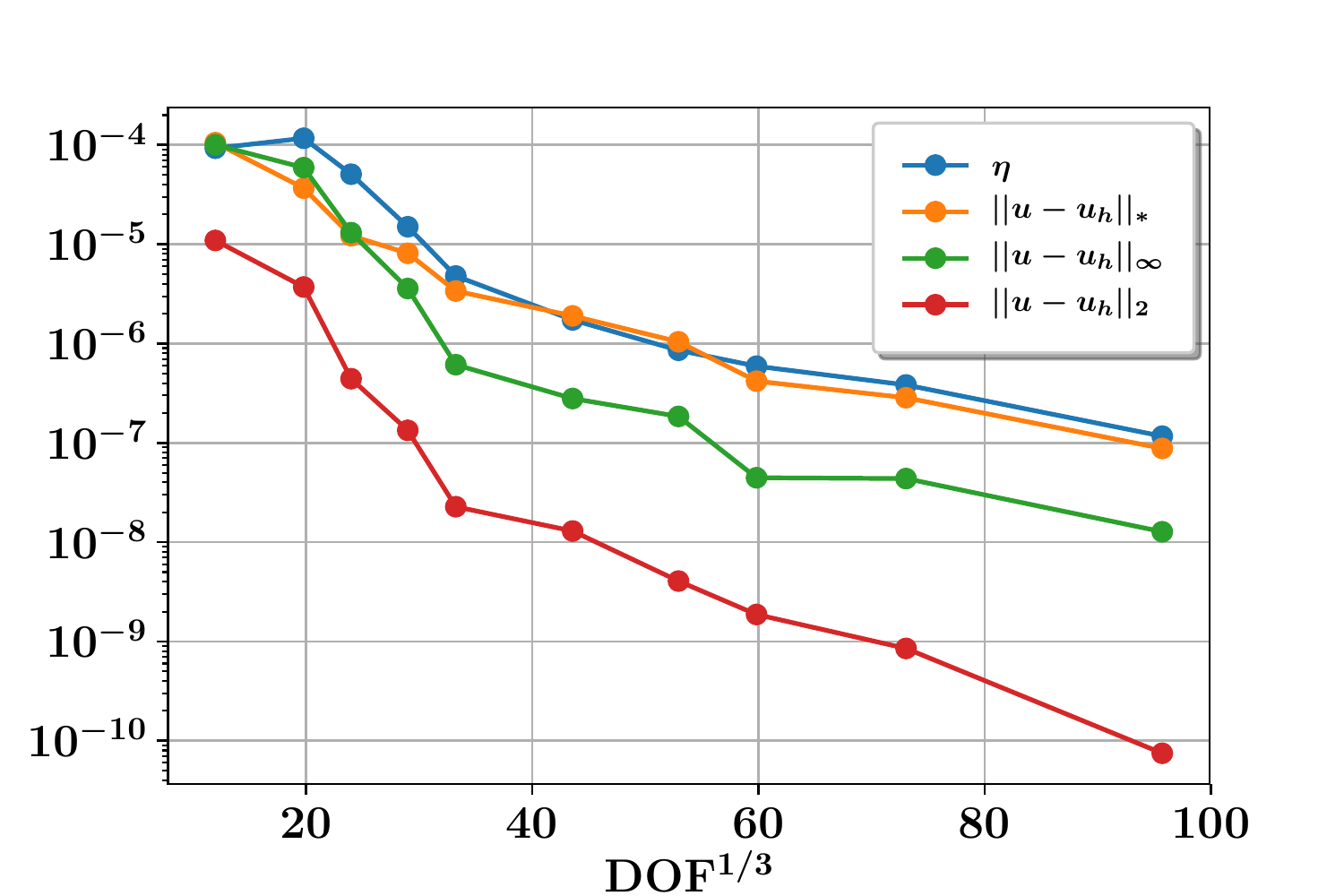}
  \caption{Problem B, Eq.~(\ref{eq:Constant_Density_Star_PDE}): Convergence of the energy norm estimator $\eta^2$ and the error between the numerical solution $u_h$ and the analytic solution $u$ in the energy norm $||\cdot||_*$ and the $||\cdot||_2$ norm}
  \label{fig:problem_b_convergence} 
\end{figure}

\begin{figure}
  \centering
 \includegraphics[width=.47\textwidth
    ]{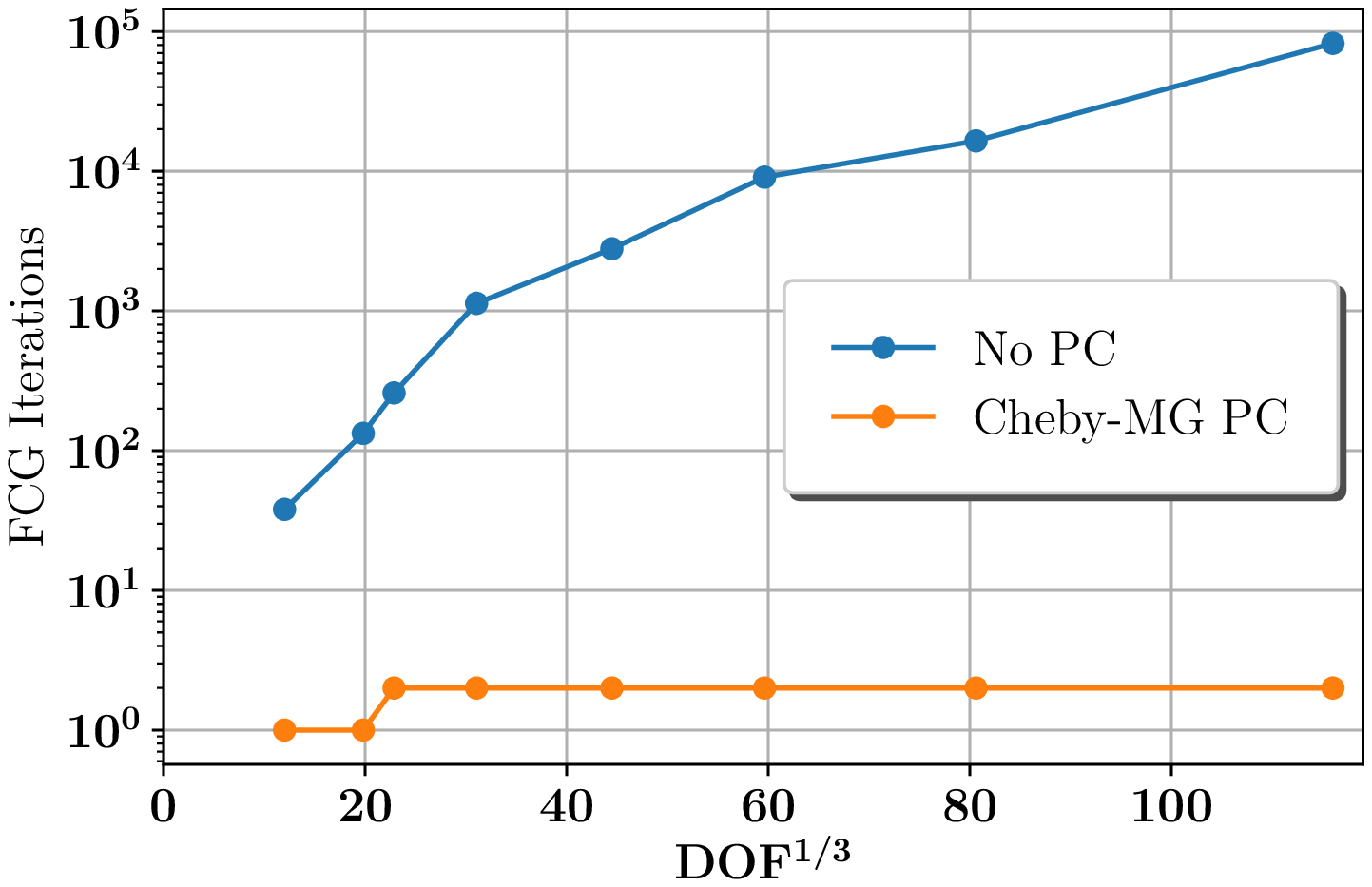}
  \caption{Comparison of the average number of iterations per Newton-Raphson step when using a Chebyshev smoothed Multigrid preconditioner and no preconditioner.}
  \label{fig:problem_b_pc_comparison} 
\end{figure}

\begin{figure}
  \centering
  \includegraphics[width=.47\textwidth,trim=1500 400 800 520,clip=true]{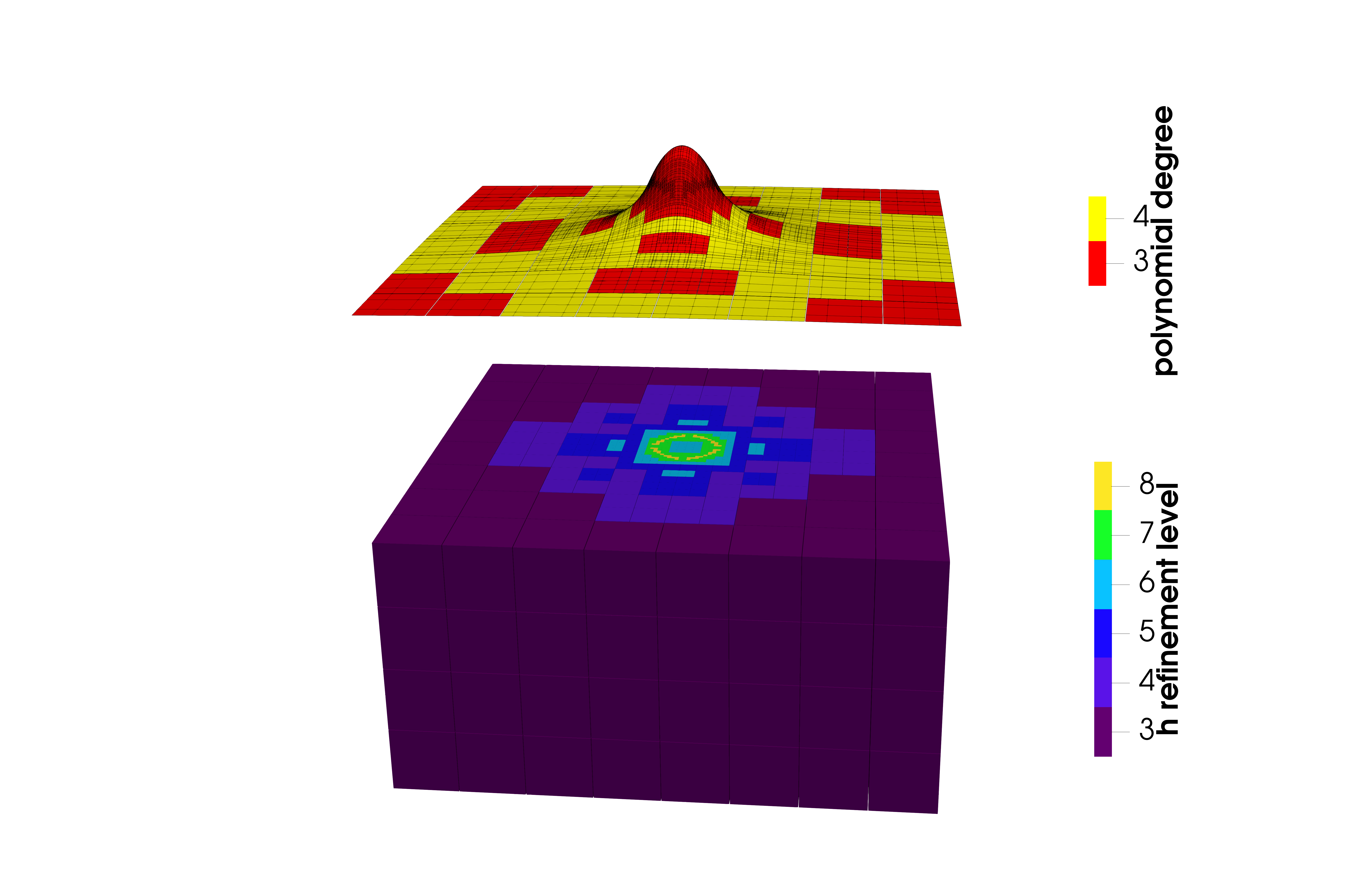}
  \caption{ \label{fig:problem_b_mesh} Visualization of
      hp-refined computational mesh for Problem B,
      Eq.~(\ref{eq:Constant_Density_Star_PDE}).  {\bf Top portion:}
      The $z=0$ cross-section of the computational grid, color-coded
      by the polynomial degree and with height representing the
      solution $\psi$.  {\bf Bottom portion:} Volume rendering of the
      $z\le 0$ part of the computational domain, color-coded by the
      $h$-refinement level.  A cell on level $l$ has size $2^{-l}$ of
      the overall computational domain. }
\end{figure}

We solve the above problem on a cubic domain with $\rho_0 =
0.001$ and analytic Dirichlet boundary conditions on a boundary at
$8r_0$. The non-linear term in Eqn.~\ref{eq:Constant_Density_Star_PDE} is handled by first interpolating $\psi$ onto the GL quadrature points of an element, evaluating $\psi^{5}$ at the GL quadrature points and then performing the necessary Gaussian quadrature sum on each element.
Figure~\ref{fig:problem_b_convergence} showcases the
convergence of the solution and the nice agreement between the
energy norm estimator $\eta^2$ and the energy norm of the analytic
error. To achieve this convergence we used the following AMR parameters, $\gamma_h = 0.25$, $\gamma_p = 0.1$.  At each AMR refinement iteration, 10\% of the elements are refined. In Figure~\ref{fig:problem_b_pc_comparison} we show a comparison between multigrid-preconditioned FCG iterations and unpreconditioned FCG iterations. We notice that for the preconditioned case, the iteration count is roughly constant with increases in DOF, whereas the unpreconditioned iterations grow with DOF. Finally, in Figure~\ref{fig:problem_b_mesh} we show the mesh after the last AMR step. This mesh showcases the highest h-refinement around the star boundary (yellow circle) as expected, since this area is not smooth. 
 
\subsection{Cubed sphere Meshes and Stretched Boundary Elements}
  \label{sec:CompactifiedLorentizan}

\begin{figure}
  \centering
  \includegraphics[width=.45\textwidth,trim=700 450 470 420, clip=true]{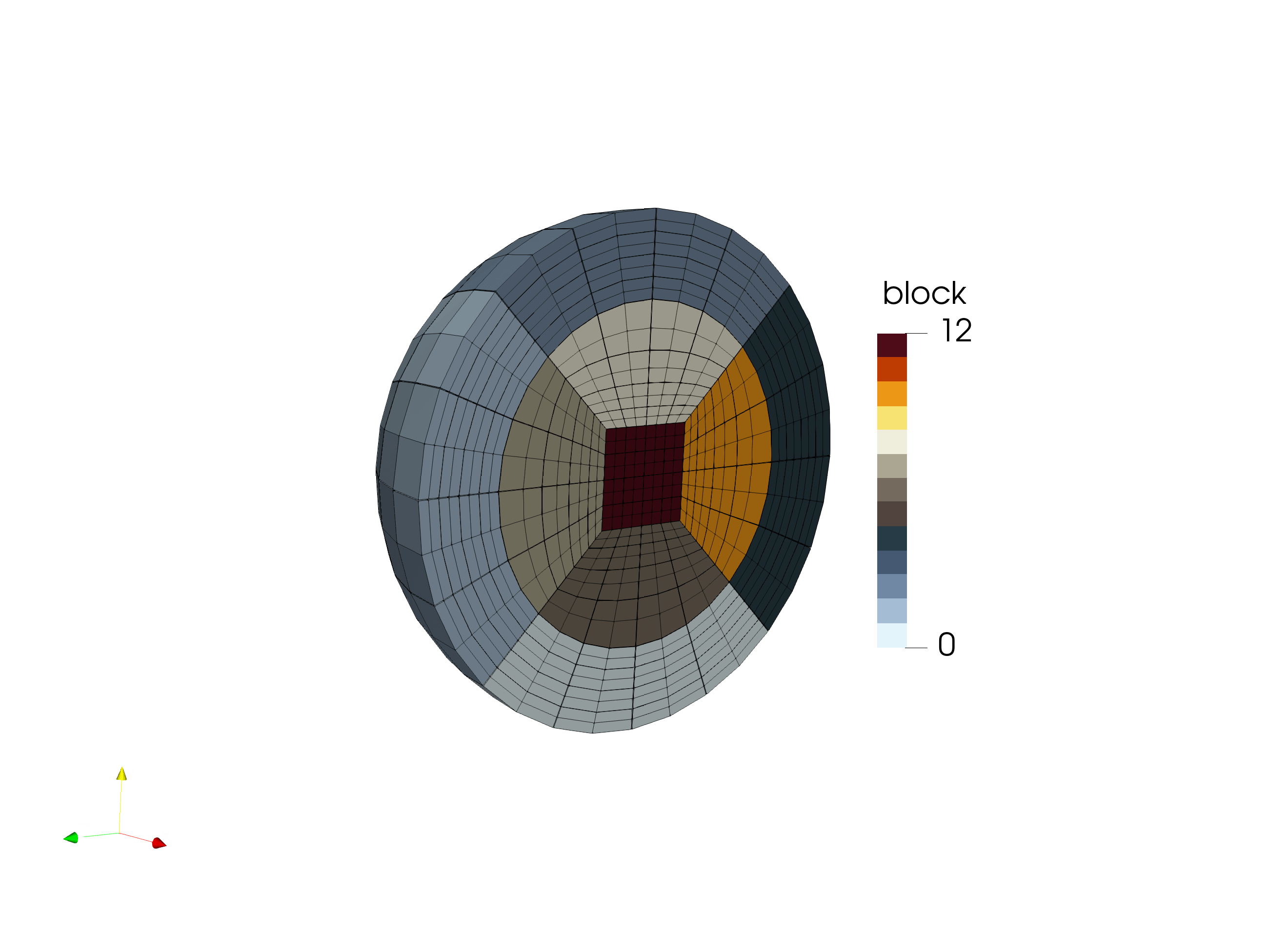}
  \caption{
The mesh structure for a computational domain with spherical outer boundary.  This structure consists of 13 macro-meshes (shown in different colors).  For clarity, only the $z\le 0$ part of the mesh is shown. }
  \label{fig:cubed_sphere_mesh}
\end{figure}

So far, we have only investigated meshes with a regular,
  Cartesian structure and with a rectangular outer boundary at close
  distance.  We will now investigate scenarios with a spherical outer
  boundary, where we use a macro-mesh arising from the 3-dimensional
  generalization of Fig.~\ref{fig:macromesh} (a).  We will also place
  the outer boundary at very large radius, typically $10^9$, to
  approximate boundary conditions at infinity. Typically, such problems have solutions which fall off as a
power series in $1/r$.

Figure~\ref{fig:cubed_sphere_mesh} shows the structure of the mesh we
will use: a central cube, surrounded by \emph{two} layers of six
deformed macro-elements each.  The inner layer interpolates from the
cubical inner region to a spherical outer region.  The outer layer has
spherical boundaries both at its inner and outer surface, and thus
radial coordinate lines that are always orthogonal to the angular
coordinate lines.  This allows to apply a \textit{radial coordinate
  transformation} in the outer layer, to move its outer boundary to
near infinity. Because we know the solution will fall off as $1/r$ we
use an inverse mapping in the outer six spherical wedges of the
cubed-sphere (blocks 0-5 in Fig.~\ref{fig:cubed_sphere_mesh}). This
mapping is defined as follows. Denote the physical grid variable with
$r\in[r_1,r_2]$ and the collocation-point integration variable as
$x\in[x_1,x_2]$, then the inverse mapping is defined by
\begin{equation}
  \label{eqn:inverse_mapping}
  r = \frac{m}{x-t},
\end{equation}
where
\begin{equation}
  m = \frac{x_2-x_1}{\frac{1}{r_2} -\frac{1}{r1}},\qquad
  t = \frac{x_1r_1-x_2r_2}{r_1 - r_2}.
\end{equation}

We will investigate the discontinuous Galerkin method on
the following test problem which captures many of the above features:
\begin{equation}\label{eq:Lorentzian}
\nabla^2 u = 3 (1 + x^2 + y^2 + z^2)^{-\frac{5}{2}},
\end{equation}
with $\Omega = \{(x,y,z) : x^2 + y^2 + z^2 < R\}$ and Dirichlet
boundary conditions given by the analytical solution, which is a
Lorentzian function $u = (1 + r^2)^{-1/2}$. The Lorentzian function
falls off as a power series in $1/r$ as $r \rightarrow \infty$.

To solve this test problem we run two schemes: uniform $p$-refinement
and adaptive p-refinement.
We run only with p-AMR because the underlying
solution is smooth everywhere so there is very little to no benefit in
running with hp-AMR for this problem (which we also found to be the
case empirically).
For the uniformly refined run, we start with refinement level $l=4$, i.e.\ with $2^{3l}=4096$ elements in each of the 13 macro-elements, and increase $p$ from $2$ to $11$.

\begin{figure}
  \centering
  \includegraphics[width=.47\textwidth,trim=12 14 8 14,clip=true]{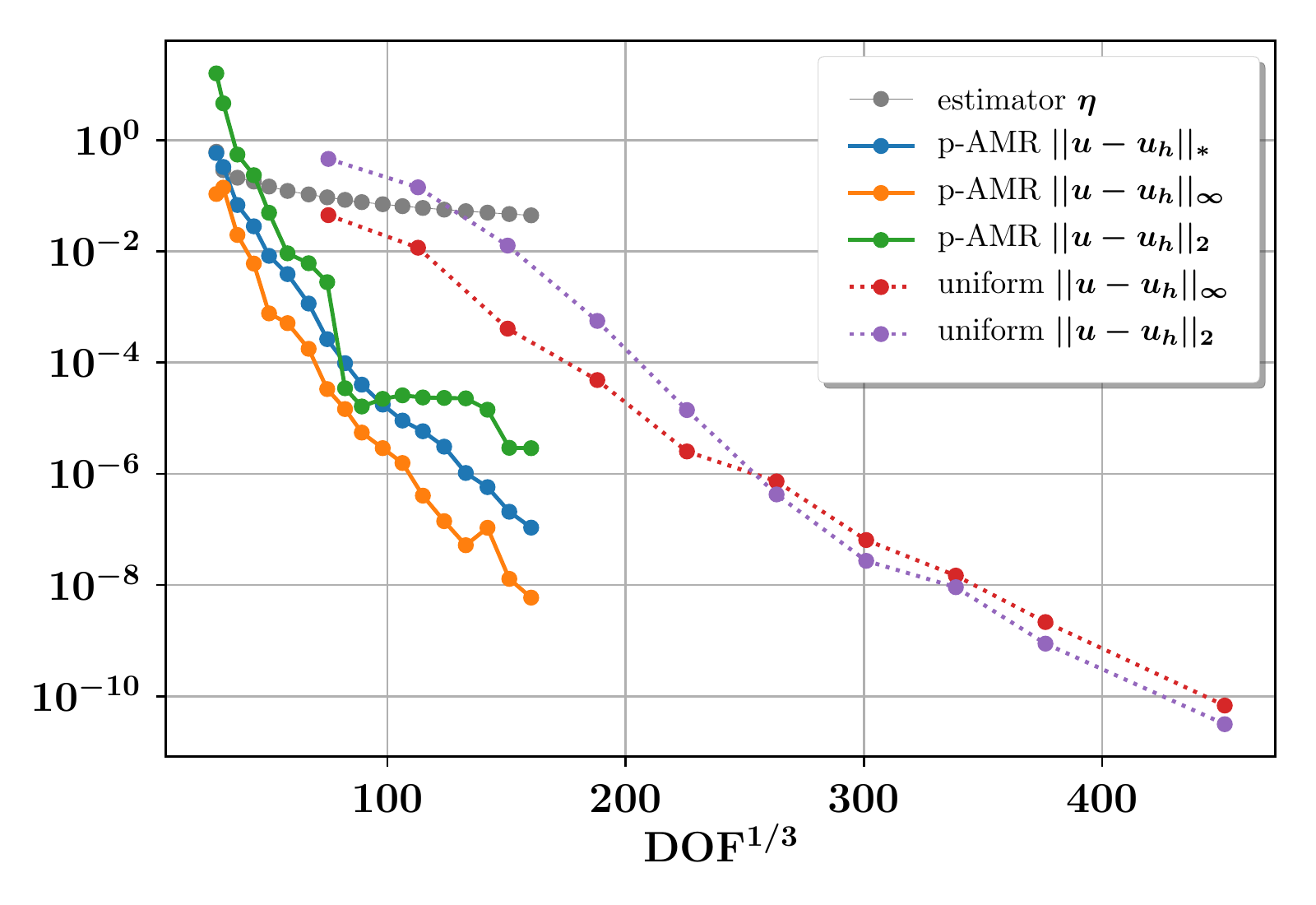}
  \caption{ \label{fig:prob_c_convergence} Problem C,
    Eq.~(\ref{eq:Lorentzian}): Convergence of the solution as the degree p is uniformly
    increased across all elements for uniform case and as the degree p
    is adaptively increased in the amr case.  (Dirichlet BC at
    R=1000).
  }
\end{figure}

\begin{figure}
  \centering
\includegraphics[width=.47\textwidth,trim=2800 1420 3000 1120,clip=true]{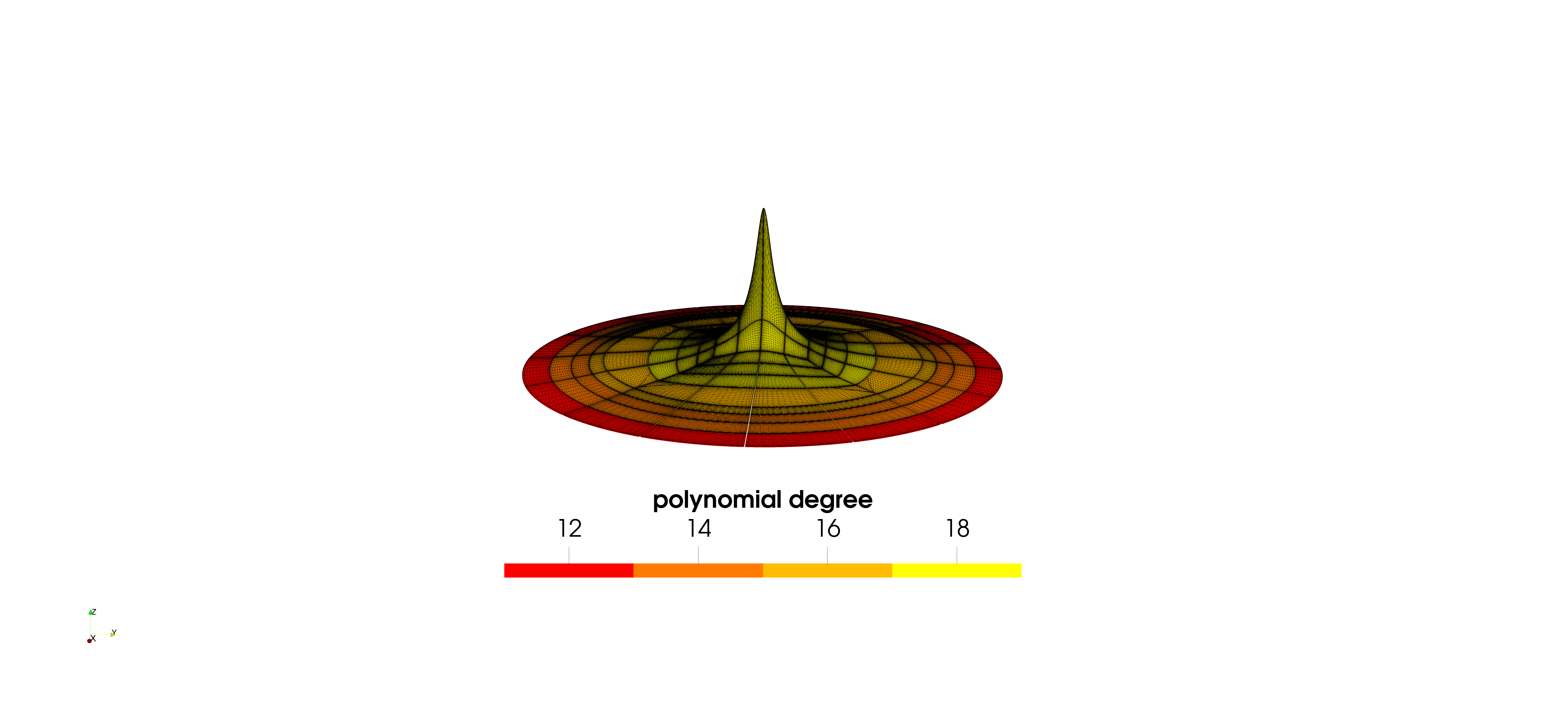}
    \includegraphics[width=.47\textwidth,trim=2220 800 2580 2630,clip=true]{./prob_c_mesh.png}
  \caption{ Visualization of the final mesh for problem C. Shown is a
    xy-plane slice of the mesh which has been warped so that the
    height of the surface corresponds to the value $u$ of the
    solution, color coded by the polynomial degree. For ease of visualization, grid-points are mapped onto the compactified grid on which the estimator $\eta_e$ is computed; the compactified outer radius $R=3$ corresponds to the physical outer radius $R=1000$.}
  \label{fig:problem_c_mesh}
\end{figure}

To achieve pure p-AMR with the hp-AMR scheme
outlined in Sec.~\ref{sec:hp-adaptivity}, we
start with refinement level $l=2$, i.e. with $2^{3 l}=64$ elements in each of the 13 macro-elements, and with polynomial order $p_e=1$ in all elements. We
set $\gamma_h\! =\! 10^6$,
$\gamma_p\! =\! 10^6$ and in each AMR iteration, we refine the $25\%$ elements with the
largest estimator. Figure \ref{fig:prob_c_convergence} shows the
convergence of the p-AMR scheme versus the p-uniform scheme.  We stop
at the maximum polynomial order currently allowed in the
code, $p\!=\!19$. Comparing the $L_\infty$ norm between the p-AMR and p-uniform
runs, the number of degrees of freedom per dimension,
$\mbox{DOF}^{1/3}$, is roughly cut in half.
The $L_2$ norm is slower
to converge because it is dominated by the contributions of the outer
stretched wedges, and this region is less aggressively refined by the
AMR.
Figure~\ref{fig:problem_c_mesh} shows the
final mesh for the problem.  The AMR algorithm increases the polynomial order $p$ predominantly in the centre, where the solution has the most structure.

  Strong coordinate stretching, as performed via the inverse map Eq.~(\ref{eqn:inverse_mapping}) leads to the following problem:
 The error estimator $\eta^2(e)$ utilizes integrals in
  physical coordinates in Eqs.~(\ref{eq:Eta2_Ingredients}).  For
  strongly stretched grids (e.g. with inverse mappings where $R$ is
  orders of magnitude larger than other length scales in the problem)
  these volume integrals will place a large emphasis on the regions at
  large distance.  AMR will then aggressively refine in the stretched
region despite the pointwise errors (as measured by the $L_\infty$ norm for individual elements) being very small.
Such regions tend to have a very low $L_\infty$ because the numerical solution is very
accurate there, but a high $L_2$ because of the large volume in the stretched region, thus also explaining why the $L_2$ norm of the p-AMR run in Fig.~\ref{fig:prob_c_convergence} fails to converge as smoothly as the $L_\infty$ norm.

We solve this problem by introducing a ``compactified grid'',
which has the same structure as the physical grid, but without the
compactification in the physical grid.  We then compute the integrals for the estimator in
Eqs.~(\ref{eq:Eta2_Integrals}) on this compactified grid.  The
integrands, Eqs.~(\ref{eq:Eta2_Ingredients}), are as before computed
on the physical grid.  In essence, this procedure merely changes the
weighting of the different regions of the grid via the Jacobians $J$
in the integrals and the parameters $h_e, h_m$ which are now computed on the compactified grid. 
In practice, we use as compactified grid
a cubed-sphere mesh where the outer spherical shell extends from radius $2$ to radius $3$, and where the middle cube has a side-length of $2/\sqrt{3}$.

\begin{figure}
  \centering
  \includegraphics[width=.47\textwidth,trim=0 2 40 37,clip=true]{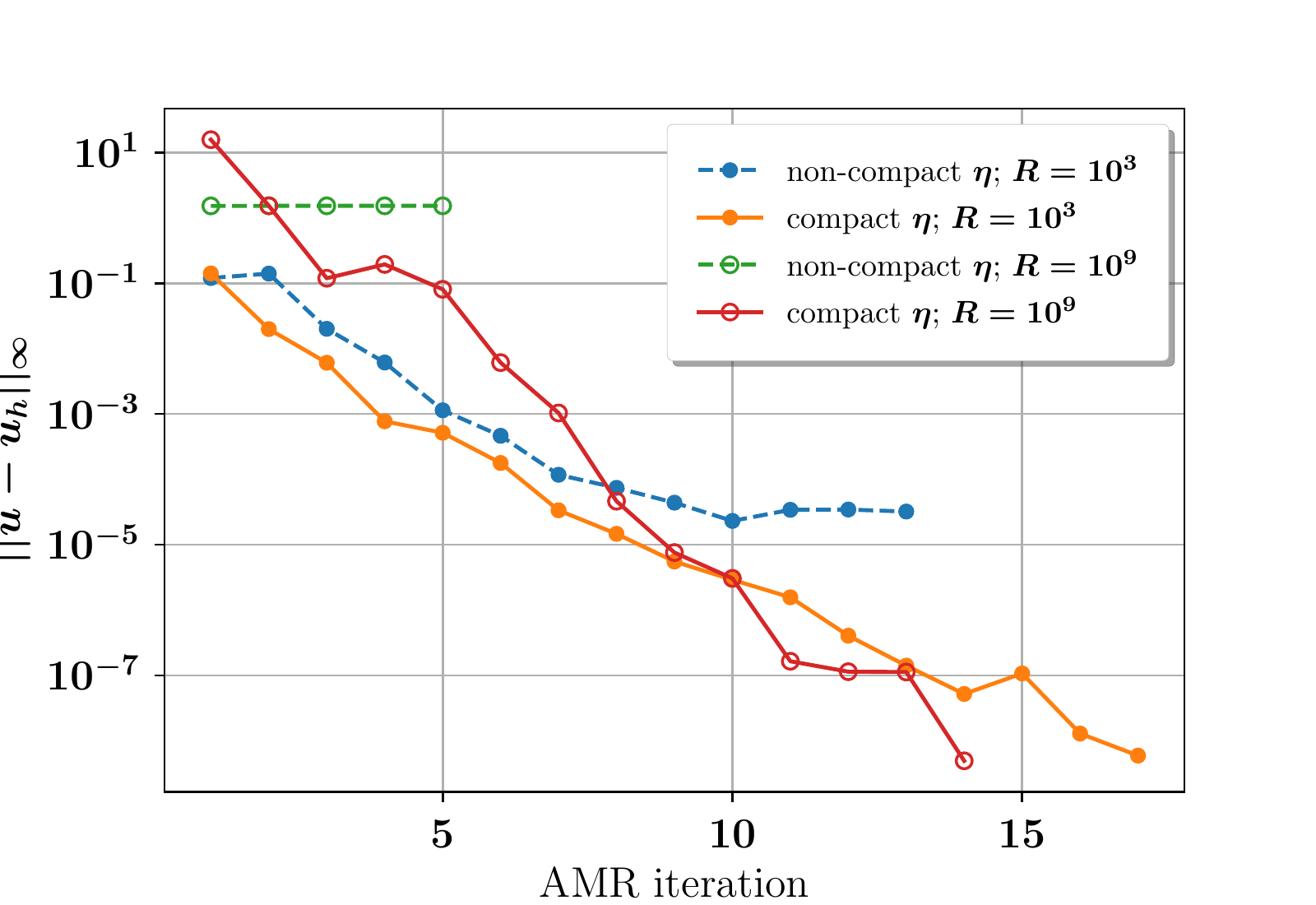}
  \caption{ \label{fig:prob_c_noncompact_vs_compact} Problem C,
    (Eq.~\ref{eq:Lorentzian}): Convergence of the solution in the
    $L_2$ norm (See Eq. \ref{eq:l2norm}) as the degree p is adaptively
    refined with a compact and non-compact estimator.  (Dirichlet BC
    at R=1000 and $R=10^9$).  All runs use the AMR/dG parameters of
    the p-amr run in Figure~\ref{fig:prob_c_convergence}.  }
\end{figure}

Figure~\ref{fig:prob_c_noncompact_vs_compact} shows the convergence of
the $L_\infty$ norm using the compactified grid versus the
non-compactified grid.  For $R = 1000$, the standard
  estimator (``non-compact $\eta$'') converges well for the first 10
  AMR iterations, but then stalls.  For $R=10^9$, the standard
  estimator fails to yield convergence of the solution at all.  For
  both cases, the new estimator (``compact $\eta$'') results in good
  convergence.  We stress that the compactified estimator is only used
  in driving the AMR refinement.  The DG-scheme is always formulated
  in the physical domain, and also the error plotted in
  Fig.~\ref{fig:prob_c_noncompact_vs_compact} is computed from the
  solution on the physical domain with outer boundary $R=1000$ or
  $10^9$.  The p-AMR run shown in Fig.~\ref{fig:prob_c_convergence}
  uses the compactified estimator.

\begin{figure}
  \centering
  \includegraphics[width=.47\textwidth,trim=10 14 10 10,clip=true]{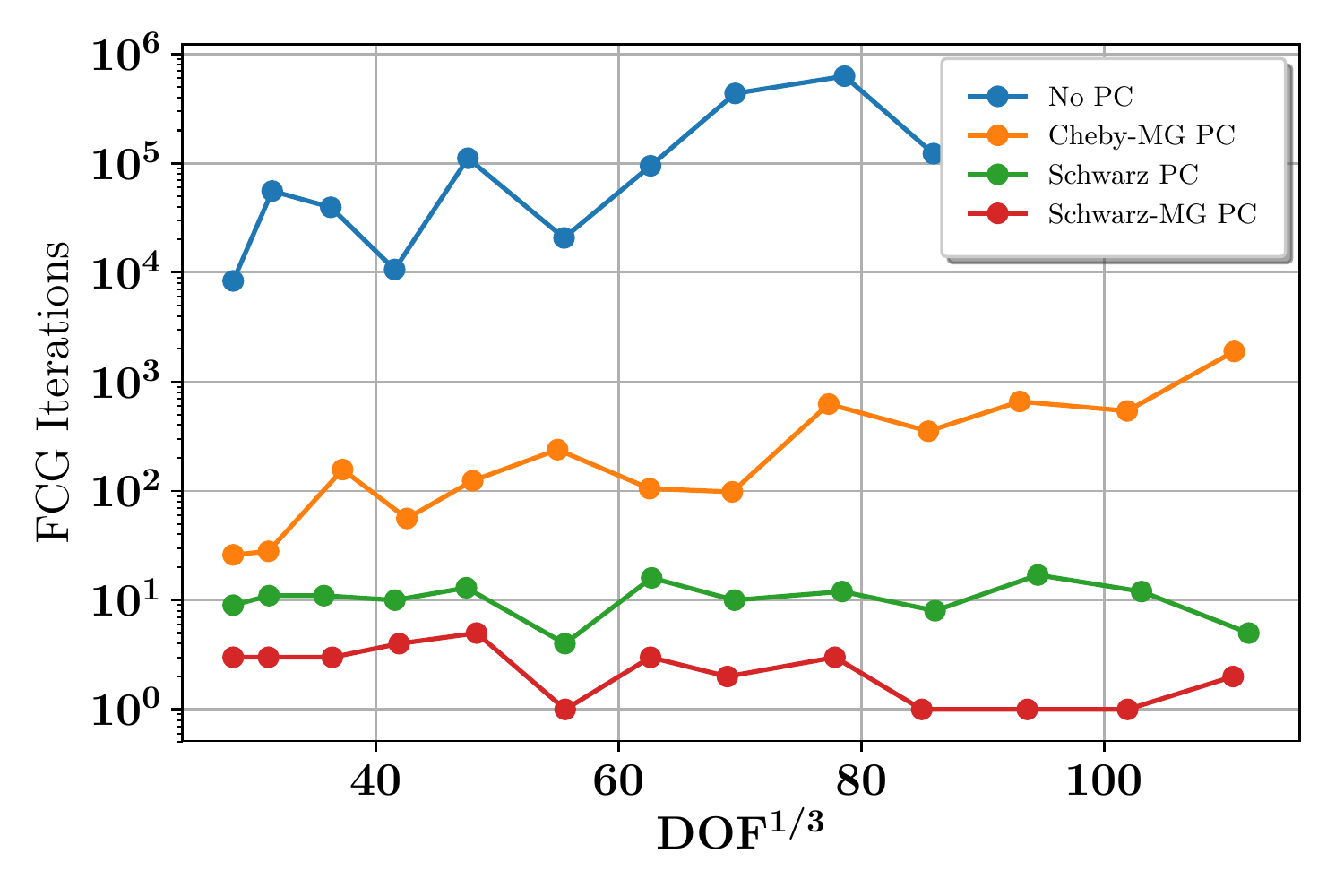}
  \caption{ \label{fig:prob_c_pc_comparison} Comparison of four
    different methods of preconditioning FCG.  All runs use the AMR/dG
    parameters of the p-amr run in
    Figure~\ref{fig:prob_c_convergence}.
  }
\end{figure}

Finally, we investigate the efficiency of preconditioners for the
p-AMR run with $R=1000$ and the compactified estimator, i.e.\ the run plotted in orange in Figs.~\ref{fig:prob_c_convergence}
  and~\ref{fig:prob_c_noncompact_vs_compact}.
Figure~\ref{fig:prob_c_pc_comparison} presents the iteration counts
for four different kinds of preconditioning.
Chebyshev-smoothed multigrid
preconditioner loses
  efficiency on cubed spheres with stretched boundaries, possibly due
to a poorly estimated upper eigenvalue in
Alg.~\ref{alg:CG_Spectral_Bound_Solver}. However, using the more
powerful domain decomposition additive Schwarz method, we can regain
the efficiency of the multigrid-preconditioner seen in the previous
two sections.  Here, the Schwarz subdomains have $N_{\rm
    overlap}=2$, and each multi-grid iteration employs $N_{\rm iter,
    Sch}=3$ iterations of Schwarz-smoothing with rtol=1e-3.  Chebyshev
  uses $N_{\rm iter,eigs}=15$ iterations for the eigenvalue estimate,
  and $N_{\rm iter,Cheb}=15$ iterations in the Chebyshev smoother.

\subsection{Puncture Initial Data}
\label{sec:punctureinitialdata}

There are various approaches to solving for binary black hole initial
data sets and these approaches are primarily distinguished by the
initial choice of hypersurface and how the physical singularity inside
the black holes is treated. One possibility when considering two black
holes is to work on $\mathbb{R}^{3}$ with two balls excised\cite{cook1994,cook2004excision,caudill2006circular}. This approach has
been shown to work well with the spectral finite element method
(e.g. \cite{pfeiffer2003multidomain}), but is more problematic for
finite-difference codes because special stencils must be created near
the curved boundaries. Another popular approach, which is more
amenable to finite difference codes is the puncture method~\cite{brandt1997simple},
where an elliptic equation is solved on $\mathbb{R}^3$, with
two points where the solution becomes singular.  These points represent
the inner asymptotically flat infinity (Brill-Lindquist
topology).   The puncture
method simplifies the numerical method because no special inner
boundary condition has to be considered, however the solution is only
$C^4$ smooth at the puncture points~\cite{brandt1997simple}. Without using contrived
coordinate systems to remove the $C^4$-smooth nature of the
punctures (See~\cite{ansorg2004single} for example), spectral methods
cannot perform optimally, because they would require the solution to
be smooth on the computational domain in order to obtain exponential
convergence. We choose to solve for binary black hole initial-data
with the puncture method in this paper for two reasons. The first is
that the method does not couple well with traditional spectral schemes
as discussed above and this allows us to compare the discontinuous
Galerkin method to the spectral method. Secondly, the equation we must
solve is less complicated than the excision case because it only
involves a solve for a single field, the conformal factor, as opposed
to the 5 fields one must solve for with the excision method (see
e.g. \cite{pfeiffer2003multidomain}), so it is easier to implement
numerically.

Nevertheless, puncture data provides a testing ground for many
  of the new techniques developed here: Singular points which benefit
  from h-refinement; smooth regions that benefit from p-refinement; a
  spherical outer boundary requiring the cubed-sphere domain shown in
  Fig.~\ref{fig:cubed_sphere_mesh}; and a boundary at infinity (or
  near infinity) which requires a compactified radial coordinate.
  Moreover, for testing of adaptivity, it is easy to add arbitrarily
  many black holes each represented by its own singularity, at
  arbitrary coordinates with arbitrary spins.

For the case of puncture data, the initial data equations of general relativity
reduce down to a single equation~\cite{brandt1997simple}:

\begin{equation}
\label{eq:Two_Punctures_PDE}
-\nabla^2 u = \frac{1}{8}\bar A^{ij} \bar A_{ij}\psi^{-7}
\end{equation}
where $\bar A_{ij}$ is a spatially dependent function given by
\begin{equation}
  \label{eq:Two_Punctures_Aij}
\bar{A}_{ij} = \frac{3}{2}\sum_{I}\frac{1}{r^{2}_{I}}[2P^{I}_{(i}n^{l}_{j)}-(f_{ij}-n^{l}_{i}n^l_j)P^{k}_{I}n^I_k + \frac{4}{r_{I}}n^l_{(i}\epsilon_{j)kl}S^k_In^l_I].
\end{equation}
Here, $n^i_I$ are the spatially varying components of the radial unit-vector $\hat n_{I}(\vec x)=(\vec{x}-\vec c_{I})/|\vec x-\vec c_I|$ relative to the position $\vec c_I$ of the $I$-th black hole~\cite{brandt1997simple}. The constant vectors $\tv P_{I}$ and $\tv S_{I}$
quantify the momentum and spin of the $I$-th black-hole and $\psi = 1 + \sum_I \frac{m_{I}}{2r_{I}}+u$. The boundary condition is
\begin{equation}
  u\to 0,\quad |\vec x| \to \infty.
\end{equation}
We solve the above elliptic PDE using first the spectral code SpEC and
then the dG solver presented in this paper. We solve for the case of
two orbiting equal mass black holes with momenta $\pm .2$, zero spin
and initial positions $(\pm 3, 0, 0)$ in units of total mass M. This
is the test case used in \cite{ansorg2007}. Since there is no
analytical solution, we will compare the solutions between refinement
levels at four reference points on the x-axis. These are $(0,0,0)$,
$(3,0,0)$, the location of the right-most puncture, $(10,0,0)$ and
$(100,0,0)$.

\begin{figure}
\includegraphics[width=0.45\textwidth]{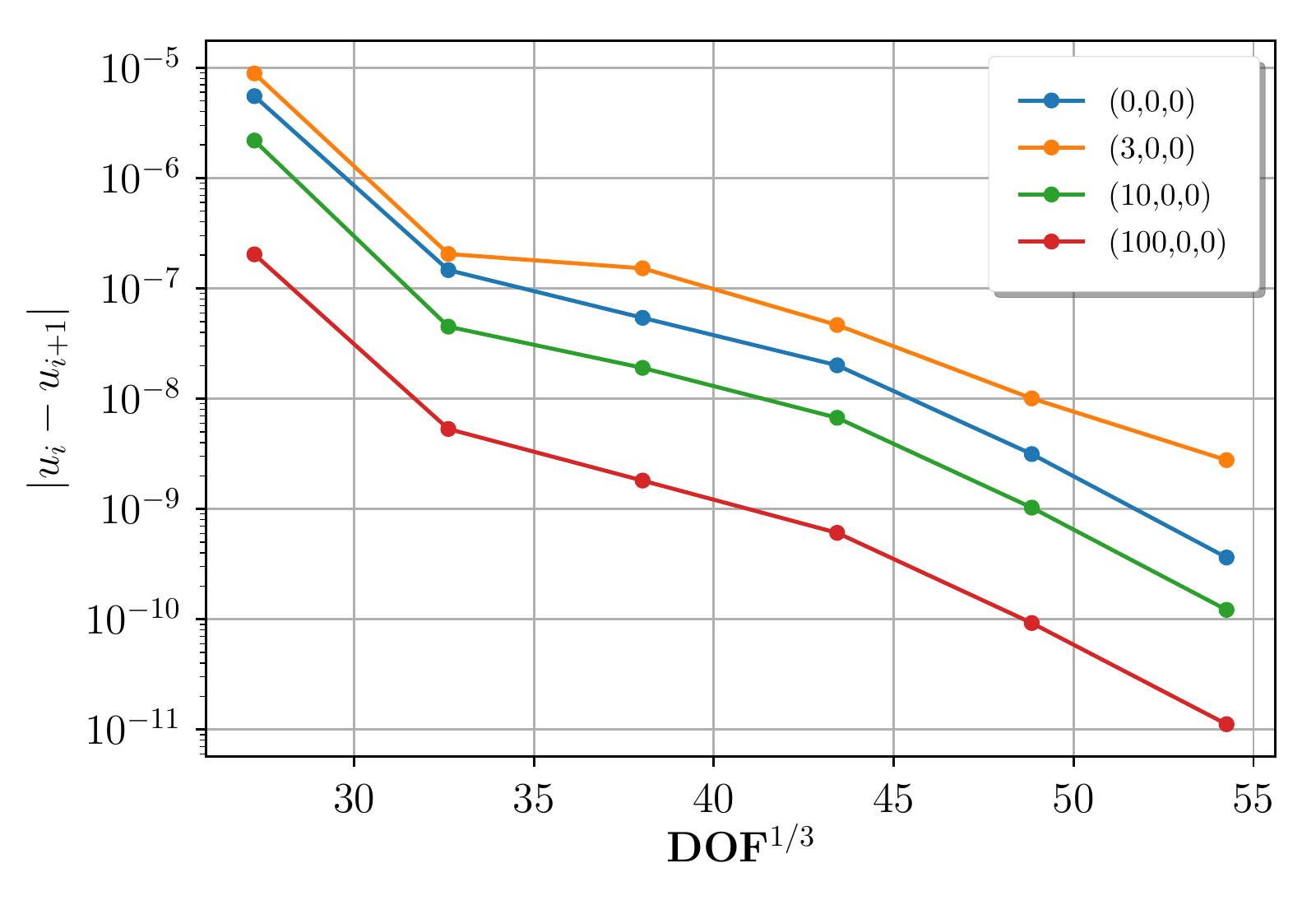}
\caption{
  \label{fig:spec_point_convergence}
  Convergence of the \texttt{SpEC}--elliptic solver with manually
  adopting the domain-decomposition and manual adjustment of
  resolution to compensate for the singularities. Plotted are differences to the
  next-\emph{lower} resolution at four points in the computational domain. }
\end{figure}

The SpEC solver was already used to solve for puncture data
in~\cite{dennison2006,lovelace2008}.  This spectral code is apriori
not well suited for puncture data which results in a non-smooth
solution $u(\vec x)$.  Because we know where the singularities of the
punctures are, this can be overcome manually, by covering the
punctures with very small rectangular blocks, at high enough
resolution, to compensate for the loss of exponential
convergence. Figure \ref{fig:spec_point_convergence} shows the
difference in the solution at the four reference points as the
resolution is manually increased. We emphasize that this
  solution obtained with SpEC depents on (i) \textit{prior knowledge}
  of the locations of the singularities; and (ii) tuning of SpEC's
  mesh and resolution \textit{by hand}.

\begin{figure}
  \centering
  \includegraphics[width=.45\textwidth,trim=0 12 0 14,clip=true]{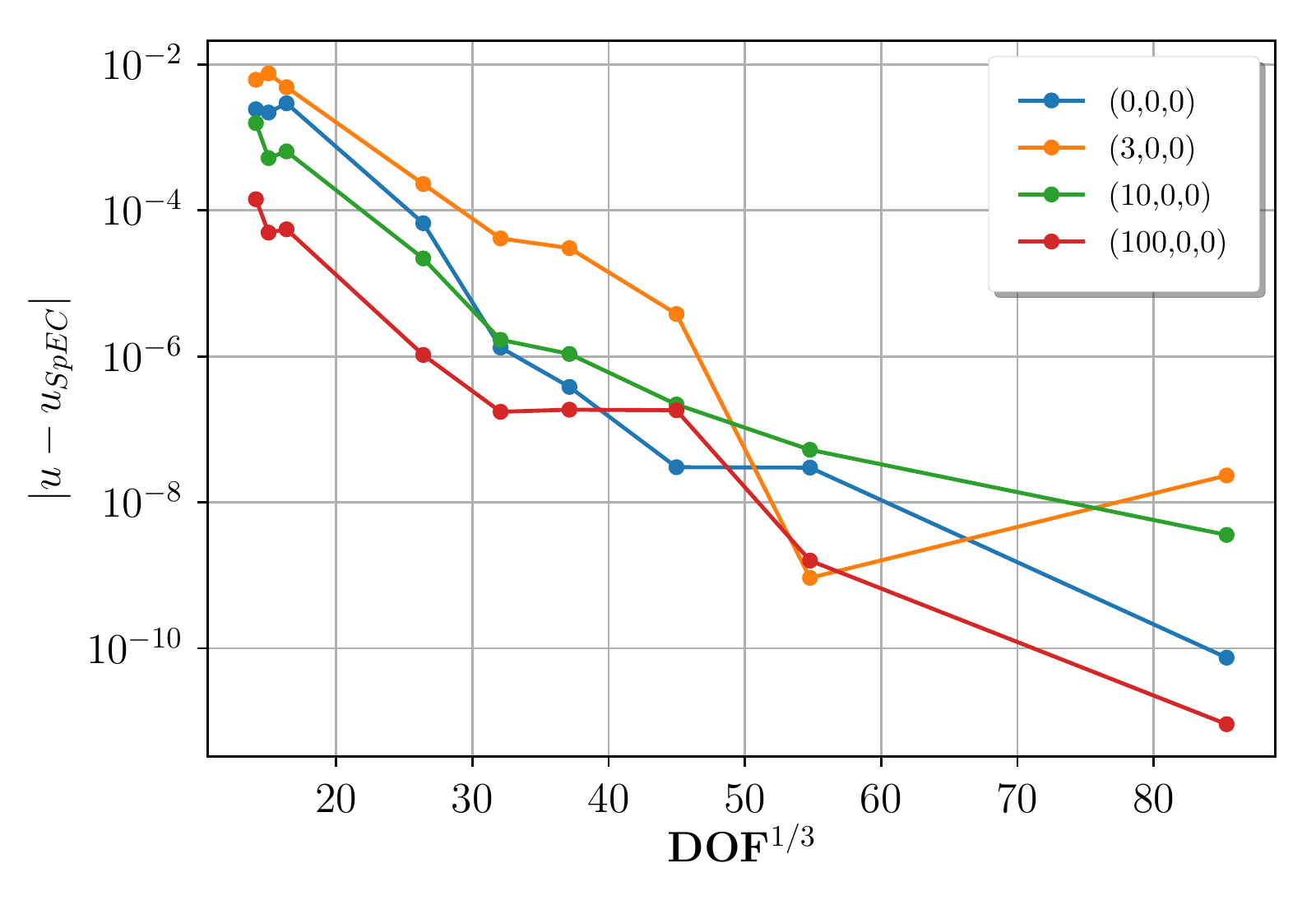}
  \caption{Problem D: Black hole initial data with two punctures.  Convergence of the error between the dG solution and the SpEC solution.
  }
  \label{fig:Two_Punctures_Mesh_Convergence} 
\end{figure}

For the dG solver, we start with a uniformly refined cubed-sphere mesh
at level $l=1$, i.e.\ 13 blocks, each consisting of eight cells.  The  outer radius at $10^{11}$M and the size of the inner cube is 10M.  We start further with elements of  polynomial
order $p=2$.  The location of the punctures is \textit{not} utilized in the dG code, and all mesh-refinement is automatic, driven by Alg.~\ref{alg:hpamr}  with parameters $\gamma_h = .25$ and $\gamma_p = .1$ and we refine the top $12.5\%$ of elements. In order
to run a Schwarz smoother for this problem we would need to transfer
ghost-data for the operator described by Eqn.~(\ref{eqn:operator_oh})
whenever a Schwarz subdomain contains a ghost element. While this is
by no means problematic, we have not yet implemented the
infrastructure to do it, so we just use a Chebyshev smoother when we
precondition this problem with
Multigrid. Figure~\ref{fig:Two_Punctures_Mesh_Convergence} shows the
convergence of the four reference points with respect to the finest
grid SpEC solution between AMR levels. We first see that in terms of
overall DOF, the dG solver doesn't do much worse than the finely tuned
SpEC solver, even though the dG solver has to adaptively find the
punctures, has a larger initial error and has no h-or-p coarsening, so
mistakes in the refinement cannot be fixed. Thus, taken all of this
into account, the convergence is highly satisfactory.  The bounce
in the $(3,0,0)$ at the second to last iteration arises because the dG-solution oscillates around the SpEC solution and coincidentally is shown near a zero-crossing. Figure~\ref{fig:Two_Punctures_Mesh_Final} shows the solution on the final mesh, which has the highest h-refinement exactly at the points of the punctures, as desired.

\begin{figure}
  \centering
\includegraphics[width=.48\textwidth,trim=830 350 270 390,clip=true]{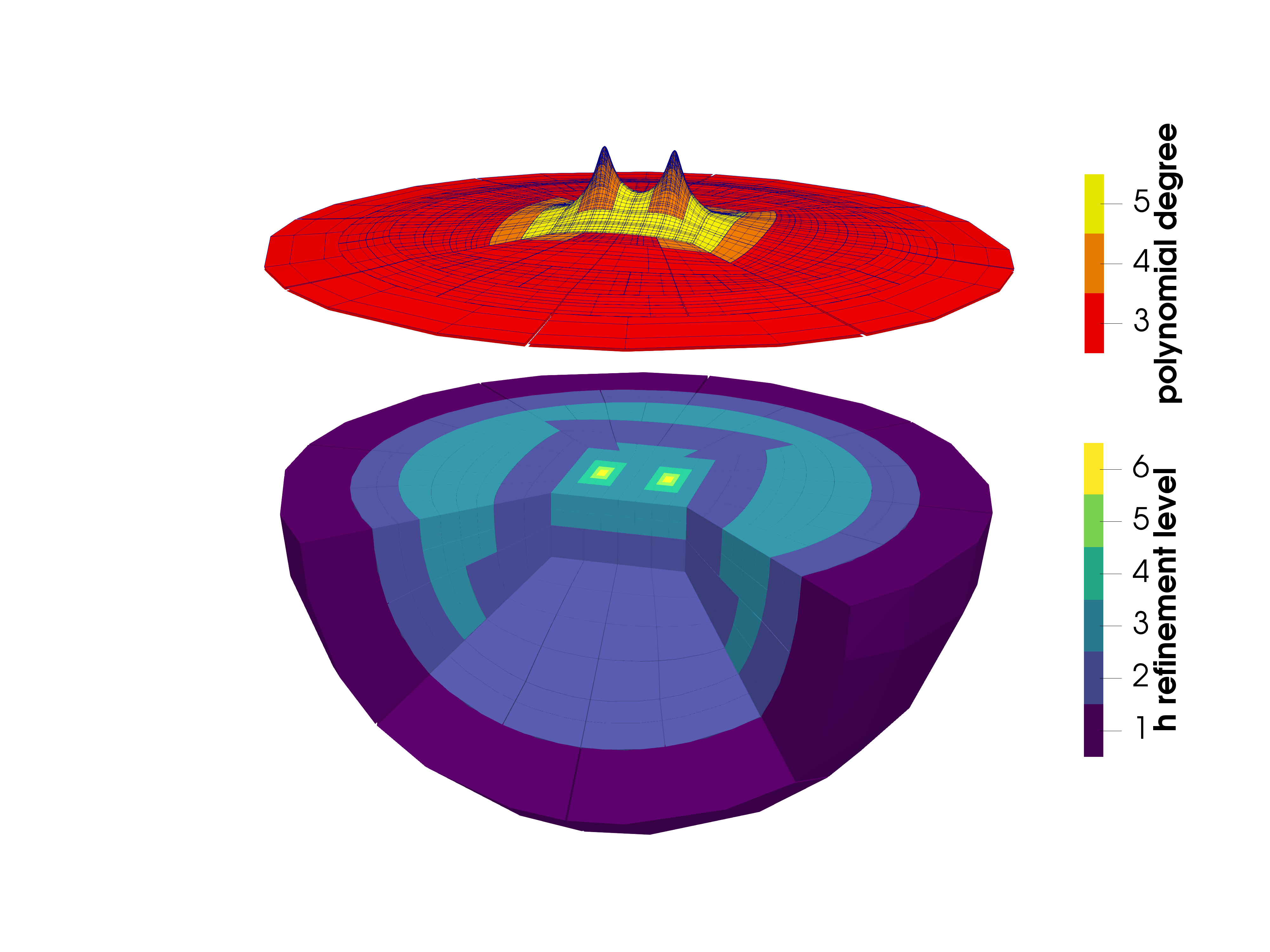}
  \caption{Problem D (Black hole initial data with two punctures):
    Visualization of the hp-refined computational mesh.  The bottom
    portion of the image shows a volume rendering of the $z<0$ portion
    of the computational domain, with two blocks removed, and
    color-coded by the h-refinement level.  The top portion of the
    image shows the $z=0$ cross-section of the computational grid, color-coded by the polynomial degree, with the height representing the solution $u$. For ease of visualization, grid-points are mapped onto the compactified grid on which the estimator $\eta_e$ is computed; the compactified outer radius $R=3$ corresponds to the physical outer radius $R=10^{11}$.}
  \label{fig:Two_Punctures_Mesh_Final}
\end{figure}

Next, we solve for the puncture initial data of three black-holes
randomly located in the xy-plane, with random spins and random
momenta. Spectral solvers such as SpEC cannot perform well when the
singular points on the grid are not known in advance. Thus, we end
this paper showcasing a problem that our discontinuous Galerkin code
can solve, but SpEC cannot. Table \ref{tab:Multi_Punctures}
illustrates the parameters for the randomly placed punctures and their
spin and momenta.

\begin{figure}
  \includegraphics[width=.48\textwidth,trim=770 100 27 40,clip=true]{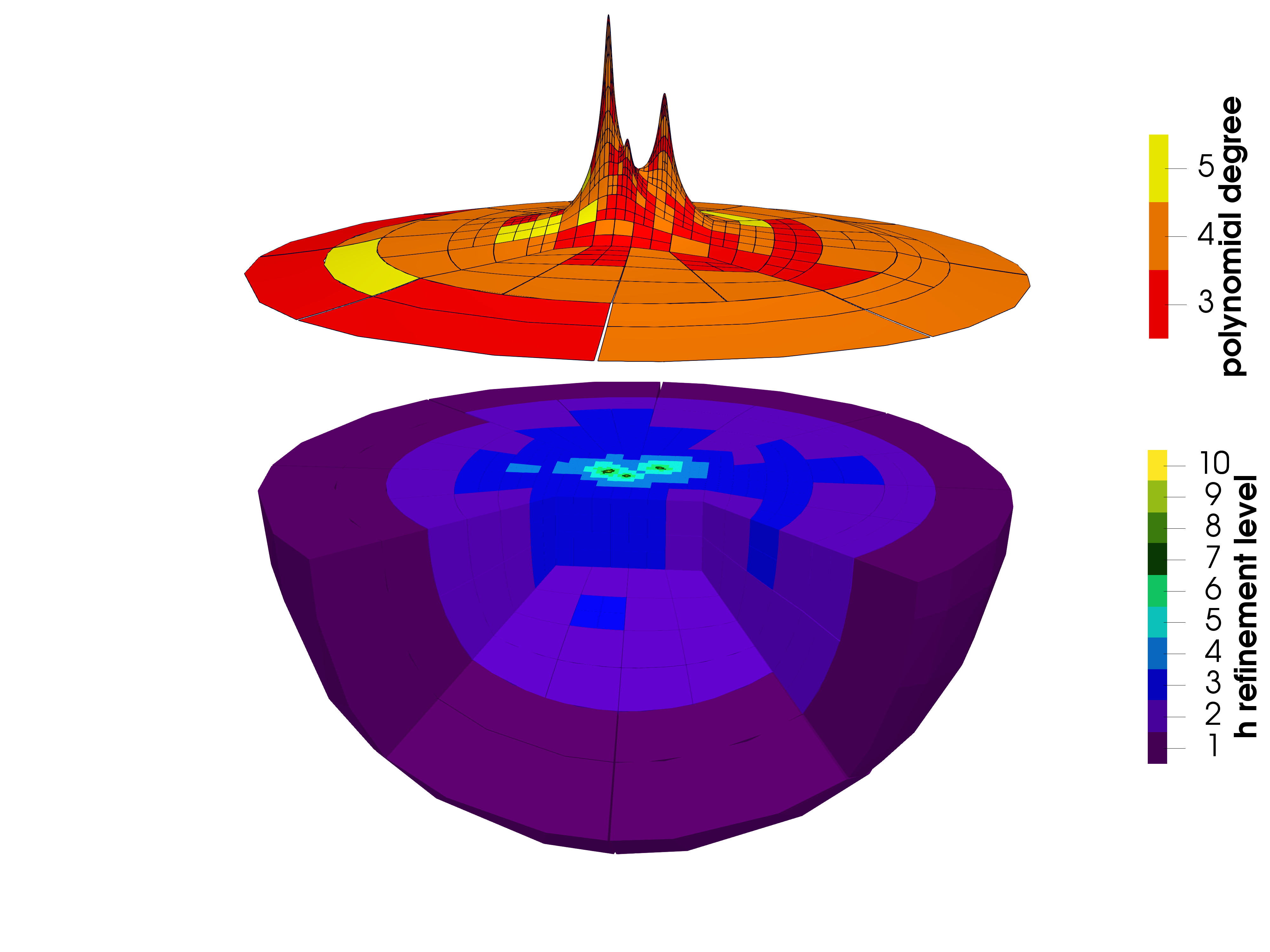}
  \\
\includegraphics[width=.46\textwidth,trim=1350 1200 1200 1200,clip=true]{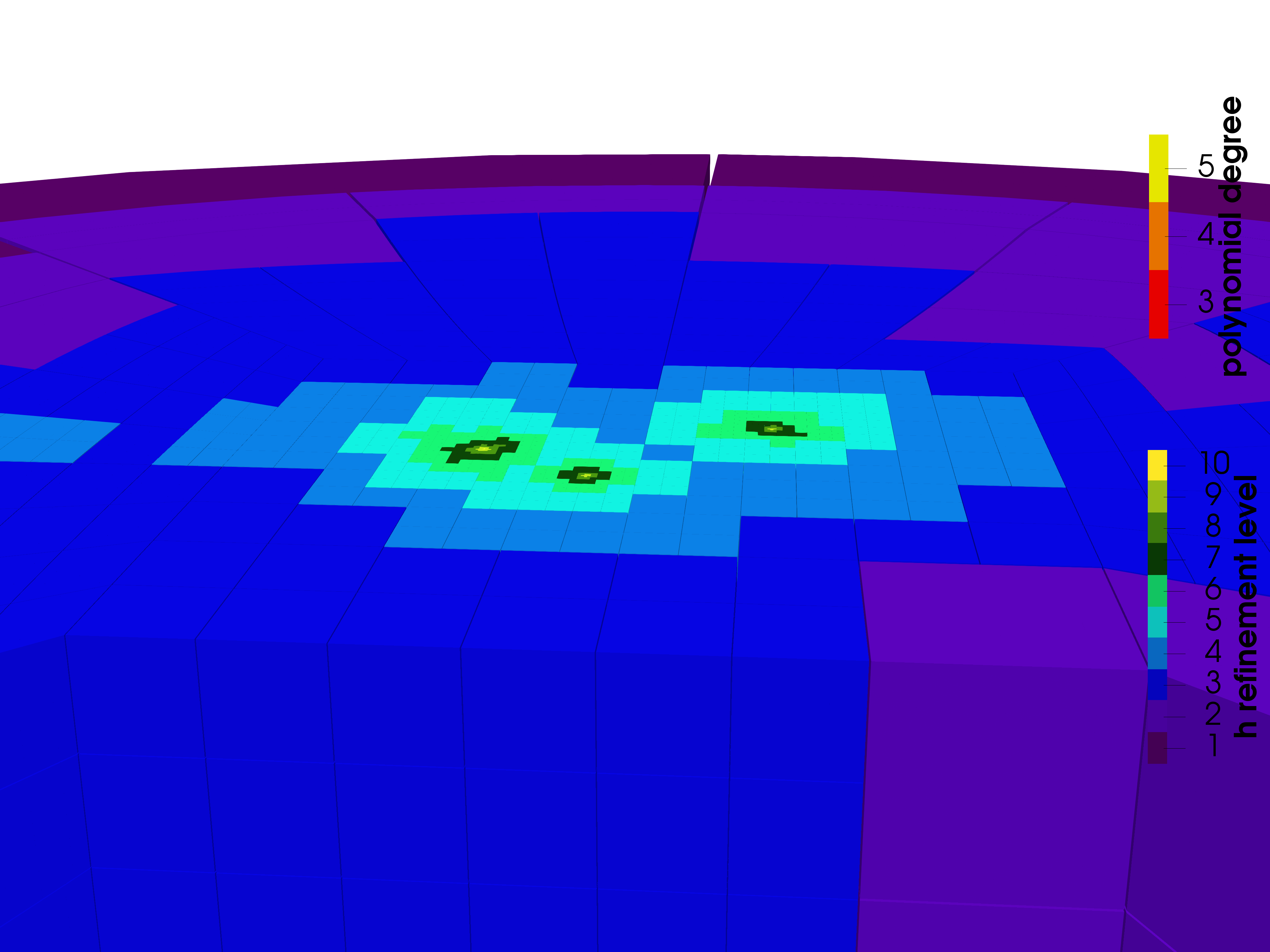}

  \caption{Problem E (Black hole initial data with three punctures):
    Visualization of the solution, and the hp-refined computational
    mesh.  {\bf Top portion:} the $z=0$ cross-section of the
    computational grid, color-coded by the polynomial degree, with the
    height representing the solution $u$.  {\bf Middle portion:}
    volume rendering of the $z<0$ part of the computational domain,
    with two blocks removed, and color-coded by the
    h-refinement level.  {\bf Bottom portion:} Zoom into the middle
    portion, highlighting the region near the three punctures with
    highest refinement level. For ease of visualization, grid-points are mapped onto the compactified grid on which the estimator $\eta_e$ is computed; the compactified outer radius $R=3$ corresponds to the physical outer radius $R=10^{11}$. }
  \label{fig:Two_Punctures_Mesh_Final}
\end{figure}

\begin{table}
\centering
\label{tab:Multi_Punctures}
\begin{tabular}{cccc}
\hline
 \,\,\,\, & Puncture 1  \,\,\,\, & Puncture 2  \,\,\,\, & Puncture 3 \\ \hline
  $m$ \,\,\,\, &0.2691  \,\,\,\, & 0.4063  \,\,\,\, & 0.3245  \\
  $x$ \,\,\,\, &0.0152  \,\,\,\, & -2.316  \,\,\,\, & -1.0279  \\
  $y$ \,\,\,\, &-0.6933  \,\,\,\, & 1.8274  \,\,\,\, &  -2.2711  \\ 
  $P_x$ \,\,\,\, &0.0585  \,\,\,\, & -0.0284  \,\,\,\, &  0.1640 \\
  $P_y$ \,\,\,\, &0.0082  \,\,\,\, &  -0.1497 \,\,\,\, &  0.0515 \\
  $S_z$ \,\,\,\, &-0.0134  \,\,\,\, &  -0.0332  \,\,\,\, & -0.0708 \\ \hline
\end{tabular}
\caption{
  The randomly generated parameters for the three black holes.  We list the mass m, the position $(x,y,0)$, the momentum $(P_x, P_y, 0)$ and the spin $(0,0,S_z)$ of the black holes.
}
\end{table}

\begin{figure}
  \centering
  \includegraphics[width=.48\textwidth,trim=0 12 0 14,clip=true]{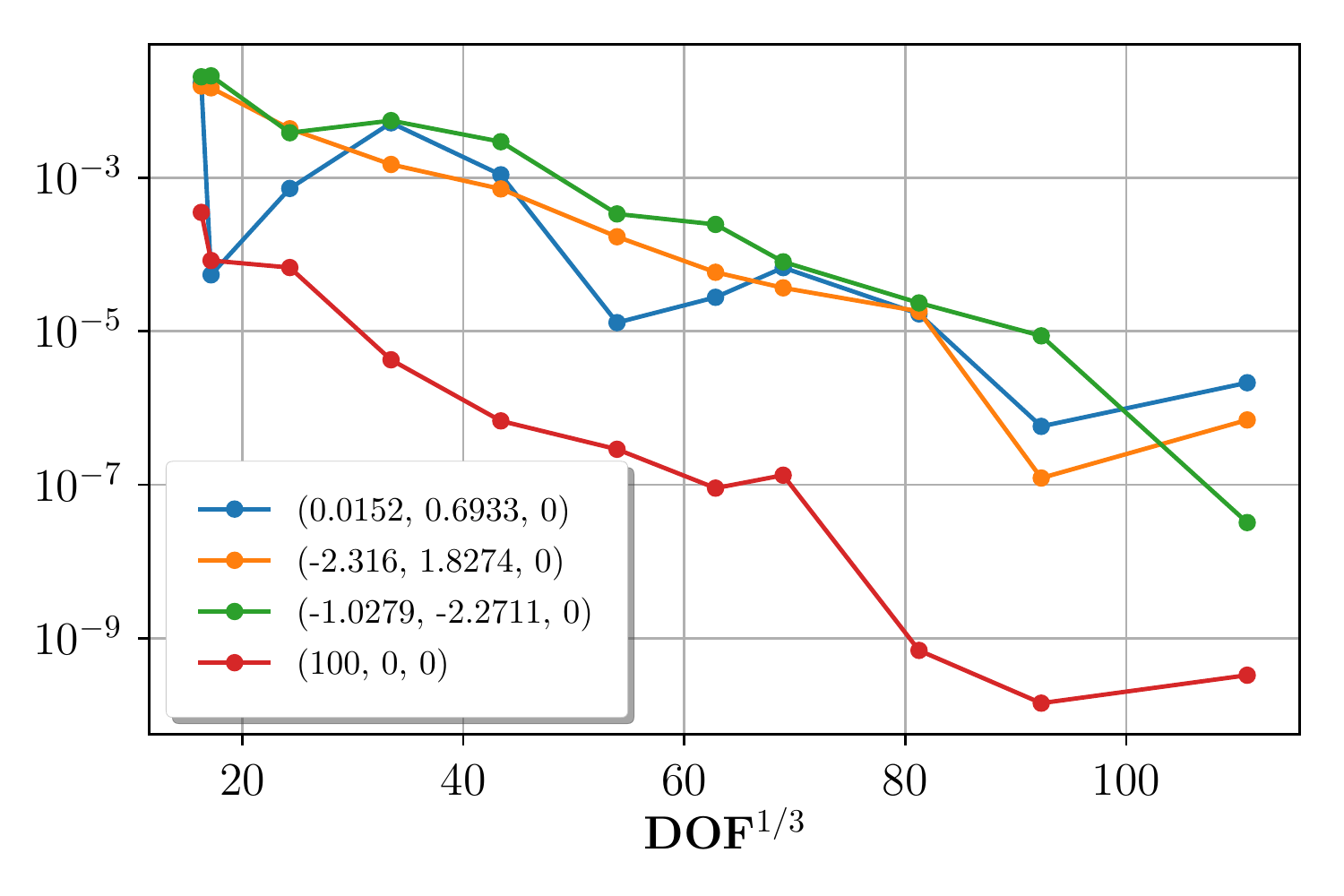}
  \caption{Problem D: Black hole initial data with three randomly generated punctures.  Convergence of the error between AMR steps at four different points, three corresponding to the location of the punctures and one at $(100,0,0)$.
  }
  \label{fig:three_punctures_convergence} 
\end{figure}

Figure~\ref{fig:three_punctures_convergence} shows the convergence of four points, three at the location of the punctures, and one at $(100,0,0)$. We use amr parameters $\gamma_h=0.25$, $\gamma_p=0.1$, $F_{\rm refined}=0.125$ for this run.

%
\section{Conclusion and Future Work}
\label{sec:Conclusions}

We presented a new code for solving
elliptic equations intended for numerical
relativity. The methodology we use differs from other codes in the
field in many important respects. In particular, we use a
discontinuous Galerkin method to discretize the equations, an
hp-adaptive mesh refinement scheme driven by an a posteriori estimator
and a matrix-free, scalable Multigrid preconditioned Newton-Krylov
solver.
Individually, many of the features of our
  code have been implemented before~\cite{kozdon2018energy,kozdon2019robust,stiller2017robust,sundar2012parallel,hesthaven2008nodal}, but they have never been combined together to create a general dG solver.  In particular, the combination of curved meshes (cf.\ Fig.~\ref{fig:cubed_sphere_mesh}) and non-conforming elements
  is novel and is crucial for generic solution-driven AMR.  Moreover,
  the compactified AMR-driver introduced in
  Sec.~\ref{sec:CompactifiedLorentizan} is also new, enabling AMR on compactified computational domains with outer boundary near infinity. Lastly, the use of a multigrid-preconditioned solver with a Schwarz (or Chebyshev) smoother on non-polygonal meshes has not been investigated in a dG setting until this paper.

For BBH puncture data, our new code approaches the accuracy of
  existing, specialized codes like SpEC.  In addition, the automatic
  AMR in the new code does not require manual tuning of the
  computational mesh, and does not utilize any prior knowledge of
  features of the solution like the location of black hole punctures.
  The new code already improves on the more specialized codes by being
  able to handle an arbitrary number of puncture-black holes.  AMR can
  also automatically resolve discontinuities without prior knowledge
  of their existence (cf. Fig.~\ref{fig:problem_b_mesh}).

  Moving forward, there are
still several areas of improvement our code could possibly benefit
from:

\begin{enumerate}
\item {\bf Load balancing}: For problems that require adaptive mesh
  refinement and multi-grid, there will naturally be a unbalanced
  number of degrees of freedom (DOF) across processors and this can
  slow down the Krylov iterations substantially. This can be amended
  by incorporating a task-based parallelism framework for load
  balancing.  The elliptic solver developed in this article will be incorporated  task-based parallel code SpECTRE, which is concurrently being developed~\cite{kidder:16}.
\item {\bf Anisotropic refinement}: Most realistic problems have some
  anisotropy and therefore a solver would benefit from anisotropic
  mesh refinement.  Indeed, most of the problems in this paper could
  have had better convergence with anisotropic refinement, for instance, Problem C is spherically symmetric, and puncture data is approximately spherically symmetric at large distance.  We use the
  p4est framework for mesh refinement and while it has support for
  anisotropic refinement, the direction of the anisotropy has to be
  known a priori. We look to go beyond this and have general
  refinement in a future edition of our code.
\item {\bf Hybridizable dG}: The discontinuous Galerkin method can be quite
  expensive in terms of the amount of DOF it requires to converge to a
  certain error. Recently, a method called Hybridizable dG has been
  coupled with matrix-free multigrid methods to solve elliptic
  problems with substantially reduced DOF over the classical dG
  method~\cite{fabien2019manycore,muralikrishnan2019multilevel}. Whether
  this method can be fully incorporated into the complex scheme
  presented in this paper, will be an area of further inquiry.
\end{enumerate}

In future work, we plan to use this solver to expand the physics in
compact object initial data, for instance, neutron star initial data
for equation of state with phase-transitions, neutron stars with
very high compactness (where current solvers fail~\cite{henriksson:2014tba}),
or compact objects in alternative theories of gravity, or with novel matter fields like boson stars.

\begin{acknowledgements}
The authors would like to acknowledge helpful discussions with
Francois Hebert, Jan Hesthaven, Saul Teukolsky and Tim Warburton.
This research was supported by NSERC of Canada.  Computations were
performed on the Minerva cluster at the Max-Planck-Institute for
Gravitational Physics, and the GPC and Niagara supercomputers at the
SciNet HPC Consortium~\cite{scinet}. SciNet is funded by: the Canada
Foundation for Innovation; the Government of Ontario; Ontario Research
Fund - Research Excellence; and the University of Toronto.
\end{acknowledgements}


\bibliography{References_final}

\begin{thebibliography}{82}%
\makeatletter
\providecommand \@ifxundefined [1]{%
 \@ifx{#1\undefined}
}%
\providecommand \@ifnum [1]{%
 \ifnum #1\expandafter \@firstoftwo
 \else \expandafter \@secondoftwo
 \fi
}%
\providecommand \@ifx [1]{%
 \ifx #1\expandafter \@firstoftwo
 \else \expandafter \@secondoftwo
 \fi
}%
\providecommand \natexlab [1]{#1}%
\providecommand \enquote  [1]{``#1''}%
\providecommand \bibnamefont  [1]{#1}%
\providecommand \bibfnamefont [1]{#1}%
\providecommand \citenamefont [1]{#1}%
\providecommand \href@noop [0]{\@secondoftwo}%
\providecommand \href [0]{\begingroup \@sanitize@url \@href}%
\providecommand \@href[1]{\@@startlink{#1}\@@href}%
\providecommand \@@href[1]{\endgroup#1\@@endlink}%
\providecommand \@sanitize@url [0]{\catcode `\\12\catcode `\$12\catcode
  `\&12\catcode `\#12\catcode `\^12\catcode `\_12\catcode `\%12\relax}%
\providecommand \@@startlink[1]{}%
\providecommand \@@endlink[0]{}%
\providecommand \url  [0]{\begingroup\@sanitize@url \@url }%
\providecommand \@url [1]{\endgroup\@href {#1}{\urlprefix }}%
\providecommand \urlprefix  [0]{URL }%
\providecommand \Eprint [0]{\href }%
\providecommand \doibase [0]{http://dx.doi.org/}%
\providecommand \selectlanguage [0]{\@gobble}%
\providecommand \bibinfo  [0]{\@secondoftwo}%
\providecommand \bibfield  [0]{\@secondoftwo}%
\providecommand \translation [1]{[#1]}%
\providecommand \BibitemOpen [0]{}%
\providecommand \bibitemStop [0]{}%
\providecommand \bibitemNoStop [0]{.\EOS\space}%
\providecommand \EOS [0]{\spacefactor3000\relax}%
\providecommand \BibitemShut  [1]{\csname bibitem#1\endcsname}%
\let\auto@bib@innerbib\@empty
\bibitem [{\citenamefont {Reed}\ and\ \citenamefont
  {Hill}(1973)}]{Reed.W;Hill.T1973}%
  \BibitemOpen
  \bibfield  {author} {\bibinfo {author} {\bibfnamefont {W.}~\bibnamefont
  {Reed}}\ and\ \bibinfo {author} {\bibfnamefont {T.}~\bibnamefont {Hill}},\
  }\href {http://www.osti.gov/scitech/servlets/purl/4491151} {\emph {\bibinfo
  {title} {Conference: National topical meeting on mathematical models and
  computational techniques for analysis of nuclear systems, Ann Arbor,
  Michigan, USA, 8 Apr 1973}}},\ \bibinfo {type} {{Conference: National topical
  meeting on mathematical models and computational techniques for analysis of
  nuclear systems, Ann Arbor, Michigan, USA, 8 Apr 1973}}\ \bibinfo {number}
  {LA-UR--73-479, CONF-730414--2}\ (\bibinfo {year} {1973})\BibitemShut
  {NoStop}%
\bibitem [{\citenamefont {Hesthaven}\ and\ \citenamefont
  {Warburton}(2008)}]{hesthaven2008nodal}%
  \BibitemOpen
  \bibfield  {author} {\bibinfo {author} {\bibfnamefont {J.}~\bibnamefont
  {Hesthaven}}\ and\ \bibinfo {author} {\bibfnamefont {T.}~\bibnamefont
  {Warburton}},\ }\href {https://books.google.ca/books?id=APQkDOmwyksC} {\emph
  {\bibinfo {title} {Nodal Discontinuous Galerkin Methods: Algorithms,
  Analysis, and Applications}}},\ Texts in Applied Mathematics\ (\bibinfo
  {publisher} {Springer},\ \bibinfo {year} {2008})\BibitemShut {NoStop}%
\bibitem [{\citenamefont {{Cockburn}}(2001)}]{Cock01}%
  \BibitemOpen
  \bibfield  {author} {\bibinfo {author} {\bibfnamefont {B.}~\bibnamefont
  {{Cockburn}}},\ }\href {\doibase 10.1016/S0377-0427(00)00512-4} {\bibfield
  {journal} {\bibinfo  {journal} {J. Comput. Appl. Math.}\ }\textbf {\bibinfo
  {volume} {128}},\ \bibinfo {pages} {187} (\bibinfo {year}
  {2001})}\BibitemShut {NoStop}%
\bibitem [{\citenamefont {{Cockburn}}\ and\ \citenamefont
  {{Shu}}(1998)}]{cockburn1998runge}%
  \BibitemOpen
  \bibfield  {author} {\bibinfo {author} {\bibfnamefont {B.}~\bibnamefont
  {{Cockburn}}}\ and\ \bibinfo {author} {\bibfnamefont {C.-W.}\ \bibnamefont
  {{Shu}}},\ }\href {\doibase 10.1006/jcph.1998.5892} {\bibfield  {journal}
  {\bibinfo  {journal} {J.\ Comput.\ Phys.}\ }\textbf {\bibinfo {volume}
  {141}},\ \bibinfo {pages} {199} (\bibinfo {year} {1998})}\BibitemShut
  {NoStop}%
\bibitem [{\citenamefont {Cockburn}(1998)}]{Cockburn.B1998}%
  \BibitemOpen
  \bibfield  {author} {\bibinfo {author} {\bibfnamefont {B.}~\bibnamefont
  {Cockburn}},\ }in\ \href {\doibase 10.1007/BFb0096353} {\emph {\bibinfo
  {booktitle} {{An introduction to the Discontinuous Galerkin method for
  convection-dominated problems; in Advanced Numerical Approximation of
  Nonlinear Hyperbolic Equations: Lectures given at the 2nd Session of the
  Centro Internazionale Matematico Estivo (C.I.M.E.) held in Cetraro, Italy,
  June 23--28, 1997}}}},\ \bibinfo {series and number} {Lecture Notes in
  Physics},\ \bibinfo {editor} {edited by\ \bibinfo {editor} {\bibfnamefont
  {A.}~\bibnamefont {Quarteroni}}}\ (\bibinfo  {publisher} {Springer},\
  \bibinfo {address} {Berlin, New York},\ \bibinfo {year} {1998})\ pp.\
  \bibinfo {pages} {150--268}\BibitemShut {NoStop}%
\bibitem [{\citenamefont {Cockburn}\ \emph {et~al.}(2000)\citenamefont
  {Cockburn}, \citenamefont {Karniadakis},\ and\ \citenamefont
  {Shu}}]{Cockburn.B;Karniadakis.G;Shu.C2000}%
  \BibitemOpen
  \bibfield  {author} {\bibinfo {author} {\bibfnamefont {B.}~\bibnamefont
  {Cockburn}}, \bibinfo {author} {\bibfnamefont {G.~E.}\ \bibnamefont
  {Karniadakis}}, \ and\ \bibinfo {author} {\bibfnamefont {C.-W.}\ \bibnamefont
  {Shu}},\ }in\ \href {\doibase 10.1007/978-3-642-59721-3_1} {\emph {\bibinfo
  {booktitle} {Discontinuous Galerkin Methods: Theory, Computation and
  Applications}}},\ \bibinfo {series and number} {Lecture Notes in
  Computational Science and Engineering},\ \bibinfo {editor} {edited by\
  \bibinfo {editor} {\bibfnamefont {B.}~\bibnamefont {Cockburn}}, \bibinfo
  {editor} {\bibfnamefont {G.~E.}\ \bibnamefont {Karniadakis}}, \ and\ \bibinfo
  {editor} {\bibfnamefont {C.-W.}\ \bibnamefont {Shu}}}\ (\bibinfo  {publisher}
  {Springer},\ \bibinfo {address} {Berlin, New York},\ \bibinfo {year} {2000})\
  pp.\ \bibinfo {pages} {3--50}\BibitemShut {NoStop}%
\bibitem [{\citenamefont {Baumgarte}\ and\ \citenamefont
  {Shapiro}(2010)}]{baumgarte2010numerical}%
  \BibitemOpen
  \bibfield  {author} {\bibinfo {author} {\bibfnamefont {T.~W.}\ \bibnamefont
  {Baumgarte}}\ and\ \bibinfo {author} {\bibfnamefont {S.~L.}\ \bibnamefont
  {Shapiro}},\ }\href@noop {} {\emph {\bibinfo {title} {Numerical relativity:
  solving Einstein's equations on the computer}}}\ (\bibinfo  {publisher}
  {Cambridge University Press},\ \bibinfo {year} {2010})\BibitemShut {NoStop}%
\bibitem [{\citenamefont {Field}\ \emph {et~al.}(2010)\citenamefont {Field},
  \citenamefont {Hesthaven}, \citenamefont {Lau},\ and\ \citenamefont
  {Mroue}}]{Field:2010mn}%
  \BibitemOpen
  \bibfield  {author} {\bibinfo {author} {\bibfnamefont {S.~E.}\ \bibnamefont
  {Field}}, \bibinfo {author} {\bibfnamefont {J.~S.}\ \bibnamefont
  {Hesthaven}}, \bibinfo {author} {\bibfnamefont {S.~R.}\ \bibnamefont {Lau}},
  \ and\ \bibinfo {author} {\bibfnamefont {A.~H.}\ \bibnamefont {Mroue}},\
  }\href {\doibase 10.1103/PhysRevD.82.104051} {\bibfield  {journal} {\bibinfo
  {journal} {Phys.\ Rev.\ D}\ }\textbf {\bibinfo {volume} {82}},\ \bibinfo
  {pages} {104051} (\bibinfo {year} {2010})},\ \Eprint
  {http://arxiv.org/abs/1008.1820} {arXiv:1008.1820 [gr-qc]} \BibitemShut
  {NoStop}%
\bibitem [{\citenamefont {{Brown}}\ \emph {et~al.}(2012)\citenamefont
  {{Brown}}, \citenamefont {{Diener}}, \citenamefont {{Field}}, \citenamefont
  {{Hesthaven}}, \citenamefont {{Herrmann}}, \citenamefont {{Mrou{\'e}}},
  \citenamefont {{Sarbach}}, \citenamefont {{Schnetter}}, \citenamefont
  {{Tiglio}},\ and\ \citenamefont {{Wagman}}}]{brown2012numerical}%
  \BibitemOpen
  \bibfield  {author} {\bibinfo {author} {\bibfnamefont {J.~D.}\ \bibnamefont
  {{Brown}}}, \bibinfo {author} {\bibfnamefont {P.}~\bibnamefont {{Diener}}},
  \bibinfo {author} {\bibfnamefont {S.~E.}\ \bibnamefont {{Field}}}, \bibinfo
  {author} {\bibfnamefont {J.~S.}\ \bibnamefont {{Hesthaven}}}, \bibinfo
  {author} {\bibfnamefont {F.}~\bibnamefont {{Herrmann}}}, \bibinfo {author}
  {\bibfnamefont {A.~H.}\ \bibnamefont {{Mrou{\'e}}}}, \bibinfo {author}
  {\bibfnamefont {O.}~\bibnamefont {{Sarbach}}}, \bibinfo {author}
  {\bibfnamefont {E.}~\bibnamefont {{Schnetter}}}, \bibinfo {author}
  {\bibfnamefont {M.}~\bibnamefont {{Tiglio}}}, \ and\ \bibinfo {author}
  {\bibfnamefont {M.}~\bibnamefont {{Wagman}}},\ }\href {\doibase
  10.1103/PhysRevD.85.084004} {\bibfield  {journal} {\bibinfo  {journal}
  {Phys.\ Rev.\ D}\ }\textbf {\bibinfo {volume} {85}},\ \bibinfo {eid} {084004}
  (\bibinfo {year} {2012})},\ \Eprint {http://arxiv.org/abs/1202.1038}
  {arXiv:1202.1038 [gr-qc]} \BibitemShut {NoStop}%
\bibitem [{\citenamefont {{Field}}\ \emph {et~al.}(2009)\citenamefont
  {{Field}}, \citenamefont {{Hesthaven}},\ and\ \citenamefont
  {{Lau}}}]{field2009discontinuous}%
  \BibitemOpen
  \bibfield  {author} {\bibinfo {author} {\bibfnamefont {S.~E.}\ \bibnamefont
  {{Field}}}, \bibinfo {author} {\bibfnamefont {J.~S.}\ \bibnamefont
  {{Hesthaven}}}, \ and\ \bibinfo {author} {\bibfnamefont {S.~R.}\ \bibnamefont
  {{Lau}}},\ }\href {\doibase 10.1088/0264-9381/26/16/165010} {\bibfield
  {journal} {\bibinfo  {journal} {Class.\ Quantum Grav.}\ }\textbf {\bibinfo
  {volume} {26}},\ \bibinfo {eid} {165010} (\bibinfo {year} {2009})},\ \Eprint
  {http://arxiv.org/abs/0902.1287} {arXiv:0902.1287 [gr-qc]} \BibitemShut
  {NoStop}%
\bibitem [{\citenamefont {{Zumbusch}}(2009)}]{zumbusch2009}%
  \BibitemOpen
  \bibfield  {author} {\bibinfo {author} {\bibfnamefont {G.}~\bibnamefont
  {{Zumbusch}}},\ }\href {\doibase 10.1088/0264-9381/26/17/175011} {\bibfield
  {journal} {\bibinfo  {journal} {Class.\ Quantum Grav.}\ }\textbf {\bibinfo
  {volume} {26}},\ \bibinfo {eid} {175011} (\bibinfo {year} {2009})},\ \Eprint
  {http://arxiv.org/abs/0901.0851} {arXiv:0901.0851 [gr-qc]} \BibitemShut
  {NoStop}%
\bibitem [{\citenamefont {Radice}\ and\ \citenamefont
  {Rezzolla}(2011)}]{Radice:2011qr}%
  \BibitemOpen
  \bibfield  {author} {\bibinfo {author} {\bibfnamefont {D.}~\bibnamefont
  {Radice}}\ and\ \bibinfo {author} {\bibfnamefont {L.}~\bibnamefont
  {Rezzolla}},\ }\href {\doibase 10.1103/PhysRevD.84.024010} {\bibfield
  {journal} {\bibinfo  {journal} {Phys.\ Rev.\ D}\ }\textbf {\bibinfo {volume}
  {84}},\ \bibinfo {pages} {024010} (\bibinfo {year} {2011})},\ \Eprint
  {http://arxiv.org/abs/1103.2426} {arXiv:1103.2426 [gr-qc]} \BibitemShut
  {NoStop}%
\bibitem [{\citenamefont {{Mocz}}\ \emph {et~al.}(2014)\citenamefont {{Mocz}},
  \citenamefont {{Vogelsberger}}, \citenamefont {{Sijacki}}, \citenamefont
  {{Pakmor}},\ and\ \citenamefont {{Hernquist}}}]{mocz:14}%
  \BibitemOpen
  \bibfield  {author} {\bibinfo {author} {\bibfnamefont {P.}~\bibnamefont
  {{Mocz}}}, \bibinfo {author} {\bibfnamefont {M.}~\bibnamefont
  {{Vogelsberger}}}, \bibinfo {author} {\bibfnamefont {D.}~\bibnamefont
  {{Sijacki}}}, \bibinfo {author} {\bibfnamefont {R.}~\bibnamefont {{Pakmor}}},
  \ and\ \bibinfo {author} {\bibfnamefont {L.}~\bibnamefont {{Hernquist}}},\
  }\href {\doibase 10.1093/mnras/stt1890} {\bibfield  {journal} {\bibinfo
  {journal} {Mon.\ Not.\ Roy.\ Astr.\ Soc.}\ }\textbf {\bibinfo {volume}
  {437}},\ \bibinfo {pages} {397} (\bibinfo {year} {2014})},\ \Eprint
  {http://arxiv.org/abs/1305.5536} {arXiv:1305.5536 [physics.comp-ph]}
  \BibitemShut {NoStop}%
\bibitem [{\citenamefont {{Zanotti}}\ \emph {et~al.}(2014)\citenamefont
  {{Zanotti}}, \citenamefont {{Dumbser}}, \citenamefont {{Hidalgo}},\ and\
  \citenamefont {{Balsara}}}]{zanotti:14}%
  \BibitemOpen
  \bibfield  {author} {\bibinfo {author} {\bibfnamefont {O.}~\bibnamefont
  {{Zanotti}}}, \bibinfo {author} {\bibfnamefont {M.}~\bibnamefont
  {{Dumbser}}}, \bibinfo {author} {\bibfnamefont {A.}~\bibnamefont
  {{Hidalgo}}}, \ and\ \bibinfo {author} {\bibfnamefont {D.}~\bibnamefont
  {{Balsara}}},\ }in\ \href@noop {} {\emph {\bibinfo {booktitle} {8th
  International Conference of Numerical Modeling of Space Plasma Flows
  (ASTRONUM 2013)}}},\ \bibinfo {series} {Astronomical Society of the Pacific
  Conference Series}, Vol.\ \bibinfo {volume} {488},\ \bibinfo {editor} {edited
  by\ \bibinfo {editor} {\bibfnamefont {N.~V.}\ \bibnamefont {{Pogorelov}}},
  \bibinfo {editor} {\bibfnamefont {E.}~\bibnamefont {{Audit}}}, \ and\
  \bibinfo {editor} {\bibfnamefont {G.~P.}\ \bibnamefont {{Zank}}}}\ (\bibinfo
  {year} {2014})\ pp.\ \bibinfo {pages} {285--290},\ \Eprint
  {http://arxiv.org/abs/1401.6448} {arXiv:1401.6448 [astro-ph.IM]} \BibitemShut
  {NoStop}%
\bibitem [{\citenamefont {{Endeve}}\ \emph {et~al.}(2015)\citenamefont
  {{Endeve}}, \citenamefont {{Hauck}}, \citenamefont {{Xing}},\ and\
  \citenamefont {{Mezzacappa}}}]{endeve:15}%
  \BibitemOpen
  \bibfield  {author} {\bibinfo {author} {\bibfnamefont {E.}~\bibnamefont
  {{Endeve}}}, \bibinfo {author} {\bibfnamefont {C.~D.}\ \bibnamefont
  {{Hauck}}}, \bibinfo {author} {\bibfnamefont {Y.}~\bibnamefont {{Xing}}}, \
  and\ \bibinfo {author} {\bibfnamefont {A.}~\bibnamefont {{Mezzacappa}}},\
  }\href {\doibase 10.1016/j.jcp.2015.02.005} {\bibfield  {journal} {\bibinfo
  {journal} {J. Comput. Phys.}\ }\textbf {\bibinfo {volume} {287}},\ \bibinfo
  {pages} {151} (\bibinfo {year} {2015})},\ \Eprint
  {http://arxiv.org/abs/1410.7431} {arXiv:1410.7431 [physics.comp-ph]}
  \BibitemShut {NoStop}%
\bibitem [{\citenamefont {{Teukolsky}}(2016)}]{teukolsky2015}%
  \BibitemOpen
  \bibfield  {author} {\bibinfo {author} {\bibfnamefont {S.~A.}\ \bibnamefont
  {{Teukolsky}}},\ }\href {\doibase 10.1016/j.jcp.2016.02.031} {\bibfield
  {journal} {\bibinfo  {journal} {J.\ Comput.\ Phys.}\ }\textbf {\bibinfo
  {volume} {312}},\ \bibinfo {pages} {333} (\bibinfo {year} {2016})},\ \Eprint
  {http://arxiv.org/abs/1510.01190} {arXiv:1510.01190 [gr-qc]} \BibitemShut
  {NoStop}%
\bibitem [{\citenamefont {{Bugner}}\ \emph {et~al.}(2016)\citenamefont
  {{Bugner}}, \citenamefont {{Dietrich}}, \citenamefont {{Bernuzzi}},
  \citenamefont {{Weyhausen}},\ and\ \citenamefont
  {{Br{\"u}gmann}}}]{Bugner:2015gqa}%
  \BibitemOpen
  \bibfield  {author} {\bibinfo {author} {\bibfnamefont {M.}~\bibnamefont
  {{Bugner}}}, \bibinfo {author} {\bibfnamefont {T.}~\bibnamefont
  {{Dietrich}}}, \bibinfo {author} {\bibfnamefont {S.}~\bibnamefont
  {{Bernuzzi}}}, \bibinfo {author} {\bibfnamefont {A.}~\bibnamefont
  {{Weyhausen}}}, \ and\ \bibinfo {author} {\bibfnamefont {B.}~\bibnamefont
  {{Br{\"u}gmann}}},\ }\href {\doibase 10.1103/PhysRevD.94.084004} {\bibfield
  {journal} {\bibinfo  {journal} {Phys.\ Rev.\ D}\ }\textbf {\bibinfo {volume}
  {94}},\ \bibinfo {eid} {084004} (\bibinfo {year} {2016})},\ \Eprint
  {http://arxiv.org/abs/1508.07147} {arXiv:1508.07147 [gr-qc]} \BibitemShut
  {NoStop}%
\bibitem [{\citenamefont {Schaal}\ \emph {et~al.}(2015)\citenamefont {Schaal},
  \citenamefont {Bauer}, \citenamefont {Chandrashekar}, \citenamefont {Pakmor},
  \citenamefont {Klingenberg},\ and\ \citenamefont
  {Springel}}]{schaal2015astrophysicalfixed}%
  \BibitemOpen
  \bibfield  {author} {\bibinfo {author} {\bibfnamefont {K.}~\bibnamefont
  {Schaal}}, \bibinfo {author} {\bibfnamefont {A.}~\bibnamefont {Bauer}},
  \bibinfo {author} {\bibfnamefont {P.}~\bibnamefont {Chandrashekar}}, \bibinfo
  {author} {\bibfnamefont {R.}~\bibnamefont {Pakmor}}, \bibinfo {author}
  {\bibfnamefont {C.}~\bibnamefont {Klingenberg}}, \ and\ \bibinfo {author}
  {\bibfnamefont {V.}~\bibnamefont {Springel}},\ }\href
  {https://arxiv.org/abs/1506.06140} {\bibfield  {journal} {\bibinfo  {journal}
  {Monthly Notices of the Royal Astronomical Society}\ }\textbf {\bibinfo
  {volume} {453}},\ \bibinfo {pages} {4278} (\bibinfo {year} {2015})},\ \Eprint
  {http://arxiv.org/abs/1506.06140} {arXiv:1506.06140 [astro-ph.CO]}
  \BibitemShut {NoStop}%
\bibitem [{\citenamefont {Zanotti}\ \emph {et~al.}(2015)\citenamefont
  {Zanotti}, \citenamefont {Fambri},\ and\ \citenamefont
  {Dumbser}}]{zanotti2015}%
  \BibitemOpen
  \bibfield  {author} {\bibinfo {author} {\bibfnamefont {O.}~\bibnamefont
  {Zanotti}}, \bibinfo {author} {\bibfnamefont {F.}~\bibnamefont {Fambri}}, \
  and\ \bibinfo {author} {\bibfnamefont {M.}~\bibnamefont {Dumbser}},\
  }\href@noop {} {\bibfield  {journal} {\bibinfo  {journal} {Monthly Notices of
  the Royal Astronomical Society}\ }\textbf {\bibinfo {volume} {452}},\
  \bibinfo {pages} {3010} (\bibinfo {year} {2015})}\BibitemShut {NoStop}%
\bibitem [{\citenamefont {{Miller}}\ and\ \citenamefont
  {{Schnetter}}(2017)}]{Miller:2016vik}%
  \BibitemOpen
  \bibfield  {author} {\bibinfo {author} {\bibfnamefont {J.~M.}\ \bibnamefont
  {{Miller}}}\ and\ \bibinfo {author} {\bibfnamefont {E.}~\bibnamefont
  {{Schnetter}}},\ }\href {\doibase 10.1088/1361-6382/34/1/015003} {\bibfield
  {journal} {\bibinfo  {journal} {Class.\ Quantum Grav.}\ }\textbf {\bibinfo
  {volume} {34}},\ \bibinfo {eid} {015003} (\bibinfo {year} {2017})},\ \Eprint
  {http://arxiv.org/abs/1604.00075} {arXiv:1604.00075 [gr-qc]} \BibitemShut
  {NoStop}%
\bibitem [{\citenamefont {{Kidder}}\ \emph {et~al.}(2017)\citenamefont
  {{Kidder}}, \citenamefont {{Field}}, \citenamefont {{Foucart}}, \citenamefont
  {{Schnetter}}, \citenamefont {{Teukolsky}}, \citenamefont {{Bohn}},
  \citenamefont {{Deppe}}, \citenamefont {{Diener}}, \citenamefont
  {{H{\'e}bert}}, \citenamefont {{Lippuner}}, \citenamefont {{Miller}},
  \citenamefont {{Ott}}, \citenamefont {{Scheel}},\ and\ \citenamefont
  {{Vincent}}}]{kidder:16}%
  \BibitemOpen
  \bibfield  {author} {\bibinfo {author} {\bibfnamefont {L.~E.}\ \bibnamefont
  {{Kidder}}}, \bibinfo {author} {\bibfnamefont {S.~E.}\ \bibnamefont
  {{Field}}}, \bibinfo {author} {\bibfnamefont {F.}~\bibnamefont {{Foucart}}},
  \bibinfo {author} {\bibfnamefont {E.}~\bibnamefont {{Schnetter}}}, \bibinfo
  {author} {\bibfnamefont {S.~A.}\ \bibnamefont {{Teukolsky}}}, \bibinfo
  {author} {\bibfnamefont {A.}~\bibnamefont {{Bohn}}}, \bibinfo {author}
  {\bibfnamefont {N.}~\bibnamefont {{Deppe}}}, \bibinfo {author} {\bibfnamefont
  {P.}~\bibnamefont {{Diener}}}, \bibinfo {author} {\bibfnamefont
  {F.}~\bibnamefont {{H{\'e}bert}}}, \bibinfo {author} {\bibfnamefont
  {J.}~\bibnamefont {{Lippuner}}}, \bibinfo {author} {\bibfnamefont
  {J.}~\bibnamefont {{Miller}}}, \bibinfo {author} {\bibfnamefont {C.~D.}\
  \bibnamefont {{Ott}}}, \bibinfo {author} {\bibfnamefont {M.~A.}\ \bibnamefont
  {{Scheel}}}, \ and\ \bibinfo {author} {\bibfnamefont {T.}~\bibnamefont
  {{Vincent}}},\ }\href {\doibase 10.1016/j.jcp.2016.12.059} {\bibfield
  {journal} {\bibinfo  {journal} {J. Comput. Phys.}\ }\textbf {\bibinfo
  {volume} {335}},\ \bibinfo {pages} {84} (\bibinfo {year} {2017})},\ \Eprint
  {http://arxiv.org/abs/1609.00098} {arXiv:1609.00098 [astro-ph.HE]}
  \BibitemShut {NoStop}%
\bibitem [{\citenamefont {{Fambri}}\ \emph {et~al.}(2018)\citenamefont
  {{Fambri}}, \citenamefont {{Dumbser}}, \citenamefont {{K{\"o}ppel}},
  \citenamefont {{Rezzolla}},\ and\ \citenamefont {{Zanotti}}}]{fambri2018}%
  \BibitemOpen
  \bibfield  {author} {\bibinfo {author} {\bibfnamefont {F.}~\bibnamefont
  {{Fambri}}}, \bibinfo {author} {\bibfnamefont {M.}~\bibnamefont {{Dumbser}}},
  \bibinfo {author} {\bibfnamefont {S.}~\bibnamefont {{K{\"o}ppel}}}, \bibinfo
  {author} {\bibfnamefont {L.}~\bibnamefont {{Rezzolla}}}, \ and\ \bibinfo
  {author} {\bibfnamefont {O.}~\bibnamefont {{Zanotti}}},\ }\href {\doibase
  10.1093/mnras/sty734} {\bibfield  {journal} {\bibinfo  {journal} {Mon.\ Not.\
  Roy.\ Astr.\ Soc.}\ } (\bibinfo {year} {2018}),\ 10.1093/mnras/sty734},\
  \Eprint {http://arxiv.org/abs/1801.02839} {arXiv:1801.02839
  [physics.comp-ph]} \BibitemShut {NoStop}%
\bibitem [{\citenamefont {Hébert}\ \emph {et~al.}(2018)\citenamefont
  {Hébert}, \citenamefont {Kidder},\ and\ \citenamefont
  {Teukolsky}}]{hebert:2018xbk}%
  \BibitemOpen
  \bibfield  {author} {\bibinfo {author} {\bibfnamefont {F.}~\bibnamefont
  {Hébert}}, \bibinfo {author} {\bibfnamefont {L.~E.}\ \bibnamefont {Kidder}},
  \ and\ \bibinfo {author} {\bibfnamefont {S.~A.}\ \bibnamefont {Teukolsky}},\
  }\href {\doibase 10.1103/PhysRevD.98.044041} {\bibfield  {journal} {\bibinfo
  {journal} {Phys. Rev.}\ }\textbf {\bibinfo {volume} {D98}},\ \bibinfo {pages}
  {044041} (\bibinfo {year} {2018})},\ \Eprint
  {http://arxiv.org/abs/1804.02003} {arXiv:1804.02003 [gr-qc]} \BibitemShut
  {NoStop}%
\bibitem [{\citenamefont {Arnold}\ \emph {et~al.}(2002)\citenamefont {Arnold},
  \citenamefont {Brezzi}, \citenamefont {Cockburn},\ and\ \citenamefont
  {Marini}}]{arnold.d;brezzi.f;cockburn.b;marini.l2002}%
  \BibitemOpen
  \bibfield  {author} {\bibinfo {author} {\bibfnamefont {D.~N.}\ \bibnamefont
  {Arnold}}, \bibinfo {author} {\bibfnamefont {F.}~\bibnamefont {Brezzi}},
  \bibinfo {author} {\bibfnamefont {B.}~\bibnamefont {Cockburn}}, \ and\
  \bibinfo {author} {\bibfnamefont {L.~D.}\ \bibnamefont {Marini}},\
  }\href@noop {} {\bibfield  {journal} {\bibinfo  {journal} {SIAM Journal on
  Numerical Analysis}\ }\textbf {\bibinfo {volume} {39}},\ \bibinfo {pages}
  {1749} (\bibinfo {year} {2002})}\BibitemShut {NoStop}%
\bibitem [{\citenamefont {Stiller}(2017)}]{stiller2017robust}%
  \BibitemOpen
  \bibfield  {author} {\bibinfo {author} {\bibfnamefont {J.}~\bibnamefont
  {Stiller}},\ }in\ \href {https://arxiv.org/abs/1612.04796} {\emph {\bibinfo
  {booktitle} {Spectral and High Order Methods for Partial Differential
  Equations ICOSAHOM 2016}}}\ (\bibinfo  {publisher} {Springer},\ \bibinfo
  {year} {2017})\ pp.\ \bibinfo {pages} {189--201},\ \Eprint
  {http://arxiv.org/abs/1612.04796} {arXiv:1612.04796 [cs.NA]} \BibitemShut
  {NoStop}%
\bibitem [{\citenamefont {Kronbichler}\ and\ \citenamefont
  {Wall}(2018)}]{kronbichler2018performance}%
  \BibitemOpen
  \bibfield  {author} {\bibinfo {author} {\bibfnamefont {M.}~\bibnamefont
  {Kronbichler}}\ and\ \bibinfo {author} {\bibfnamefont {W.~A.}\ \bibnamefont
  {Wall}},\ }\href {https://arxiv.org/abs/1611.03029} {\bibfield  {journal}
  {\bibinfo  {journal} {SIAM Journal on Scientific Computing}\ }\textbf
  {\bibinfo {volume} {40}},\ \bibinfo {pages} {A3423} (\bibinfo {year}
  {2018})},\ \Eprint {http://arxiv.org/abs/1611.03029} {arXiv:1611.03029
  [math.NA]} \BibitemShut {NoStop}%
\bibitem [{\citenamefont {Fick}(2014)}]{fick2014interior}%
  \BibitemOpen
  \bibfield  {author} {\bibinfo {author} {\bibfnamefont {P.~W.}\ \bibnamefont
  {Fick}},\ }\href {https://arxiv.org/abs/1401.0339} {\bibfield  {journal}
  {\bibinfo  {journal} {arXiv preprint}\ } (\bibinfo {year} {2014})},\ \Eprint
  {http://arxiv.org/abs/1401.0339} {arXiv:1401.0339 [math.NA]} \BibitemShut
  {NoStop}%
\bibitem [{\citenamefont {Kozdon}\ and\ \citenamefont
  {Wilcox}(2018)}]{kozdon2018energy}%
  \BibitemOpen
  \bibfield  {author} {\bibinfo {author} {\bibfnamefont {J.~E.}\ \bibnamefont
  {Kozdon}}\ and\ \bibinfo {author} {\bibfnamefont {L.~C.}\ \bibnamefont
  {Wilcox}},\ }\href {https://arxiv.org/abs/1706.00513} {\bibfield  {journal}
  {\bibinfo  {journal} {Journal of Scientific Computing}\ }\textbf {\bibinfo
  {volume} {76}},\ \bibinfo {pages} {1742} (\bibinfo {year} {2018})},\ \Eprint
  {http://arxiv.org/abs/1706.00513} {arXiv:1706.00513 [math.NA]} \BibitemShut
  {NoStop}%
\bibitem [{\citenamefont {Kozdon}\ \emph {et~al.}(2019)\citenamefont {Kozdon},
  \citenamefont {Wilcox}, \citenamefont {Hagstrom},\ and\ \citenamefont
  {Banks}}]{kozdon2019robust}%
  \BibitemOpen
  \bibfield  {author} {\bibinfo {author} {\bibfnamefont {J.~E.}\ \bibnamefont
  {Kozdon}}, \bibinfo {author} {\bibfnamefont {L.~C.}\ \bibnamefont {Wilcox}},
  \bibinfo {author} {\bibfnamefont {T.}~\bibnamefont {Hagstrom}}, \ and\
  \bibinfo {author} {\bibfnamefont {J.~W.}\ \bibnamefont {Banks}},\ }\href
  {https://arxiv.org/abs/1806.06103} {\bibfield  {journal} {\bibinfo  {journal}
  {Journal of Computational Physics}\ } (\bibinfo {year} {2019})},\ \Eprint
  {http://arxiv.org/abs/1806.06103} {arXiv:1806.06103 [math.NA]} \BibitemShut
  {NoStop}%
\bibitem [{\citenamefont {Pfeiffer}(2005)}]{pfeiffer:2005}%
  \BibitemOpen
  \bibfield  {author} {\bibinfo {author} {\bibfnamefont {H.~P.}\ \bibnamefont
  {Pfeiffer}},\ }\href@noop {} {\bibfield  {journal} {\bibinfo  {journal} {J.\
  Hyperbol.\ Differ.\ Eq.}\ }\textbf {\bibinfo {volume} {2}},\ \bibinfo {pages}
  {497} (\bibinfo {year} {2005})},\ \Eprint
  {http://arxiv.org/abs/gr-qc/0412002} {gr-qc/0412002} \BibitemShut {NoStop}%
\bibitem [{\citenamefont {Cook}(2000)}]{cook2000}%
  \BibitemOpen
  \bibfield  {author} {\bibinfo {author} {\bibfnamefont {G.}~\bibnamefont
  {Cook}},\ }\href {http://www.livingreviews.org/lrr-2000-5} {\bibfield
  {journal} {\bibinfo  {journal} {Living Rev.~Rel.}\ }\textbf {\bibinfo
  {volume} {3}} (\bibinfo {year} {2000})},\ \bibinfo {note} {5}\BibitemShut
  {NoStop}%
\bibitem [{\citenamefont {Lau}\ \emph {et~al.}(2009)\citenamefont {Lau},
  \citenamefont {Pfeiffer},\ and\ \citenamefont {Hesthaven}}]{laupfeiffer2008}%
  \BibitemOpen
  \bibfield  {author} {\bibinfo {author} {\bibfnamefont {S.~R.}\ \bibnamefont
  {Lau}}, \bibinfo {author} {\bibfnamefont {H.~P.}\ \bibnamefont {Pfeiffer}}, \
  and\ \bibinfo {author} {\bibfnamefont {J.~S.}\ \bibnamefont {Hesthaven}},\
  }\href@noop {} {\bibfield  {journal} {\bibinfo  {journal} {Commun. Comput.
  Phys.}\ }\textbf {\bibinfo {volume} {6}},\ \bibinfo {pages} {1063} (\bibinfo
  {year} {2009})},\ \Eprint {http://arxiv.org/abs/arXiv:0808.2597}
  {arXiv:0808.2597} \BibitemShut {NoStop}%
\bibitem [{\citenamefont {Lau}\ \emph {et~al.}(2011)\citenamefont {Lau},
  \citenamefont {Lovelace},\ and\ \citenamefont {Pfeiffer}}]{lau:2011we}%
  \BibitemOpen
  \bibfield  {author} {\bibinfo {author} {\bibfnamefont {S.~R.}\ \bibnamefont
  {Lau}}, \bibinfo {author} {\bibfnamefont {G.}~\bibnamefont {Lovelace}}, \
  and\ \bibinfo {author} {\bibfnamefont {H.~P.}\ \bibnamefont {Pfeiffer}},\
  }\href {\doibase 10.1103/PhysRevD.84.084023} {\bibfield  {journal} {\bibinfo
  {journal} {Phys.\ Rev.\ D}\ }\textbf {\bibinfo {volume} {84}},\ \bibinfo
  {pages} {084023} (\bibinfo {year} {2011})},\ \Eprint
  {http://arxiv.org/abs/1105.3922} {arXiv:1105.3922 [gr-qc]} \BibitemShut
  {NoStop}%
\bibitem [{\citenamefont {Henriksson}\ \emph {et~al.}(2016)\citenamefont
  {Henriksson}, \citenamefont {Foucart}, \citenamefont {Kidder},\ and\
  \citenamefont {Teukolsky}}]{henriksson:2014tba}%
  \BibitemOpen
  \bibfield  {author} {\bibinfo {author} {\bibfnamefont {K.}~\bibnamefont
  {Henriksson}}, \bibinfo {author} {\bibfnamefont {F.}~\bibnamefont {Foucart}},
  \bibinfo {author} {\bibfnamefont {L.~E.}\ \bibnamefont {Kidder}}, \ and\
  \bibinfo {author} {\bibfnamefont {S.~A.}\ \bibnamefont {Teukolsky}},\ }\href
  {\doibase 10.1088/0264-9381/33/10/105009} {\bibfield  {journal} {\bibinfo
  {journal} {Class.\ Quantum Grav.}\ }\textbf {\bibinfo {volume} {33}},\
  \bibinfo {pages} {105009} (\bibinfo {year} {2016})},\ \Eprint
  {http://arxiv.org/abs/1409.7159} {arXiv:1409.7159 [gr-qc]} \BibitemShut
  {NoStop}%
\bibitem [{\citenamefont {Tsokaros}\ \emph {et~al.}(2016)\citenamefont
  {Tsokaros}, \citenamefont {Mundim}, \citenamefont {Galeazzi}, \citenamefont
  {Rezzolla},\ and\ \citenamefont {Ury{\=u}}}]{tsokaros2016initialfixed}%
  \BibitemOpen
  \bibfield  {author} {\bibinfo {author} {\bibfnamefont {A.}~\bibnamefont
  {Tsokaros}}, \bibinfo {author} {\bibfnamefont {B.~C.}\ \bibnamefont
  {Mundim}}, \bibinfo {author} {\bibfnamefont {F.}~\bibnamefont {Galeazzi}},
  \bibinfo {author} {\bibfnamefont {L.}~\bibnamefont {Rezzolla}}, \ and\
  \bibinfo {author} {\bibfnamefont {K.}~\bibnamefont {Ury{\=u}}},\ }\href
  {https://arxiv.org/abs/1605.07205} {\bibfield  {journal} {\bibinfo  {journal}
  {Physical Review D}\ }\textbf {\bibinfo {volume} {94}},\ \bibinfo {pages}
  {044049} (\bibinfo {year} {2016})},\ \Eprint
  {http://arxiv.org/abs/1605.07205} {arXiv:1605.07205 [gr-qc]} \BibitemShut
  {NoStop}%
\bibitem [{\citenamefont {Pfeiffer}\ \emph
  {et~al.}(2003{\natexlab{a}})\citenamefont {Pfeiffer}, \citenamefont {Kidder},
  \citenamefont {Scheel},\ and\ \citenamefont {Teukolsky}}]{pfeiffer2003}%
  \BibitemOpen
  \bibfield  {author} {\bibinfo {author} {\bibfnamefont {H.~P.}\ \bibnamefont
  {Pfeiffer}}, \bibinfo {author} {\bibfnamefont {L.~E.}\ \bibnamefont
  {Kidder}}, \bibinfo {author} {\bibfnamefont {M.~A.}\ \bibnamefont {Scheel}},
  \ and\ \bibinfo {author} {\bibfnamefont {S.~A.}\ \bibnamefont {Teukolsky}},\
  }\href {\doibase 10.1016/S0010-4655(02)00847-0} {\bibfield  {journal}
  {\bibinfo  {journal} {Comput.\ Phys.\ Commun.}\ }\textbf {\bibinfo {volume}
  {152}},\ \bibinfo {pages} {253} (\bibinfo {year} {2003}{\natexlab{a}})},\
  \Eprint {http://arxiv.org/abs/gr-qc/0202096} {gr-qc/0202096} \BibitemShut
  {NoStop}%
\bibitem [{\citenamefont {Sherwin}\ \emph {et~al.}(2006)\citenamefont
  {Sherwin}, \citenamefont {Kirby}, \citenamefont {Peir{\'o}}, \citenamefont
  {Taylor},\ and\ \citenamefont {Zienkiewicz}}]{sherwin20062d}%
  \BibitemOpen
  \bibfield  {author} {\bibinfo {author} {\bibfnamefont {S.}~\bibnamefont
  {Sherwin}}, \bibinfo {author} {\bibfnamefont {R.}~\bibnamefont {Kirby}},
  \bibinfo {author} {\bibfnamefont {J.}~\bibnamefont {Peir{\'o}}}, \bibinfo
  {author} {\bibfnamefont {R.}~\bibnamefont {Taylor}}, \ and\ \bibinfo {author}
  {\bibfnamefont {O.}~\bibnamefont {Zienkiewicz}},\ }\href@noop {} {\bibfield
  {journal} {\bibinfo  {journal} {International journal for numerical methods
  in engineering}\ }\textbf {\bibinfo {volume} {65}},\ \bibinfo {pages} {752}
  (\bibinfo {year} {2006})}\BibitemShut {NoStop}%
\bibitem [{\citenamefont {Di~Pietro}\ and\ \citenamefont
  {Ern}(2011)}]{di2011mathematical}%
  \BibitemOpen
  \bibfield  {author} {\bibinfo {author} {\bibfnamefont {D.~A.}\ \bibnamefont
  {Di~Pietro}}\ and\ \bibinfo {author} {\bibfnamefont {A.}~\bibnamefont
  {Ern}},\ }\href@noop {} {\emph {\bibinfo {title} {Mathematical aspects of
  discontinuous Galerkin methods}}},\ Vol.~\bibinfo {volume} {69}\ (\bibinfo
  {publisher} {Springer Science \&amp; Business Media},\ \bibinfo {year}
  {2011})\BibitemShut {NoStop}%
\bibitem [{\citenamefont {Mengaldo}\ \emph {et~al.}(2015)\citenamefont
  {Mengaldo}, \citenamefont {De~Grazia}, \citenamefont {Moxey}, \citenamefont
  {Vincent},\ and\ \citenamefont {Sherwin}}]{mengaldo2015dealiasing}%
  \BibitemOpen
  \bibfield  {author} {\bibinfo {author} {\bibfnamefont {G.}~\bibnamefont
  {Mengaldo}}, \bibinfo {author} {\bibfnamefont {D.}~\bibnamefont {De~Grazia}},
  \bibinfo {author} {\bibfnamefont {D.}~\bibnamefont {Moxey}}, \bibinfo
  {author} {\bibfnamefont {P.}~\bibnamefont {Vincent}}, \ and\ \bibinfo
  {author} {\bibfnamefont {S.}~\bibnamefont {Sherwin}},\ }\href@noop {}
  {\bibfield  {journal} {\bibinfo  {journal} {Journal of Computational
  Physics}\ }\textbf {\bibinfo {volume} {299}},\ \bibinfo {pages} {56}
  (\bibinfo {year} {2015})}\BibitemShut {NoStop}%
\bibitem [{\citenamefont {Saad}(2003)}]{saad2003iterative}%
  \BibitemOpen
  \bibfield  {author} {\bibinfo {author} {\bibfnamefont {Y.}~\bibnamefont
  {Saad}},\ }\href@noop {} {\emph {\bibinfo {title} {Iterative methods for
  sparse linear systems}}}\ (\bibinfo  {publisher} {Siam},\ \bibinfo {year}
  {2003})\BibitemShut {NoStop}%
\bibitem [{\citenamefont {Notay}(2000)}]{notay2000flexible}%
  \BibitemOpen
  \bibfield  {author} {\bibinfo {author} {\bibfnamefont {Y.}~\bibnamefont
  {Notay}},\ }\href@noop {} {\bibfield  {journal} {\bibinfo  {journal} {SIAM
  Journal on Scientific Computing}\ }\textbf {\bibinfo {volume} {22}},\
  \bibinfo {pages} {1444} (\bibinfo {year} {2000})}\BibitemShut {NoStop}%
\bibitem [{\citenamefont {Antonietti}\ and\ \citenamefont
  {Houston}(2011)}]{antonietti2011class}%
  \BibitemOpen
  \bibfield  {author} {\bibinfo {author} {\bibfnamefont {P.~F.}\ \bibnamefont
  {Antonietti}}\ and\ \bibinfo {author} {\bibfnamefont {P.}~\bibnamefont
  {Houston}},\ }\href@noop {} {\bibfield  {journal} {\bibinfo  {journal}
  {Journal of Scientific Computing}\ }\textbf {\bibinfo {volume} {46}},\
  \bibinfo {pages} {124} (\bibinfo {year} {2011})}\BibitemShut {NoStop}%
\bibitem [{\citenamefont {Briggs}\ \emph {et~al.}(2000)\citenamefont {Briggs},
  \citenamefont {Henson},\ and\ \citenamefont
  {McCormick}}]{briggs2000multigrid}%
  \BibitemOpen
  \bibfield  {author} {\bibinfo {author} {\bibfnamefont {W.}~\bibnamefont
  {Briggs}}, \bibinfo {author} {\bibfnamefont {V.}~\bibnamefont {Henson}}, \
  and\ \bibinfo {author} {\bibfnamefont {S.}~\bibnamefont {McCormick}},\ }\href
  {https://books.google.ca/books?id=oSTGBm64o1AC} {\emph {\bibinfo {title} {A
  Multigrid Tutorial: Second Edition}}}\ (\bibinfo  {publisher} {Society for
  Industrial and Applied Mathematics},\ \bibinfo {year} {2000})\BibitemShut
  {NoStop}%
\bibitem [{\citenamefont {Sampath}\ \emph {et~al.}(2008)\citenamefont
  {Sampath}, \citenamefont {Adavani}, \citenamefont {Sundar}, \citenamefont
  {Lashuk},\ and\ \citenamefont {Biros}}]{sampath2008dendro}%
  \BibitemOpen
  \bibfield  {author} {\bibinfo {author} {\bibfnamefont {R.~S.}\ \bibnamefont
  {Sampath}}, \bibinfo {author} {\bibfnamefont {S.~S.}\ \bibnamefont
  {Adavani}}, \bibinfo {author} {\bibfnamefont {H.}~\bibnamefont {Sundar}},
  \bibinfo {author} {\bibfnamefont {I.}~\bibnamefont {Lashuk}}, \ and\ \bibinfo
  {author} {\bibfnamefont {G.}~\bibnamefont {Biros}},\ }in\ \href@noop {}
  {\emph {\bibinfo {booktitle} {Proceedings of the 2008 ACM/IEEE conference on
  Supercomputing}}}\ (\bibinfo {organization} {IEEE Press},\ \bibinfo {year}
  {2008})\ p.~\bibinfo {pages} {18}\BibitemShut {NoStop}%
\bibitem [{\citenamefont {Sundar}\ \emph {et~al.}(2012)\citenamefont {Sundar},
  \citenamefont {Biros}, \citenamefont {Burstedde}, \citenamefont {Rudi},
  \citenamefont {Ghattas},\ and\ \citenamefont {Stadler}}]{sundar2012parallel}%
  \BibitemOpen
  \bibfield  {author} {\bibinfo {author} {\bibfnamefont {H.}~\bibnamefont
  {Sundar}}, \bibinfo {author} {\bibfnamefont {G.}~\bibnamefont {Biros}},
  \bibinfo {author} {\bibfnamefont {C.}~\bibnamefont {Burstedde}}, \bibinfo
  {author} {\bibfnamefont {J.}~\bibnamefont {Rudi}}, \bibinfo {author}
  {\bibfnamefont {O.}~\bibnamefont {Ghattas}}, \ and\ \bibinfo {author}
  {\bibfnamefont {G.}~\bibnamefont {Stadler}},\ }in\ \href@noop {} {\emph
  {\bibinfo {booktitle} {Proceedings of the International Conference on High
  Performance Computing, Networking, Storage and Analysis}}}\ (\bibinfo
  {organization} {IEEE Computer Society Press},\ \bibinfo {year} {2012})\
  p.~\bibinfo {pages} {43}\BibitemShut {NoStop}%
\bibitem [{\citenamefont {Shang}(2009)}]{shang2009distributed}%
  \BibitemOpen
  \bibfield  {author} {\bibinfo {author} {\bibfnamefont {Y.}~\bibnamefont
  {Shang}},\ }\href@noop {} {\bibfield  {journal} {\bibinfo  {journal}
  {Computers \&amp; Mathematics with Applications}\ }\textbf {\bibinfo {volume}
  {57}},\ \bibinfo {pages} {1369} (\bibinfo {year} {2009})}\BibitemShut
  {NoStop}%
\bibitem [{\citenamefont {Li}\ \emph {et~al.}(2011)\citenamefont {Li},
  \citenamefont {Huang}, \citenamefont {Zhang}, \citenamefont {Liu},\ and\
  \citenamefont {Gu}}]{li2011chebyshev}%
  \BibitemOpen
  \bibfield  {author} {\bibinfo {author} {\bibfnamefont {H.-B.}\ \bibnamefont
  {Li}}, \bibinfo {author} {\bibfnamefont {T.-Z.}\ \bibnamefont {Huang}},
  \bibinfo {author} {\bibfnamefont {Y.}~\bibnamefont {Zhang}}, \bibinfo
  {author} {\bibfnamefont {X.-P.}\ \bibnamefont {Liu}}, \ and\ \bibinfo
  {author} {\bibfnamefont {T.-X.}\ \bibnamefont {Gu}},\ }\href@noop {}
  {\bibfield  {journal} {\bibinfo  {journal} {Applied Mathematics and
  Computation}\ }\textbf {\bibinfo {volume} {218}},\ \bibinfo {pages} {260}
  (\bibinfo {year} {2011})}\BibitemShut {NoStop}%
\bibitem [{\citenamefont {Ghysels}\ \emph {et~al.}(2012)\citenamefont
  {Ghysels}, \citenamefont {K{\l}osiewicz},\ and\ \citenamefont
  {Vanroose}}]{ghysels2012improving}%
  \BibitemOpen
  \bibfield  {author} {\bibinfo {author} {\bibfnamefont {P.}~\bibnamefont
  {Ghysels}}, \bibinfo {author} {\bibfnamefont {P.}~\bibnamefont
  {K{\l}osiewicz}}, \ and\ \bibinfo {author} {\bibfnamefont {W.}~\bibnamefont
  {Vanroose}},\ }\href@noop {} {\bibfield  {journal} {\bibinfo  {journal}
  {Numerical Linear Algebra with Applications}\ }\textbf {\bibinfo {volume}
  {19}},\ \bibinfo {pages} {253} (\bibinfo {year} {2012})}\BibitemShut
  {NoStop}%
\bibitem [{\citenamefont {Scales}(1989)}]{scales1989use}%
  \BibitemOpen
  \bibfield  {author} {\bibinfo {author} {\bibfnamefont {J.~A.}\ \bibnamefont
  {Scales}},\ }\href@noop {} {\bibfield  {journal} {\bibinfo  {journal}
  {Geophysical Journal International}\ }\textbf {\bibinfo {volume} {97}},\
  \bibinfo {pages} {179} (\bibinfo {year} {1989})}\BibitemShut {NoStop}%
\bibitem [{\citenamefont {Bell}(1965)}]{bell1965gershgorin}%
  \BibitemOpen
  \bibfield  {author} {\bibinfo {author} {\bibfnamefont {H.~E.}\ \bibnamefont
  {Bell}},\ }\href@noop {} {\bibfield  {journal} {\bibinfo  {journal} {The
  American Mathematical Monthly}\ }\textbf {\bibinfo {volume} {72}},\ \bibinfo
  {pages} {292} (\bibinfo {year} {1965})}\BibitemShut {NoStop}%
\bibitem [{\citenamefont {Elman}\ \emph {et~al.}(2001)\citenamefont {Elman},
  \citenamefont {Ernst},\ and\ \citenamefont {O'leary}}]{elman2001multigrid}%
  \BibitemOpen
  \bibfield  {author} {\bibinfo {author} {\bibfnamefont {H.~C.}\ \bibnamefont
  {Elman}}, \bibinfo {author} {\bibfnamefont {O.~G.}\ \bibnamefont {Ernst}}, \
  and\ \bibinfo {author} {\bibfnamefont {D.~P.}\ \bibnamefont {O'leary}},\
  }\href@noop {} {\bibfield  {journal} {\bibinfo  {journal} {SIAM Journal on
  scientific computing}\ }\textbf {\bibinfo {volume} {23}},\ \bibinfo {pages}
  {1291} (\bibinfo {year} {2001})}\BibitemShut {NoStop}%
\bibitem [{\citenamefont {Houston}\ \emph {et~al.}(2007)\citenamefont
  {Houston}, \citenamefont {Sch{\"o}tzau},\ and\ \citenamefont
  {Wihler}}]{houston2007energy}%
  \BibitemOpen
  \bibfield  {author} {\bibinfo {author} {\bibfnamefont {P.}~\bibnamefont
  {Houston}}, \bibinfo {author} {\bibfnamefont {D.}~\bibnamefont
  {Sch{\"o}tzau}}, \ and\ \bibinfo {author} {\bibfnamefont {T.~P.}\
  \bibnamefont {Wihler}},\ }\href@noop {} {\bibfield  {journal} {\bibinfo
  {journal} {Mathematical Models and Methods in Applied Sciences}\ }\textbf
  {\bibinfo {volume} {17}},\ \bibinfo {pages} {33} (\bibinfo {year}
  {2007})}\BibitemShut {NoStop}%
\bibitem [{\citenamefont {Houston}\ \emph {et~al.}(2008)\citenamefont
  {Houston}, \citenamefont {S{\"u}li},\ and\ \citenamefont
  {Wihler}}]{houston2008posteriori}%
  \BibitemOpen
  \bibfield  {author} {\bibinfo {author} {\bibfnamefont {P.}~\bibnamefont
  {Houston}}, \bibinfo {author} {\bibfnamefont {E.}~\bibnamefont {S{\"u}li}}, \
  and\ \bibinfo {author} {\bibfnamefont {T.~P.}\ \bibnamefont {Wihler}},\
  }\href@noop {} {\bibfield  {journal} {\bibinfo  {journal} {IMA journal of
  numerical analysis}\ }\textbf {\bibinfo {volume} {28}},\ \bibinfo {pages}
  {245} (\bibinfo {year} {2008})}\BibitemShut {NoStop}%
\bibitem [{\citenamefont {Houston}\ \emph
  {et~al.}(2005{\natexlab{a}})\citenamefont {Houston}, \citenamefont {Robson},\
  and\ \citenamefont {S{\"u}li}}]{houston2005discontinuous}%
  \BibitemOpen
  \bibfield  {author} {\bibinfo {author} {\bibfnamefont {P.}~\bibnamefont
  {Houston}}, \bibinfo {author} {\bibfnamefont {J.}~\bibnamefont {Robson}}, \
  and\ \bibinfo {author} {\bibfnamefont {E.}~\bibnamefont {S{\"u}li}},\
  }\href@noop {} {\bibfield  {journal} {\bibinfo  {journal} {IMA journal of
  numerical analysis}\ }\textbf {\bibinfo {volume} {25}},\ \bibinfo {pages}
  {726} (\bibinfo {year} {2005}{\natexlab{a}})}\BibitemShut {NoStop}%
\bibitem [{\citenamefont {Bi}\ \emph {et~al.}(2015)\citenamefont {Bi},
  \citenamefont {Wang},\ and\ \citenamefont {Lin}}]{bi2015posteriori}%
  \BibitemOpen
  \bibfield  {author} {\bibinfo {author} {\bibfnamefont {C.}~\bibnamefont
  {Bi}}, \bibinfo {author} {\bibfnamefont {C.}~\bibnamefont {Wang}}, \ and\
  \bibinfo {author} {\bibfnamefont {Y.}~\bibnamefont {Lin}},\ }\href@noop {}
  {\bibfield  {journal} {\bibinfo  {journal} {Computer Methods in Applied
  Mechanics and Engineering}\ }\textbf {\bibinfo {volume} {297}},\ \bibinfo
  {pages} {140} (\bibinfo {year} {2015})}\BibitemShut {NoStop}%
\bibitem [{\citenamefont {Sch{\"o}tzau}\ \emph {et~al.}(2014)\citenamefont
  {Sch{\"o}tzau}, \citenamefont {Schwab}, \citenamefont {Wihler},\ and\
  \citenamefont {Wirz}}]{schotzau2014exponential}%
  \BibitemOpen
  \bibfield  {author} {\bibinfo {author} {\bibfnamefont {D.}~\bibnamefont
  {Sch{\"o}tzau}}, \bibinfo {author} {\bibfnamefont {C.}~\bibnamefont
  {Schwab}}, \bibinfo {author} {\bibfnamefont {T.}~\bibnamefont {Wihler}}, \
  and\ \bibinfo {author} {\bibfnamefont {M.}~\bibnamefont {Wirz}},\ }\href@noop
  {} {\emph {\bibinfo {title} {Exponential convergence of hp-DGFEM for elliptic
  problems in polyhedral domains}}}\ (\bibinfo  {publisher} {Springer},\
  \bibinfo {year} {2014})\BibitemShut {NoStop}%
\bibitem [{\citenamefont {Houston}\ \emph
  {et~al.}(2005{\natexlab{b}})\citenamefont {Houston}, \citenamefont
  {Sch\"{o}tzau},\ and\ \citenamefont {Wihler}}]{houstonschotzau05}%
  \BibitemOpen
  \bibfield  {author} {\bibinfo {author} {\bibfnamefont {P.}~\bibnamefont
  {Houston}}, \bibinfo {author} {\bibfnamefont {D.}~\bibnamefont
  {Sch\"{o}tzau}}, \ and\ \bibinfo {author} {\bibfnamefont {T.}~\bibnamefont
  {Wihler}},\ }\href@noop {} {\bibfield  {journal} {\bibinfo  {journal} {J.
  Sci. Comput.}\ }\textbf {\bibinfo {volume} {22}},\ \bibinfo {pages} {347}
  (\bibinfo {year} {2005}{\natexlab{b}})}\BibitemShut {NoStop}%
\bibitem [{\citenamefont {Hansbo}\ and\ \citenamefont
  {Larson}(2011)}]{hansbo2011energy}%
  \BibitemOpen
  \bibfield  {author} {\bibinfo {author} {\bibfnamefont {P.}~\bibnamefont
  {Hansbo}}\ and\ \bibinfo {author} {\bibfnamefont {M.~G.}\ \bibnamefont
  {Larson}},\ }\href@noop {} {\bibfield  {journal} {\bibinfo  {journal}
  {Computer Methods in Applied Mechanics and Engineering}\ }\textbf {\bibinfo
  {volume} {200}},\ \bibinfo {pages} {3026} (\bibinfo {year}
  {2011})}\BibitemShut {NoStop}%
\bibitem [{\citenamefont {Zhu}\ \emph {et~al.}(2011)\citenamefont {Zhu},
  \citenamefont {Giani}, \citenamefont {Houston},\ and\ \citenamefont
  {Sch{\"o}tzau}}]{zhu2011energy}%
  \BibitemOpen
  \bibfield  {author} {\bibinfo {author} {\bibfnamefont {L.}~\bibnamefont
  {Zhu}}, \bibinfo {author} {\bibfnamefont {S.}~\bibnamefont {Giani}}, \bibinfo
  {author} {\bibfnamefont {P.}~\bibnamefont {Houston}}, \ and\ \bibinfo
  {author} {\bibfnamefont {D.}~\bibnamefont {Sch{\"o}tzau}},\ }\href@noop {}
  {\bibfield  {journal} {\bibinfo  {journal} {Mathematical Models and Methods
  in Applied Sciences}\ }\textbf {\bibinfo {volume} {21}},\ \bibinfo {pages}
  {267} (\bibinfo {year} {2011})}\BibitemShut {NoStop}%
\bibitem [{\citenamefont {Sch{\"o}tzau}\ and\ \citenamefont
  {Zhu}(2009)}]{schotzau.d;zhu.l2009}%
  \BibitemOpen
  \bibfield  {author} {\bibinfo {author} {\bibfnamefont {D.}~\bibnamefont
  {Sch{\"o}tzau}}\ and\ \bibinfo {author} {\bibfnamefont {L.}~\bibnamefont
  {Zhu}},\ }\href@noop {} {\bibfield  {journal} {\bibinfo  {journal} {Applied
  Numerical Mathematics}\ }\textbf {\bibinfo {volume} {59}},\ \bibinfo {pages}
  {2236} (\bibinfo {year} {2009})}\BibitemShut {NoStop}%
\bibitem [{\citenamefont {Houston}\ \emph
  {et~al.}(2005{\natexlab{c}})\citenamefont {Houston}, \citenamefont
  {Perugia},\ and\ \citenamefont {Sch\"{o}tzau}}]{houstonperugia05}%
  \BibitemOpen
  \bibfield  {author} {\bibinfo {author} {\bibfnamefont {P.}~\bibnamefont
  {Houston}}, \bibinfo {author} {\bibfnamefont {I.}~\bibnamefont {Perugia}}, \
  and\ \bibinfo {author} {\bibfnamefont {D.}~\bibnamefont {Sch\"{o}tzau}},\
  }\href@noop {} {\bibfield  {journal} {\bibinfo  {journal} {Comput. Methods
  Appl. Mech. Eng.}\ }\textbf {\bibinfo {volume} {194}},\ \bibinfo {pages}
  {499} (\bibinfo {year} {2005}{\natexlab{c}})}\BibitemShut {NoStop}%
\bibitem [{\citenamefont {Lovadina}\ and\ \citenamefont
  {Marini}(2009)}]{lovadina.c;marini.l2009}%
  \BibitemOpen
  \bibfield  {author} {\bibinfo {author} {\bibfnamefont {C.}~\bibnamefont
  {Lovadina}}\ and\ \bibinfo {author} {\bibfnamefont {L.}~\bibnamefont
  {Marini}},\ }\href@noop {} {\bibfield  {journal} {\bibinfo  {journal} {J.
  Sci. Comput.}\ }\textbf {\bibinfo {volume} {40}},\ \bibinfo {pages} {340}
  (\bibinfo {year} {2009})}\BibitemShut {NoStop}%
\bibitem [{\citenamefont {Mitchell}\ and\ \citenamefont
  {McClain}(2011)}]{mitchell2011survey}%
  \BibitemOpen
  \bibfield  {author} {\bibinfo {author} {\bibfnamefont {W.~F.}\ \bibnamefont
  {Mitchell}}\ and\ \bibinfo {author} {\bibfnamefont {M.~A.}\ \bibnamefont
  {McClain}},\ }in\ \href@noop {} {\emph {\bibinfo {booktitle} {Recent advances
  in computational and applied mathematics}}}\ (\bibinfo  {publisher}
  {Springer},\ \bibinfo {year} {2011})\ pp.\ \bibinfo {pages}
  {227--258}\BibitemShut {NoStop}%
\bibitem [{\citenamefont {Melenk}\ and\ \citenamefont
  {Wohlmuth}(2001)}]{melenk2001residual}%
  \BibitemOpen
  \bibfield  {author} {\bibinfo {author} {\bibfnamefont {J.~M.}\ \bibnamefont
  {Melenk}}\ and\ \bibinfo {author} {\bibfnamefont {B.~I.}\ \bibnamefont
  {Wohlmuth}},\ }\href@noop {} {\bibfield  {journal} {\bibinfo  {journal}
  {Advances in Computational Mathematics}\ }\textbf {\bibinfo {volume} {15}},\
  \bibinfo {pages} {311} (\bibinfo {year} {2001})}\BibitemShut {NoStop}%
\bibitem [{\citenamefont {Feng}\ \emph {et~al.}(2015)\citenamefont {Feng},
  \citenamefont {Straka}, \citenamefont {Di~Matteo},\ and\ \citenamefont
  {Croft}}]{feng2015mp}%
  \BibitemOpen
  \bibfield  {author} {\bibinfo {author} {\bibfnamefont {Y.}~\bibnamefont
  {Feng}}, \bibinfo {author} {\bibfnamefont {M.}~\bibnamefont {Straka}},
  \bibinfo {author} {\bibfnamefont {T.}~\bibnamefont {Di~Matteo}}, \ and\
  \bibinfo {author} {\bibfnamefont {R.}~\bibnamefont {Croft}},\ }\href@noop {}
  {\  (\bibinfo {year} {2015})}\BibitemShut {NoStop}%
\bibitem [{\citenamefont {Buis}\ and\ \citenamefont
  {Dyksen}(1996)}]{buis1996efficient}%
  \BibitemOpen
  \bibfield  {author} {\bibinfo {author} {\bibfnamefont {P.~E.}\ \bibnamefont
  {Buis}}\ and\ \bibinfo {author} {\bibfnamefont {W.~R.}\ \bibnamefont
  {Dyksen}},\ }\href@noop {} {\bibfield  {journal} {\bibinfo  {journal} {ACM
  Transactions on Mathematical Software (TOMS)}\ }\textbf {\bibinfo {volume}
  {22}},\ \bibinfo {pages} {18} (\bibinfo {year} {1996})}\BibitemShut {NoStop}%
\bibitem [{\citenamefont {Burstedde}\ \emph {et~al.}(2011)\citenamefont
  {Burstedde}, \citenamefont {Wilcox},\ and\ \citenamefont
  {Ghattas}}]{burstedde2011p4est}%
  \BibitemOpen
  \bibfield  {author} {\bibinfo {author} {\bibfnamefont {C.}~\bibnamefont
  {Burstedde}}, \bibinfo {author} {\bibfnamefont {L.~C.}\ \bibnamefont
  {Wilcox}}, \ and\ \bibinfo {author} {\bibfnamefont {O.}~\bibnamefont
  {Ghattas}},\ }\href@noop {} {\bibfield  {journal} {\bibinfo  {journal} {SIAM
  Journal on Scientific Computing}\ }\textbf {\bibinfo {volume} {33}},\
  \bibinfo {pages} {1103} (\bibinfo {year} {2011})}\BibitemShut {NoStop}%
\bibitem [{\citenamefont {Balay}\ \emph {et~al.}(2001)\citenamefont {Balay},
  \citenamefont {Buschelman}, \citenamefont {Gropp}, \citenamefont {Kaushik},
  \citenamefont {McInnes},\ and\ \citenamefont {Smith}}]{petsc_home_page}%
  \BibitemOpen
  \bibfield  {author} {\bibinfo {author} {\bibfnamefont {S.}~\bibnamefont
  {Balay}}, \bibinfo {author} {\bibfnamefont {K.}~\bibnamefont {Buschelman}},
  \bibinfo {author} {\bibfnamefont {W.~D.}\ \bibnamefont {Gropp}}, \bibinfo
  {author} {\bibfnamefont {D.}~\bibnamefont {Kaushik}}, \bibinfo {author}
  {\bibfnamefont {L.~C.}\ \bibnamefont {McInnes}}, \ and\ \bibinfo {author}
  {\bibfnamefont {B.~F.}\ \bibnamefont {Smith}},\ }\href@noop {} {\enquote
  {\bibinfo {title} {{PETSc} home page},}\ } (\bibinfo {year} {2001}),\
  \bibinfo {note} {\url{http://www.mcs.anl.gov/petsc}}\BibitemShut {NoStop}%
\bibitem [{\citenamefont {Stamm}\ and\ \citenamefont
  {Wihler}(2010)}]{stamm2010hp}%
  \BibitemOpen
  \bibfield  {author} {\bibinfo {author} {\bibfnamefont {B.}~\bibnamefont
  {Stamm}}\ and\ \bibinfo {author} {\bibfnamefont {T.}~\bibnamefont {Wihler}},\
  }\href@noop {} {\bibfield  {journal} {\bibinfo  {journal} {Mathematics of
  Computation}\ }\textbf {\bibinfo {volume} {79}},\ \bibinfo {pages} {2117}
  (\bibinfo {year} {2010})}\BibitemShut {NoStop}%
\bibitem [{\citenamefont {Baumgarte}\ \emph {et~al.}(2007)\citenamefont
  {Baumgarte}, \citenamefont {O'Murchadha},\ and\ \citenamefont
  {Pfeiffer}}]{baumgarte2007}%
  \BibitemOpen
  \bibfield  {author} {\bibinfo {author} {\bibfnamefont {T.~W.}\ \bibnamefont
  {Baumgarte}}, \bibinfo {author} {\bibfnamefont {N.}~\bibnamefont
  {O'Murchadha}}, \ and\ \bibinfo {author} {\bibfnamefont {H.~P.}\ \bibnamefont
  {Pfeiffer}},\ }\href@noop {} {\bibfield  {journal} {\bibinfo  {journal}
  {Phys.\ Rev.\ D}\ }\textbf {\bibinfo {volume} {75}},\ \bibinfo {pages}
  {044009} (\bibinfo {year} {2007})},\ \Eprint
  {http://arxiv.org/abs/gr-qc/0610120} {gr-qc/0610120} \BibitemShut {NoStop}%
\bibitem [{\citenamefont {Cook}(1994)}]{cook1994}%
  \BibitemOpen
  \bibfield  {author} {\bibinfo {author} {\bibfnamefont {G.~B.}\ \bibnamefont
  {Cook}},\ }\href@noop {} {\bibfield  {journal} {\bibinfo  {journal} {Phys.\
  Rev.\ D}\ }\textbf {\bibinfo {volume} {50}},\ \bibinfo {pages} {5025}
  (\bibinfo {year} {1994})},\ \Eprint {http://arxiv.org/abs/gr-qc/9404043}
  {gr-qc/9404043} \BibitemShut {NoStop}%
\bibitem [{\citenamefont {Cook}\ and\ \citenamefont
  {Pfeiffer}(2004)}]{cook2004excision}%
  \BibitemOpen
  \bibfield  {author} {\bibinfo {author} {\bibfnamefont {G.~B.}\ \bibnamefont
  {Cook}}\ and\ \bibinfo {author} {\bibfnamefont {H.~P.}\ \bibnamefont
  {Pfeiffer}},\ }\href@noop {} {\bibfield  {journal} {\bibinfo  {journal}
  {Physical Review D}\ }\textbf {\bibinfo {volume} {70}},\ \bibinfo {pages}
  {104016} (\bibinfo {year} {2004})},\ \Eprint
  {http://arxiv.org/abs/gr-qc/0407078} {gr-qc/0407078} \BibitemShut {NoStop}%
\bibitem [{\citenamefont {Caudill}\ \emph {et~al.}(2006)\citenamefont
  {Caudill}, \citenamefont {Cook}, \citenamefont {Grigsby},\ and\ \citenamefont
  {Pfeiffer}}]{caudill2006circular}%
  \BibitemOpen
  \bibfield  {author} {\bibinfo {author} {\bibfnamefont {M.}~\bibnamefont
  {Caudill}}, \bibinfo {author} {\bibfnamefont {G.~B.}\ \bibnamefont {Cook}},
  \bibinfo {author} {\bibfnamefont {J.~D.}\ \bibnamefont {Grigsby}}, \ and\
  \bibinfo {author} {\bibfnamefont {H.~P.}\ \bibnamefont {Pfeiffer}},\ }\href
  {\doibase 10.1103/PhysRevD.74.064011} {\bibfield  {journal} {\bibinfo
  {journal} {Physical Review D}\ }\textbf {\bibinfo {volume} {74}},\ \bibinfo
  {pages} {064011} (\bibinfo {year} {2006})},\ \Eprint
  {http://arxiv.org/abs/1410.7431} {arXiv:1410.7431 [physics.comp-ph]}
  \BibitemShut {NoStop}%
\bibitem [{\citenamefont {Pfeiffer}\ \emph
  {et~al.}(2003{\natexlab{b}})\citenamefont {Pfeiffer}, \citenamefont {Kidder},
  \citenamefont {Scheel},\ and\ \citenamefont
  {Teukolsky}}]{pfeiffer2003multidomain}%
  \BibitemOpen
  \bibfield  {author} {\bibinfo {author} {\bibfnamefont {H.~P.}\ \bibnamefont
  {Pfeiffer}}, \bibinfo {author} {\bibfnamefont {L.~E.}\ \bibnamefont
  {Kidder}}, \bibinfo {author} {\bibfnamefont {M.~A.}\ \bibnamefont {Scheel}},
  \ and\ \bibinfo {author} {\bibfnamefont {S.~A.}\ \bibnamefont {Teukolsky}},\
  }\href@noop {} {\bibfield  {journal} {\bibinfo  {journal} {Computer physics
  communications}\ }\textbf {\bibinfo {volume} {152}},\ \bibinfo {pages} {253}
  (\bibinfo {year} {2003}{\natexlab{b}})},\ \Eprint
  {http://arxiv.org/abs/gr-qc/0202096} {gr-qc/0202096} \BibitemShut {NoStop}%
\bibitem [{\citenamefont {Brandt}\ and\ \citenamefont
  {Br{\"u}gmann}(1997)}]{brandt1997simple}%
  \BibitemOpen
  \bibfield  {author} {\bibinfo {author} {\bibfnamefont {S.}~\bibnamefont
  {Brandt}}\ and\ \bibinfo {author} {\bibfnamefont {B.}~\bibnamefont
  {Br{\"u}gmann}},\ }\href@noop {} {\bibfield  {journal} {\bibinfo  {journal}
  {Physical Review Letters}\ }\textbf {\bibinfo {volume} {78}},\ \bibinfo
  {pages} {3606} (\bibinfo {year} {1997})},\ \Eprint
  {http://arxiv.org/abs/gr-qc/9703066} {gr-qc/9703066} \BibitemShut {NoStop}%
\bibitem [{\citenamefont {Ansorg}\ \emph {et~al.}(2004)\citenamefont {Ansorg},
  \citenamefont {Br{\"u}gmann},\ and\ \citenamefont
  {Tichy}}]{ansorg2004single}%
  \BibitemOpen
  \bibfield  {author} {\bibinfo {author} {\bibfnamefont {M.}~\bibnamefont
  {Ansorg}}, \bibinfo {author} {\bibfnamefont {B.}~\bibnamefont
  {Br{\"u}gmann}}, \ and\ \bibinfo {author} {\bibfnamefont {W.}~\bibnamefont
  {Tichy}},\ }\href@noop {} {\bibfield  {journal} {\bibinfo  {journal}
  {Physical Review D}\ }\textbf {\bibinfo {volume} {70}},\ \bibinfo {pages}
  {064011} (\bibinfo {year} {2004})},\ \Eprint
  {http://arxiv.org/abs/gr-qc/0404056} {gr-qc/0404056} \BibitemShut {NoStop}%
\bibitem [{\citenamefont {Ansorg}(2007)}]{ansorg2007}%
  \BibitemOpen
  \bibfield  {author} {\bibinfo {author} {\bibfnamefont {M.}~\bibnamefont
  {Ansorg}},\ }\href@noop {} {\bibfield  {journal} {\bibinfo  {journal}
  {Class.\ Quantum Grav.}\ }\textbf {\bibinfo {volume} {24}},\ \bibinfo {pages}
  {S1} (\bibinfo {year} {2007})},\ \Eprint {http://arxiv.org/abs/gr-qc/0612081}
  {gr-qc/0612081} \BibitemShut {NoStop}%
\bibitem [{\citenamefont {Dennison}\ \emph {et~al.}(2006)\citenamefont
  {Dennison}, \citenamefont {Baumgarte},\ and\ \citenamefont
  {Pfeiffer}}]{dennison2006}%
  \BibitemOpen
  \bibfield  {author} {\bibinfo {author} {\bibfnamefont {K.~A.}\ \bibnamefont
  {Dennison}}, \bibinfo {author} {\bibfnamefont {T.~W.}\ \bibnamefont
  {Baumgarte}}, \ and\ \bibinfo {author} {\bibfnamefont {H.~P.}\ \bibnamefont
  {Pfeiffer}},\ }\href@noop {} {\bibfield  {journal} {\bibinfo  {journal}
  {Phys.\ Rev.\ D}\ }\textbf {\bibinfo {volume} {74}},\ \bibinfo {pages}
  {064016} (\bibinfo {year} {2006})},\ \Eprint
  {http://arxiv.org/abs/gr-qc/0606037} {gr-qc/0606037} \BibitemShut {NoStop}%
\bibitem [{\citenamefont {Lovelace}\ \emph {et~al.}(2008)\citenamefont
  {Lovelace}, \citenamefont {Owen}, \citenamefont {Pfeiffer},\ and\
  \citenamefont {Chu}}]{lovelace2008}%
  \BibitemOpen
  \bibfield  {author} {\bibinfo {author} {\bibfnamefont {G.}~\bibnamefont
  {Lovelace}}, \bibinfo {author} {\bibfnamefont {R.}~\bibnamefont {Owen}},
  \bibinfo {author} {\bibfnamefont {H.~P.}\ \bibnamefont {Pfeiffer}}, \ and\
  \bibinfo {author} {\bibfnamefont {T.}~\bibnamefont {Chu}},\ }\href {\doibase
  10.1103/PhysRevD.78.084017} {\bibfield  {journal} {\bibinfo  {journal} {Phys.
  Rev.}\ }\textbf {\bibinfo {volume} {D78}},\ \bibinfo {pages} {084017}
  (\bibinfo {year} {2008})},\ \Eprint {http://arxiv.org/abs/0805.4192}
  {arXiv:0805.4192 [gr-qc]} \BibitemShut {NoStop}%
\bibitem [{\citenamefont {Fabien}\ \emph {et~al.}(2019)\citenamefont {Fabien},
  \citenamefont {Knepley}, \citenamefont {Mills},\ and\ \citenamefont
  {Riviere}}]{fabien2019manycore}%
  \BibitemOpen
  \bibfield  {author} {\bibinfo {author} {\bibfnamefont {M.~S.}\ \bibnamefont
  {Fabien}}, \bibinfo {author} {\bibfnamefont {M.~G.}\ \bibnamefont {Knepley}},
  \bibinfo {author} {\bibfnamefont {R.~T.}\ \bibnamefont {Mills}}, \ and\
  \bibinfo {author} {\bibfnamefont {B.~M.}\ \bibnamefont {Riviere}},\
  }\href@noop {} {\bibfield  {journal} {\bibinfo  {journal} {SIAM Journal on
  Scientific Computing}\ }\textbf {\bibinfo {volume} {41}},\ \bibinfo {pages}
  {C73} (\bibinfo {year} {2019})}\BibitemShut {NoStop}%
\bibitem [{\citenamefont {Muralikrishnan}\ \emph {et~al.}(2019)\citenamefont
  {Muralikrishnan}, \citenamefont {Bui-Thanh},\ and\ \citenamefont
  {Shadid}}]{muralikrishnan2019multilevel}%
  \BibitemOpen
  \bibfield  {author} {\bibinfo {author} {\bibfnamefont {S.}~\bibnamefont
  {Muralikrishnan}}, \bibinfo {author} {\bibfnamefont {T.}~\bibnamefont
  {Bui-Thanh}}, \ and\ \bibinfo {author} {\bibfnamefont {J.~N.}\ \bibnamefont
  {Shadid}},\ }\href {https://arxiv.org/abs/1903.11045} {\bibfield  {journal}
  {\bibinfo  {journal} {arXiv preprint}\ } (\bibinfo {year} {2019})},\ \Eprint
  {http://arxiv.org/abs/1903.11045} {arXiv:1903.11045 [math.NA]} \BibitemShut
  {NoStop}%
\bibitem [{\citenamefont {Loken}\ \emph {et~al.}(2010)\citenamefont {Loken},
  \citenamefont {Gruner}, \citenamefont {Groer}, \citenamefont {Peltier},
  \citenamefont {Bunn}, \citenamefont {Craig}, \citenamefont {Henriques},
  \citenamefont {Dempsey}, \citenamefont {Yu}, \citenamefont {Chen},
  \citenamefont {Dursi}, \citenamefont {Chong}, \citenamefont {Northrup},
  \citenamefont {Pinto}, \citenamefont {Knecht},\ and\ \citenamefont
  {Zon}}]{scinet}%
  \BibitemOpen
  \bibfield  {author} {\bibinfo {author} {\bibfnamefont {C.}~\bibnamefont
  {Loken}}, \bibinfo {author} {\bibfnamefont {D.}~\bibnamefont {Gruner}},
  \bibinfo {author} {\bibfnamefont {L.}~\bibnamefont {Groer}}, \bibinfo
  {author} {\bibfnamefont {R.}~\bibnamefont {Peltier}}, \bibinfo {author}
  {\bibfnamefont {N.}~\bibnamefont {Bunn}}, \bibinfo {author} {\bibfnamefont
  {M.}~\bibnamefont {Craig}}, \bibinfo {author} {\bibfnamefont
  {T.}~\bibnamefont {Henriques}}, \bibinfo {author} {\bibfnamefont
  {J.}~\bibnamefont {Dempsey}}, \bibinfo {author} {\bibfnamefont {C.-H.}\
  \bibnamefont {Yu}}, \bibinfo {author} {\bibfnamefont {J.}~\bibnamefont
  {Chen}}, \bibinfo {author} {\bibfnamefont {L.~J.}\ \bibnamefont {Dursi}},
  \bibinfo {author} {\bibfnamefont {J.}~\bibnamefont {Chong}}, \bibinfo
  {author} {\bibfnamefont {S.}~\bibnamefont {Northrup}}, \bibinfo {author}
  {\bibfnamefont {J.}~\bibnamefont {Pinto}}, \bibinfo {author} {\bibfnamefont
  {N.}~\bibnamefont {Knecht}}, \ and\ \bibinfo {author} {\bibfnamefont {R.~V.}\
  \bibnamefont {Zon}},\ }\href {\doibase 10.1088/1742-6596/256/1/012026}
  {\bibfield  {journal} {\bibinfo  {journal} {J. Phys.: Conf. Ser.}\ }\textbf
  {\bibinfo {volume} {256}},\ \bibinfo {pages} {012026} (\bibinfo {year}
  {2010})}\BibitemShut {NoStop}%
\end{thebibliography}%

\end{document}